\documentclass[twocolumn, nofootinbib, showpacs, superscriptaddress]{revtex4-1} 
\usepackage{amsfonts}
\usepackage{amsmath}
\usepackage{dsfont}
\usepackage[pdftex]{graphicx}
\usepackage{epstopdf}
\usepackage{ulem}
\usepackage{textcomp} 
\usepackage{wasysym} 
\usepackage{placeins}
\usepackage{graphicx}
\usepackage{bm}
\usepackage[varg]{txfonts} 
\usepackage{esint}
\usepackage{xcolor}
\usepackage{cancel}
\usepackage{comment}
\usepackage{empheq}
\usepackage{collectbox}
\usepackage{hyperref}
\hypersetup{
    colorlinks=true,
    linkcolor=red,
    citecolor=blue,
} 
\specialcomment{supplement}{\begingroup\color{cyan}\footnotesize}{\endgroup}
\specialcomment{formula}{\begingroup\footnotesize}{\endgroup}
\excludecomment{details}
\usepackage{accents}

\newcommand{\NEW}[1]{#1}
\renewcommand{\H}[0]{{\textsc{h}}}
\newcommand{\W}[0]{{\textsc{w}}}
\usepackage{pifont}
\newcommand{\cmark}{\ding{51}}%
\newcommand{\xmark}{\ding{55}}%
\newcommand{\sigu}{{\sigma_{\! u}}}
\newcommand{\sigx}{{\sigma_{\! x}}}
\newcommand{\thbar}{\tilde\hbar}

\renewcommand{\d}[0]{{\rm{d}}}
\renewcommand{\c}[0]{{\rm{c}}}
\newcommand{\del}[0]{\partial }

\newcommand{\vol}[2]{\hspace{-0.8mm}\mbox{$\text{d}^{\hspace{-0.0mm}#1}$}\hspace{-0.2mm}#2\hspace{0.8mm}\ }
\renewcommand{\v}[1]{\bm{#1} }
\newcommand{\vx}[0]{\bm{x} }
\newcommand{\vX}[0]{\bm{X} }
\newcommand{\vq}[0]{\bm{q} }
\newcommand{\vQ}[0]{\bm{Q} }

\newcommand{\vu}[0]{\bm{u} }
\newcommand{\vU}[0]{\bm{U} }
\newcommand{\vnabla}[0]{\bm{\nabla} }
\newcommand{\npsi}{n_\psi}
\newcommand{\vupsi}{\vu_\psi}



\definecolor{myblue}{rgb}{.9, .94, 1}

\newcommand*\mybluebox[1]{%
{\setlength{\fboxsep}{6pt}\colorbox{myblue}{#1}}}

\newcommand*\myotherbluebox[1]{%
{\setlength{\fboxsep}{1pt}\colorbox{myblue}{#1}}}

\makeatletter
\g@addto@macro\bfseries{\boldmath}
\makeatother

\begin{document}
\title{Solving the Vlasov equation in two spatial dimensions with the Schr\"odinger method} 
\author{Michael Kopp} 
\email{kopp@fzu.cz}
\affiliation{Department of Physics, University of Cyprus, 1, Panepistimiou Street,
2109, Aglantzia, Cyprus}
\affiliation{CEICO, Institute of Physics of the Czech Academy of Sciences, Na Slovance 2, 18221 Praha 8
Czech Republic}
\author{Kyriakos Vattis} 
\email{kyriakos\_vattis@brown.edu}
\affiliation{Department of Physics, University of Cyprus, 1, Panepistimiou Street,
2109, Aglantzia, Cyprus}
\affiliation{Department of Physics, Brown University, 182 Hope St., Providence, RI 02912, USA}
\author{Constantinos Skordis} 
\email{skordis@fzu.cz}
\affiliation{Department of Physics, University of Cyprus, 1, Panepistimiou Street,
2109, Aglantzia, Cyprus}
\affiliation{CEICO, Institute of Physics of the Czech Academy of Sciences, Na Slovance 2, 18221 Praha 8
Czech Republic}
\date{\today}

\begin{abstract}
We demonstrate that the Vlasov equation describing collisionless self-gravitating matter
 may be solved with the so-called Schr\"odinger method (ScM).  
With the ScM, one solves the Schr\"odinger-Poisson system of equations for a complex wave function in $d$ dimensions, 
rather than the Vlasov equation for a $2d$-dimensional phase space density.
The ScM also allows calculating the $d$-dimensional cumulants directly through quasi-local manipulations of the 
wave function, avoiding the complexity of $2d$-dimensional phase space.

We perform for the first time a quantitative comparison of the ScM and a conventional Vlasov solver in $d=2$ dimensions.
Our numerical tests were carried out using two types of cold cosmological initial conditions: the classic collapse of a sine wave 
and those of a gaussian random field as commonly used in cosmological cold dark matter N-body simulations. 
 We compare the first three cumulants, that is, the density, velocity and velocity dispersion, 
 to those obtained by solving the Vlasov equation using the publicly available code \texttt{ColDICE}. 
 We find excellent qualitative and quantitative agreement between these codes, demonstrating the feasibility 
and advantages of the ScM as an alternative to N-body simulations.
We discuss, the emergence of effective vorticity in the ScM through the winding number around the points where the wave function vanishes.
As an application we evaluate the background pressure induced by the non-linearity of large scale structure formation, thereby estimating the magnitude of cosmological backreaction. We find that it is negligibly small and has time dependence 
and magnitude compatible with expectations from the effective field theory of large scale structure. 
\end{abstract}

\maketitle
{\small 
\tableofcontents
}

\maketitle
\section{Introduction}

Cold dark matter (CDM)  is one of the necessary ingredients of the concordance model of modern cosmology, the  $\Lambda$CDM model.
The CDM model explains the amount of Large-Scale Structure (LSS) in the Universe 
and the formation of halos where galaxies develop and evolve. 
According to the $\Lambda$CDM model initially small density perturbations evolve into bound structures, for instance CDM halos, 
that are themselves distributed within the cosmic web composed of superclusters, filaments and sheets \cite{P80,SWF05,TSM13}. 
The morphology of the cosmic web and the clustering of CDM depends sensitively on the cosmological parameters.
Thus, accurate modeling and theoretical understanding of the CDM dynamics is required to infer these parameters from observations.

The microphysical nature of CDM is still unknown and it is generally assumed to be composed of a particle species (or perhaps many)
 which otherwise remains undetected  at the present time.
In this article we are interested in the modeling of the dynamics of self-gravitating collisionless (= dark)  matter (DM), 
 that may be used to describe a wide variety of particle candidates. In particular, we are interested in cold DM (CDM), 
where ``cold'' from the point of view of LSS formation simply means that  a smooth density $n(\vx)$ and velocity $\vu(\vx)$ field suffice to describe the initial DM matter phase space distribution $f(\vx,\v{p})$. In other words, $f(\vx,\v{p})$ is non-zero only on a three-dimensional sheet defined by $\v{p}=m\v{u}(\vx)$, and moreover, $\vu(\vx)$ is single valued.
This situation is usually referred to as the single stream regime of CDM.

If cold initial conditions are used at an initial time when linear perturbation theory still applies, 
the Poisson equation approximates the Einstein equations on all scales relevant 
 for LSS and halo formation (see for instance \cite{CZ11,GW12}) justifying the usual use of Newtonian gravity 
and non-relativistic equations in general.  Furthermore,  gravitational two-body collisions are suppressed 
due to the presence of a large number of particles in the systems of interest, so that the phase space dynamics 
of the 1-particle distribution function is collisionless  \cite{G68}.
Therefore, the time evolution of the phase space density $f(t,\v{x},\v{p})$ is governed by the Vlasov (the collisionless Boltzmann) equation, 
also known as Jeans equation \cite{Jeans1915}.

A solution to the Vlasov equation is equivalent to a solution of the coupled infinite hierarchy of equations 
for the cumulants of the phase space density.  In the linear regime of LSS formation, before multi-streaming occurs, 
the Boltzmann hierarchy of cumulants of the Vlasov equation can be consistently truncated, so that
the Vlasov system may be solved by the dust model \cite{P80} where the curl part of the 
velocity and all higher cumulants vanish identically.
The dust description breaks down soon after the density contrast evolves into the non-linear regime, 
as this is generally followed by caustic formation, that is, `shell-crossing' singularities. Consequently,
  multiple streams occur, vorticity and higher cumulants are generated, 
and solving the full Vlasov equation is warranted from that point onwards.
A popular way of providing approximate solutions to the Vlasov equation, and thus,
 determining the non-linear evolution of CDM, is through the use of N-body simulations, 
see for instance \cite{T02,SWF05,SWF06,SW09,AHK12,HAK13}.

\begin{figure}[t!]
\includegraphics[width=0.48\textwidth]{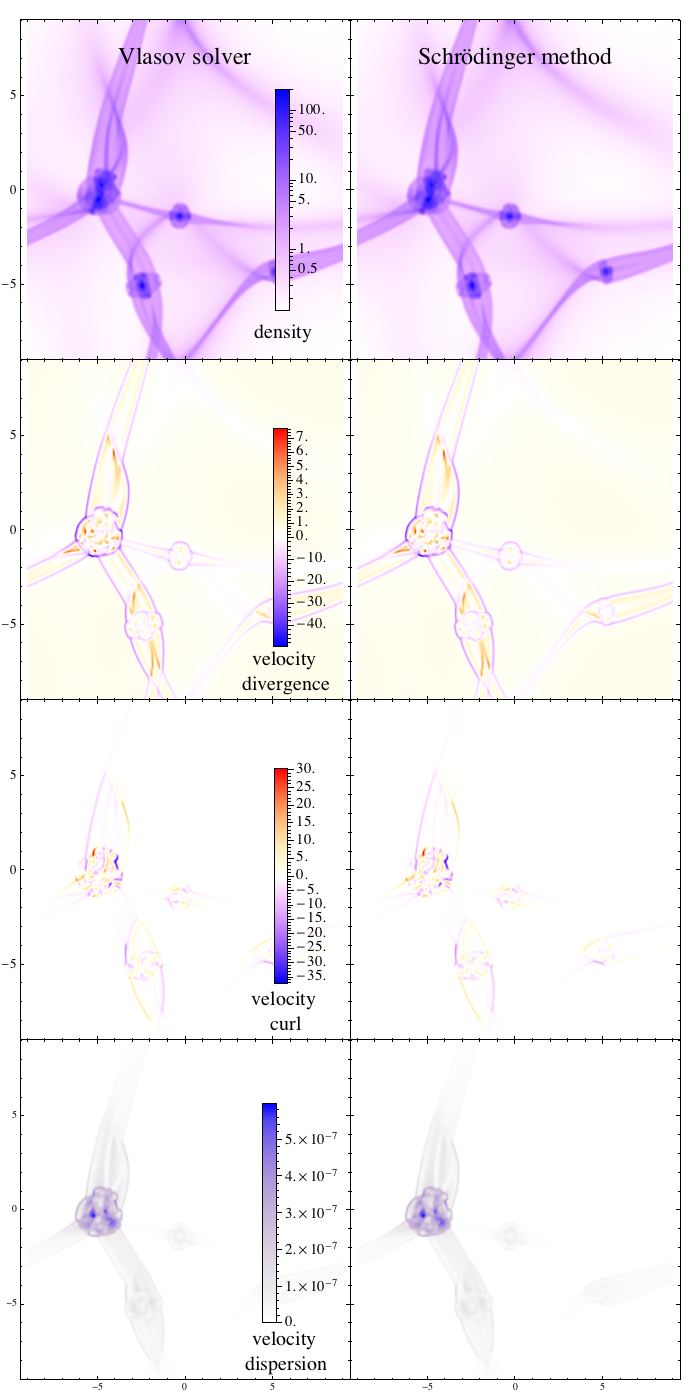}\\
\caption{Time snapshots of a two-dimensional cosmological simulation evolved to the present time $a=1$.
 Both codes were started with the same single-stream initial conditions, set up using the Zel'dovich approximation at $a_{\rm ini}=1/51$.
From top to bottom: density, velocity divergence, velocity curl and trace of the velocity dispersion tensor. 
The left column shows the smoothed results of the 2D-version of the Vlasov solver \texttt{ColDICE}~\cite{SousbieColombi2016}.
The result of the Schr\"odinger method is shown in the right column.
 The differences are barely visible by eye. A quantitative comparison is presented in Sec.\,\ref{sec:numerics}.}
\label{fig:ScMteaser}
\end{figure}
  
In most applications, one is not interested in the fine-grained distribution of particles, or the fine details of the distribution function,
but rather at the lowest cumulants: density, velocity and velocity dispersion (perhaps even higher cumulants could be of interest).
The route to the cumulants via the high-dimensional phase space seems, however, unavoidable if the hierarchy cannot be truncated. 
Fortunately, this is unnecessary if a different method of solving the Vlasov equation is used, as we describe below in more detail.

The Schr\"odinger method (ScM), originally proposed in the context of plasma physics in \cite{BertrandVanTuanGrosEtal1980} and later independently in the context of gravity by Widrow and Kaiser~\cite{WK93,DW96},
is a numerical technique for approximating solutions to the Vlasov equation while avoiding the difficulties of dealing with phase space.
With this method, CDM is modeled as a complex scalar field obeying the coupled Schr\"odinger-Poisson 
equations (SPE) \cite{SC02,SC06,G95}, where $\hbar$ is treated as a free parameter determining the desired phase space resolution. 
The ScM comprises two steps; (1) solving the SPE  with suitably chosen initial conditions
and (2) constructing the cumulants of choice. The second step may be performed in
 two mutually independent ways. Either (2a) by taking the Husimi transform \cite{H40} in order to construct a phase space distribution 
from the wave function and use it to calculate the cumulants.  Or, (2b) construct the cumulants directly through 
quasi-local manipulations of the wave function, avoiding the complexity of $2d$-dimensional phase space.
In this work we follow the second route, that is (2b). 

 The first three cumulants of a two-dimensional toy cosmological simulation are shown in Fig.\,\ref{fig:ScMteaser}. 
From top to bottom, we show the density, the two components of the velocity vector field and the trace of the velocity dispersion tensor.
The correspondence between the Vlasov solver \texttt{ColDICE}~\cite{SousbieColombi2016} and our implementation of the  ScM is depicted 
by the left and right column.
 The same initial linear gaussian random field at $a_{\rm ini}=1/51$ was used in both cases.
The Vlasov solver evolves the two-dimensional phase space sheet in four-dimensional phase space, and we obtained the figures on the left column 
by subsequent projection of that phase space sheet weighted by suitable powers of $\vu$ onto the two-dimensional Eulerian space, 
followed by  gaussian smoothing.  The figures on the right column have been obtained by evolving a complex wave function 
in two-dimensional Eulerian space and subsequently constructing the corresponding coarse-grained cumulants through simple 
spatial differentiation of the wave function, again followed by spatial coarse graining.  Although these methods are vastly 
different both in implementation and also conceptually, the results are barely distinguishable by eye.

\paragraph*{Objective}
The main result of this article is the extension of all previous one-dimensional tests of the 
 ScM \cite{T89,SRvB89,WK93,DW96,UhlemannKoppHaugg2014,GarnyKonstandin2017},\footnote{A method  different from ScM  where $N$ wave functions are coupled via the Poisson equation, was shown to accurately solve the Vlasov equation in one-dimensional situations in \cite{SBRB13} 
without the necessity of coarse graining, see also \cite{MarkowichMauser1993,ZhangZhengMauser2002} for a mathematical treatment.} to two-dimensional configurations, 
using  a  single smooth wave function on a cosmological background with nearly cold initial conditions, 
similarly to \cite{UhlemannKoppHaugg2014}. 
In contrast to \cite{UhlemannKoppHaugg2014}, we also quantitatively show that the resulting first three 
cumulants -- density, velocity and velocity dispersion -- are in good correspondence 
with the same quantities extracted from a conventional state-of-the-art Vlasov code, \texttt{ColDICE} \cite{SousbieColombi2016}.
This is highly non-trivial, since as outlined in Table \ref{tab:tabScM}, the Vlasov equation with cold initial 
conditions requires 2$d$ degrees of freedom (d.o.f.) to encode the phase space sheet, whereas the ScM requires only {\emph two}, 
independent of the spatial dimension $d$. Therefore, while the success of the ScM for $d=1$ might have been a coincidence 
due to the matching numbers of d.o.f., its success in $d=2$ gives strong evidence that the ScM will work for any dimension $d$.
Furthermore, we extend the work of \cite{UhlemannKoppHaugg2014} by considering non-symmetric initial conditions, 
in particular, a gaussian random field akin to standard cosmological simulations.
This is another important milestone in showing the generality of the ScM as a method for solving the Vlasov equation.
In addition, our work outlines an efficient algorithm for determining moments and cumulants from snapshots of the wave function at 
any desired time.

\paragraph*{Outline}
This article is organized as follows. In Sec.\,\ref{sec:PS-CDM} we review the connection
 between the gravitational N-body problem, the phase space description in terms of the Vlasov equation and its connection to the dust fluid and Lagrangian formulation of the Vlasov equation.
 We review the connection between the Husimi distribution and the coarse-grained Vlasov distribution function. 
Furthermore we review the algorithm to directly construct moments and cumulants that approximately satisfy the Boltzmann hierarchy 
in a way that circumvents the full phase space and is therefore easy to implement. 

In Sec.\,\ref{sec:Implementation} we describe the numerical implementation of our experimental two-dimensional Schr\"odinger code.
 In Sec.\,\ref{sec:numerics} we investigate cosmological simulations of two different types of initial conditions, 
 a two-dimensional pancake collapse as well as a more realistic gaussian random field, to provide further empirical 
proof of the accuracy and feasibility of the ScM.  We quantitatively compare cumulants to the corresponding quantities obtained with a 
state of the art Vlasov code \texttt{ColDICE}. 

In our discussion section, Sec.\,\ref{sec:Discussion}, we explain how vorticity arises 
in the ScM without the seemingly necessary vortical degrees of freedom, 
 a qualitative new phenomenon in $d=2$ dimensions which goes beyond the previous one-dimensional studies.
 We also demonstrate the advantage of the ScM, with two applications.
 Firstly, we evaluate the dynamical pressure induced by the non-linear structure formation -- an easy task in the ScM -- allowing
an estimate of the magnitude of backreaction on the background cosmology.
Secondly, we outline how to calculate the entropy and the free energy.
 Since the phase space density comes with a fixed coarse-graining scale, the entropy is well defined and will shed new light on the nature of LSS formation.
 
The reader may also find useful the appendices. In App.\,\ref{sec:ZAini} we present Zel'dovich initial conditions. 
In App.\,\ref{sec:numericalimplementation} we give some further details regarding our numerical implementation.
In App.\,\ref{sec:CDM1D} we describe the one-dimensional continuum formulation of CDM, as it is not readily available 
in the literature and propose an iterative improvement of the Zel'dovich approximation in the multi-stream regime.
In App.\,\ref{sec:DerivationofCDMtrajectories} and \ref{sec:CDMsolvesVlasovProof} we collect derivations and proofs, and in
 App.\,\ref{Diffoffs} we collect various results from \cite{UhlemannKoppHaugg2014} in order to assist the reader.

\begin{table*}
\centering
\begin{tabular}{|l|c|c|c|}
\hline
 & ScM:\ \ $f_\H(t, \vx,\vu)$\,,\ \ Sec.\,\ref{sec:ScMtoVlasov} & Dust:\ \ $f_\d(t, \vx,\vu)$\,,\ \ Sec.\,\ref{sec:DusttoVlasov}    & CDM:\ \  $f_{\rm c}(t,\vx,\vu)$\,,\ \  Sec.\,\ref{sec:CDMtoVlasov}\\\hline
 Degrees of freedom (d.o.f.) type & 1$\times\mathbb C$:\quad $\psi(t,\vx)$ &2$\times\mathbb R$:\quad $n_\d(t,\vx),\, \phi_\d(t,\vx)$ &$2\times\mathbb R^d$:\quad $\vX(t,\vq),\, \vU(t,\vq)$\\
  Number of d.o.f. & 2 &2 &2$d$\\
Equations of motion & 1st order SPE (\ref{schrPoissEqFRW}) & 1st order fluid equations (\ref{Fluideq}) & 1st order trajectories \eqref{CDMtrajectoriesEOM}\\
Singularity-free d.o.fs  & \cmark & \xmark & \cmark \\
$f(\vx,\vu)$ constructed from & \qquad \  \ $\psi(\vx)$, quasi-local \eqref{Husimi}  & $n_\d(\vx), \vnabla\phi_\d(\vx)$,\, local \eqref{fdust}& $\vX(\vq), \vU(\vq)$, non-local \eqref{fCDM} \\
$M^{(n)}(\vx)$ constructed from & $\partial^{0\leq m \leq n}_x \psi(\vx)$, quasi-local \eqref{momentscgw}  & $n_\d(\vx), \vnabla\phi_\d(\vx)$,\, local \phantom{blab} & $\vX(\vq), \vU(\vq)$, non-local \eqref{CDMMoments} \\
Vlasov equation (\ref{VlasovPoissonEq}) solved & approximately ($\thbar,\sigx$) &  exactly until shell-crossing &  exactly\\
Multi-streaming and virialization & \cmark & \xmark  & \cmark\\
 Initial conditions type & arbitrary, incl. cold \eqref{inipsi} & cold \eqref{fdust} &  cold \eqref{CDMInitialX}  \\\hline
\end{tabular}
\caption{Comparison between the Schr\"odinger method, and two other continuum formulations of CDM: the dust fluid that ceases to describe CDM after shell-crossing and the Lagrangian formulation of the Vlasov equation with cold initial conditions, that can be seen as the continuum definition of CDM. The mathematical correspondence of $f_\H$ with cold initial conditions and the coarse-grained CDM phase space density $\bar f_\c$ \eqref{fcgCDM} is established by virtue of \eqref{fHandfcgDiff} and numerically verified in the two-dimensional case in Sec.\,\ref{sec:numerics}.}
\label{tab:tabScM}
\end{table*}

\section{Modeling cold dark matter}
\label{sec:PS-CDM}
The reader already familiar with the ScM or not interested in its details might want to skip this section. 
For quick reference, the most important equations are highlighted and a summary of the construction algorithm of the phase space density is given at the end of Sec.\,\ref{sec:Husimidist}, and for moments and cumulants these can be found at the end of Sec.\,\ref{sec:HusHierarchy}. It is only the second approach that we explicitly use in this article.

\subsection{N-body to Vlasov}
The gravitational N-body problem is defined by the Hamiltonian system 
\begin{align} 
E_N &= \sum_{i=1}^N\left(\frac{\vu_i^2}{2a^2}- \frac{m G}{2 a }\sum_{j=1, j\neq i}^N\frac{1}{|\vx_i - \vx_j |} \right) \notag \\
\dot \vx_i &= \frac{\partial E_N}{\partial \vu_i}\,,\qquad \dot{\vu}_i = - \frac{\partial E_N}{\partial \v{x}_i} \label{HamiltonSystem}
\end{align}
where $a$ is the scale factor, $E_N$ is the energy per particle mass $m$,  $\vx_i$ is the comoving spatial coordinate of the particle $i$ with associated conjugate momentum $\v{p}_i= m \vu_i$.\footnote{For simplicity and without loss of much generality we assumed that the mass $m$  is the same for all particles. It is therefore convenient to use $\vu$ instead of $\v{p}$ as phase space coordinate because many equations appear less cluttered.} 
Defining the exact microscopic, or Klimontovich, phase space density
\begin{align} \label{Klimonotovichf}
f_{\rm K} (t,\vx,\vu) = \frac{m}{\rho_0}\sum_{i=1}^N \delta_{\rm D}\big[\vx-\vx_i(t)\big]\,\delta_{\rm D}\big[\vu-\vu_i(t)\big] \,,
\end{align}
using the three-dimensional Dirac delta function $\delta_{\rm D}$, the $N$ Hamiltonian equations \eqref{HamiltonSystem} that determine the phase space trajectories $\{\vx_i(t),  \v{u}_i(t)\}$  
can be neatly expressed in form of  the Klimontovich equation \cite{Klimontovich67} 
\begin{subequations}
\label{KlimoPoissonEq}
\begin{align}
\label{KlimoEq}
\partial_t f_{\rm K}&=   -\frac{\vu}{a^2}\cdot \vnabla_{\! \!  x} f_{\rm K} +  \vnabla_{\! \!  x} \varPhi_{\rm K} \cdot \vnabla_{\! \!  u} f_{\rm K} \ \\
\Delta \varPhi_{\rm K} &= \frac{4\pi G\rho_0}{a} \left(\int \vol{3}{u} \!f_{\rm K} - 1 \right) \,,\label{PoissonEq}
\end{align}
\end{subequations}
which has the simple interpretation that $f_{\rm K}$ is conserved along phase space trajectories, 
  $\,0=Df_{\rm K} /dt = \left(\partial_t f_{\rm  K} + \dot{\v{x}}\cdot\vnabla_{\! \!  x} f_{\rm K} + \dot{\vu} \cdot \vnabla_{\! \!  u} 
 f_{\rm K}\right) \big|_{\v{x}=\v{x}_i , \v{u}=\v{u}_i} $ for all $i$.\footnote{$f_{\rm K}$ remains non-zero 
 (the height of the six-dimensional $\delta_{\rm D}$)  along each particle's phase space trajectory and 
vanishes  everywhere else.} 
This follows from the fact that for Hamiltonian systems the phase space trajectories never cross.

For convenience we introduced the  (constant) comoving matter background density $\rho_0 = m\,N/V$, such that background, or spatial average value of $f_{\rm K}$ over some large volume $V \rightarrow \infty$, is normalized $\langle \int \vol{3}{u}\!f_{\rm K}\rangle_{V}=1$. 

Although it is exactly \eqref{KlimoPoissonEq} that N-body simulations solve, it is not feasible to use values of $N$ and $m$ that we expect from particle DM candidates.
Even if DM were primordial black holes with $m\simeq15 M_\odot$, possibly the largest $m$ encountered for DM in the literature, current cosmological N-body simulations have particle masses that are 8 orders larger than this value and therefore can only approximate the physical $N$-body problem.
The continuum limit $f(\vx,\vu) = \lim_{N \rightarrow \infty}  f_{\rm K} (\vx,\vu)$ or `pulverization,' where $N \rightarrow \infty $ and $m\rightarrow 0$ with $\rho_0 = m N/V$ constant, although an idealization, comes closer to modeling collisonless particle DM. 
The such defined smooth phase space density  satisfies the Vlasov equation 

\begin{subequations}
\label{VlasovPoissonEq}
\begin{empheq}[box=\mybluebox]{align}
\label{VlasovEq}
\partial_t f&=   -\frac{\vu}{a^2}\cdot\vnabla_{\! \!  x} f +  \vnabla_{\! \!  x} \varPhi \cdot\vnabla_{\! \!  u} f \\
\Delta \varPhi &= \frac{4\pi G\,\rho_0}{a}  \left(\int \vol{3}{u}\!\!f - 1\ \right) \,,
\end{empheq}
\end{subequations}
that is, the collisionless Boltzmann equation with a long range gravitational force $\vnabla_{\! \!  x} \varPhi$ determined by $f$ itself. 
The form of Klimontovich and Vlasov equations, \eqref{KlimoPoissonEq} and \eqref{VlasovPoissonEq}, is identical and therefore the Vlasov equation simply states the conservation of the now continuous phase space density along phase space trajectories of the continuum. 
While \eqref{KlimoPoissonEq} is in effect short-hand notation for the Newtonian equations of motion 
of $N$ particles, \eqref{VlasovPoissonEq} is an evolution equation for a smooth phase space distribution.

The Vlasov equation is often taken as the defining equation of purely self-gravitating, and thus collisionless, non-relativistic matter. 
It also has a straightforward relativistic generalization, see for instance \cite{Choquet-Bruhat2008} for a detailed treatment. 
Although the Vlasov equation is a very useful definition for collisionless DM in the context of LSS formation, it is not sufficient
when the discreteness matters, as is the case for the internal dynamics of globular clusters 
where the number $N$ of particles\footnote{For a globular cluster the particles are stars such that $m\simeq 1 M_\odot$.} is only a 
few thousand  \cite{G68}.
In this case a better continu\-um approximation to the $N$-body dynamics \eqref{KlimoPoissonEq} starts with a transformation of \eqref{KlimoPoissonEq} into the Bogoliubov-Born-Green-Kirkwood-Yvon (BBGKY) hierarchy for the $n$-particle distribution functions using an appropriate ensemble average without taking the limit $N\rightarrow \infty$. 
A truncation of this hierarchy at lowest order in $1/N$ gives the Vlasov equation, while the one at second order contains a collision term that takes into account some of the discreteness through the 2-particle correlation function \cite{G68,BT08,Chavanis2008}.  
Note that cosmological N-body simulations suffer from artificial discreteness effects due to the relatively small N$\,\ll N_{\rm phys}$ and therefore may fail to approximate the Vlasov equation as well as the physical N-body problem.

\paragraph*{Moments and Cumulants}
Most applications of $f$ in cosmology involve only its first few ($n=0,1,2$) moments \begin{align}
M^{(n)}_{i_1 ... i_n}(\vx) \equiv \int \vol{3}{u}  u_{i_1} \cdots u_{i_n}  f(\vx,\vu) \,.
\end{align}
In order to calculate the $n$th moment it is convenient to define the moment generating function
\begin{align}
\label{genfun}
 G(\vx,\v{J}) &\equiv\int \vol{3}{u} e^{i\vu\cdot\v{J}}   f(\vx,\vu)\,,
\end{align}
such that moments and cumulants are given by
\begin{subequations} \label{momentsandcumulants}
\begin{align}
M^{(n)}_{i_1 \cdots i_n}&= (-i)^n \left.\frac{\del^n G}{\del J_{i_1} \ldots \del J_{i_n}} \right|_{\v{J}=0}  \label{moments}\\
C^{(n)}_{i_1 \cdots i_n}&:= (-i)^n \left.\frac{\del^n \ln G}{\del J_{i_1} \ldots \del J_{i_n}} \right|_{\v{J}=0} \label{cumulants}\,.
\end{align}
\end{subequations}
Taking time derivatives of $C^{(n)}$ or $M^{(n)}$ and using the Vlasov equation \eqref{VlasovEq} to eliminate $\partial_t f$ at the right-hand side of \eqref{momentsandcumulants}, results in an infinite coupled collection of first order time evolution equations for the $C^{(n)}$ or $M^{(n)}$, the Boltzmann hierachy, see \cite{UhlemannKoppHaugg2014} for details.
The first 3 moments are
\begin{subequations} \label{momentf}
\begin{align}
 M^{(0)} &=  n \label{0momentf} \,, \\
 M^{(1)}_i&=:  n  u_i \label{1momentf} \\
 M^{(2)}_{i j}&=:  n  u_i  u_j +   n  \Sigma_{ij} \label{2momentf}  \,,
\end{align}
and the resulting cumulants are then determined by them as
\begin{align}
C^{(0)} &= \ln  n \\
C^{(1)}_i &=  u_i \\
C^{(2)}_{ij} &=  \Sigma_{ij}\,. \label{2cumulantf} 
\end{align}
\end{subequations}
The physical interpretation is as follows. The scalar field $n(t,\vx)$ is the density, the vector field $u_i(t,\vx)$ is the mass-weighted 
velocity related to the peculiar velocity as $v_i = u_i/a$ and the tensor field $\Sigma_{ij}(t,\vx)$ is the mass-weighted velocity dispersion. 

The physical significance of the moments $M^{(1)}$ and $M^{(2)}$ becomes clear when we consider the four-dimensional energy momentum tensor 
$T_{\mu \nu}=\frac{ \rho_0}{\sqrt{-g}}\, \int  \frac{du_1 du_2 du_3}{ u^0(u_j)} u_\mu u_\nu f(x^\alpha,u_{i})$ in the Newtonian limit. Its components w.r.t. coordinates $x^\alpha=(t,\vx)$ are 
\begin{subequations} \label{EMtensor}
\begin{align}
T_{00} &=\ \ \, \rho_0 \, M^{(0)}/a^3 \\
T_{0i} &= - \rho_0\, M^{(1)}_i /a^3 \\
T_{ij} &= \ \ \,\rho_0\, M^{(2)}_{i j} /a^3, \label{stresstensor}
\end{align}
\end{subequations}
where we have neglected any metric perturbations and relativistic effects, such that $g_{00}=-1$, $g_{0i} =0$, $g_{ij}= a^2 \delta_{ij}$ and $u^0=1$.\footnote{This is a good approximation in the Newtonian (or Poisson) gauge for non-relativistic matter.}
\subsection{Dust to Vlasov} \label{sec:DusttoVlasov}
The dust model is a  pressureless perfect fluid with density $n_{\rm d}(t,\vx)$ and an irrotational velocity $\vu_\d(t,\vx)=\vnabla \phi_{\rm d}(t,\vx)$\footnote{We will drop very often the argument $\vx$ from the velocity $\vu(\vx)$ and simply write $\vu$ whenever it is clear from the context that $\vu$ is not a coordinate in phase space, but a vector field in real space.}   satisfying the continuity, Euler and Poisson equations
\begin{subequations}\label{Fluideq}
\begin{align}
\del_t n_\d &= -\frac{1}{a^2}\vnabla \cdot(n_\d\vu_\d)  \label{Contidust}  \,,\\
\del_t \vu_\d &= -\frac{1}{a^2}(\vu_\d\cdot\vnabla)\vu_\d -\vnabla \varPhi_\d \,, \label{dustEuler}\\
\vnabla \times \vu_\d &= 0  \label{dustconstr}\\
\Delta \varPhi_\d&=\frac{4\pi G\,\rho_0}{a}\Big(n_\d -  1 \Big) \label{Poissdust}  \,.
\end{align}
\end{subequations}
These equations describe cold collisionless DM (CDM) in the so-called single stream regime.
It is therefore often used as the defining equation for CDM in many analytical studies of LSS formation. 
The dust phase space density is given by
\begin{align}
\label{fdust}
f_\d (t, \vx,\vu)&= n_\d(t, \vx)\, \delta_{\mathrm D}\Big[\vu-\vnabla \phi_\d(t,\vx) \Big] 
\end{align}
and solves the Vlasov equation \eqref{VlasovPoissonEq} automatically until the appearance of so-called shell-crossing singularities or caustics where $n_\d$ diverges.\footnote{The different types of singularities in the Zel'dovich approximation have been classified in \cite{ArnoldShandarinZeldovich1982}.} 
For cold initial conditions, where $f$ initially has the form \eqref{fdust}, these singularities generically occur and are harmless in the Vlasov equation \eqref{VlasovPoissonEq} but lead to the break-down of the dust fluid description  \eqref{Fluideq}.
This is because after shell crossing, $f$ can no longer be expressed  as the product \eqref{fdust}.
The product form of $f_\d$ makes all the cumulants vanish with $n>1$. After shell-crossing, however, $f$ generally produces non-zero vorticity 
invalidating \eqref{dustconstr}, and non-zero velocity dispersion $C^{(2)}_{ij}$ thus
 modifying the Euler equation \eqref{dustEuler}. The equation of motion for $C^{(2)}_{ij}$ contains $C^{(3)}_{ijk}$, and so on,
 such that a truncation of the Boltzmann hierarchy is no longer possible, see \cite{UhlemannKoppHaugg2014,PS09}. 
\subsection{CDM to Vlasov}\label{sec:CDMtoVlasov}
A useful continuum definition of CDM is a phase space density $f_{\rm c}$ that is solution  of the Vlasov equation \eqref{VlasovPoissonEq} with initial conditions of the form \eqref{fdust}. 

Although after shell crossing $f_{\rm c}$ cannot be expressed solely as function of $\vx$, it can still be expressed as function of the Lagrangian coordinate $\vq$, the continuum generalisation of the particle label $i$. 
For this we introduce the particle trajectories $\vX(t,\vq)$ which smoothly depend on the initial positions $\vq$. The equations of motions for $\vX(t,\vq)$ and $\vU(t,\vq)$ are in full analogy to \eqref{HamiltonSystem} given by
\begin{align} \label{CDMtrajectoriesEOM}
a^2 \dot{\vX}(\vq) &= \vU(\vq) \\
\dot{\vU}(\vq)& =  \frac{G\,\rho_0}{a}\int \vol{3}{q'}  \frac{\vX(\vq) - \vq'}{|\vX(\vq) - \vq'|^3}- \frac{\vX(\vq) -\vX(\vq')}{|\vX(\vq)- \vX(\vq')|^3}\,, \notag
 \end{align}
 where we have assumed that 
 \begin{equation} \label{CDMInitialX}
 \vX(\vq,t\rightarrow 0) = \vq\,,
 \end{equation}
  corresponding to growing mode initial conditions of \eqref{Fluideq}.
  We explicitly wrote $-\vnabla_{\!\!x} \varPhi$ in terms of $\vX(\vq)$ to make it manifest that  
shell-crossing infinities in the density do not necessarily lead to divergencies in $\vX$ and $\vU$. See App.\,\ref{sec:DerivationofCDMtrajectories} for a derivation.
Indeed, these variables remain continuous and sufficiently smooth at shell-crossings such that they can be numerically evolved without problems.

Although a numerical implementation of \eqref{CDMtrajectoriesEOM} necessarily discretises $\vX(t,\vq)$, the continuum formulation allows
 to keep track of the phase space sheet, which then in turn  {\it i}) allows to refine the discretisation at any time in a well 
defined manner and {\it ii}) leads to much smoother moments \eqref{CDMMomentsstreams} compared to conventional count-in-cells estimators.
 This is done in codes like \texttt{ColDICE} \cite{SousbieColombi2016} or the ones presented in \cite{ShandarinHabibHeitmann2012,AbelHahnKaehler2012,HahnAnguloAbel2015, HahnAngulo2016}, but does not occur in conventional N-body simulations.
 For illustration, we consider the one-dimensional case of \eqref{CDMtrajectoriesEOM} in App.\,\ref{sec:CDM1D}. 

The CDM phase space density $f_{\rm c}$ is the continuum limit of the Klimontovich phase space density \eqref{Klimonotovichf}. It
 can be constructed from $\vX(\vq)$ and ${\vU}(\vq)$ as
\begin{subequations}  \label{fCDM}
\begin{align}
\!\!\!f_{\rm c} (t,\vx,\vu)
&= \int \vol{3}{q} \delta_{\rm{D}} [\vx - \vX(\vq)] \, \delta_{\rm{D}} [\vu - \vU(\vq)] 
\label{fCDMqspace} 
\\
&=\sum_{\substack{\vq~\mathrm{with}\\ \vx=\vX(t,\vq)}} \frac{ \delta_{\rm{D}} [\vu - \vU(\vq)]}{|\det \partial_{q^i} X^j(\vq)|}  
 \label{fCDMstreams}
\end{align}
\end{subequations}
and satisfies the Vlasov equation \eqref{VlasovPoissonEq}. 
Thus the Vlasov equation for cold initial conditions may be solved with the help of six degrees of freedom $\vX(\vq)$ and ${\vU}(\vq)$, see  App.\,\ref{sec:CDMsolvesVlasovProof} for a derivation, while in the single stream regime  two degrees of freedom, $n_\d(\vx)$ and $\phi_\d(\vx)$, are sufficient. 

The moments $M^{\c}$ are trivially obtained from \eqref{fCDM} as
\begin{subequations}   \label{CDMMoments}
\begin{align}
M^{\mathrm{c} (n)}_{i_1,...,i_n}(\vx) 
&= \int \vol{3}{q} \delta_{\rm{D}} (\vx - \vX(\vq))\, U_{i_1}(\vq) ... U_{i_n}(\vq) 
\label{CDMMomentsqspace} 
\\
&=\sum_{\substack{\vq~\mathrm{with}\\ \vx=\vX(t,\vq)}} \frac{U_{i_1}(\vq) ... U_{i_n}(\vq)}{|\det \partial_{q^i} X^j(\vq)|} 
 \label{CDMMomentsstreams}
\end{align} 
\end{subequations}
and are, like $f_{\rm c}$, non-local in $\vq$-space. The second equalities in \eqref{fCDM} and \eqref{CDMMoments}, respectively, follow from the properties of the Dirac $\delta$-function, $\delta_{\rm D}$.

These sum over streams\footnote{Due to the assumed initial conditions \eqref{CDMInitialX}, the map $\vX(t,\vq)$ belongs to the homotopy class of the identity, whose degree is one. 
The degree of a function $\vX(\vq)$ at a regular point $\vx$  is the natural number of points $\v{q}$ for which $\vx= \vX(\vq)$. 
In our case this is the number of streams. A point $\vx$ is regular if it does not lie on a shell-crossing caustic. 
Now, according to a theorem of differential topology, see \S4 of \cite{Milnor1988}, the parity of the degree is independent of the regular point $\vx$ and also independent of the representative $\vX(\vq)$ of the homotopy class. Therefore, the number of streams is always finite and odd, excluding the zero-measure subset of   $\vx$ on caustics.} have to be performed in Vlasov solvers like \texttt{ColDICE} to obtain $M^{\c (0)} = n_\c$ in order to solve the Poisson equation at each time step.
From \eqref{fCDMstreams} and \eqref{CDMMomentsstreams} it is clear that if there is only a single stream, such that $\vx=\vX(t,\vq)$ is invertible for $\vq$, the resulting $f_\c$ is of the product form \eqref{fdust}.
 The assumed initial conditions \eqref{CDMInitialX} thus guarantee that there is an early time where $f_\c = f_\d$.

It is also worth mentioning that \eqref{CDMtrajectoriesEOM} makes it manifest that the acceleration $\dot \vU(\vq)$ of the stream at $\vq$ is determined by all streams with the same $\vX(\vq)$. 
This is not the case in conventional treatments of the Lagrangian formulation of CDM, where the invertibility of $\partial_{q^i} X^j(\vq)$ is assumed and used in reformulating \eqref{CDMtrajectoriesEOM} into an equation local in $\vq$. 
Solutions of this transformed equation lead necessarily to unphysical behavior after shell-crossing \cite{Buchert1996}.
We discuss this further in App.\,\ref{sec:CDM1D}. 

\subsection{Coarse-grained Vlasov}
\subsubsection{General case}
The coarse-grained phase space density
\begin{align}
\label{cgf}
\notag  \bar f (t, \vx,\vu)&=\int \frac{\vol{3}{x'}\vol{3}{u'}}{(2 \pi\sigx\sigu)^3 } 
   e^{-\frac{(\vx-\vx')^2}{2\sigx^2}-\frac{(\vu-\vu')^2}{2\sigu^2} } f(t, \vx',\vu')\,,
\\
&= e^{\frac{\sigx^2}{2}\Delta_x+\frac{\sigu^2}{2}\Delta_u} \{f\}\ 
\end{align}
 is obtained from $f$ through convolution with a Gaussian filter with width $\sigx$ and $\sigu$ in $\vx$ and $\vu$ space.
  For notational simplicity we will use a shorthand operator representation of the gaussian smoothing defined in the second line of \eqref{cgf}.
The corresponding coarse-grained Vlasov equation can be obtained by applying the smoothing operator on \eqref{VlasovPoissonEq}, 
see \cite{MW03,UhlemannKoppHaugg2014}.

The moments $\bar M^{(n)}$ of $\bar f$ are simply obtained from the moments of $f$ as
\begin{subequations}\label{cgmomentsfrommoments}
\begin{align}
\bar M^{(0)}& = e^{\frac{\sigx^2}{2}\Delta_x} \{ M^{(0)} \}  \\
\bar M^{(1)}_i &= e^{\frac{\sigx^2}{2}\Delta_x} \{ M^{(1)}_i \}  \\
\bar M^{(2)}_{ij}& = e^{\frac{\sigx^2}{2}\Delta_x} \{ M^{(2)}_{ij} \}  + \sigu^2 \bar M^{(0)} \delta_{ij}   \,. \label{cg2ndmomentfrom2ndmoment}
\end{align}
\end{subequations}
The cumulants $\bar C^{(n)} $ are obtained from the $\bar M^{(n)}$ in the standard fashion according to \eqref{momentf} with all quantities barred. 
Note that in particular $\bar u_i \neq e^{\frac{\sigx^2}{2}\Delta_x} \{ u_i \}  $, and instead $\bar u_i  = \bar M^{(1)}_i/ \bar n$.
The second term in \eqref{cg2ndmomentfrom2ndmoment} arises through the spreading of the velocity distribution by the gaussian with variance $\sigu$. Note that the first two moments are not affected by $\sigu$.

The coarse-grained phase space density $\bar f$ has many desirable features that $f$ does not have. 
For instance $\bar f$ is no longer exactly conserved along phase space trajectories which allows us to properly define phase mixing,
 virialization and entropy production \cite{L67, MW03}.
This also makes $\bar f$ better behaved from a numerical point of view since $\bar f$ cannot develop structures below 
the scales $\sigx$ and $\sigu$, in contrast to $f$ which continues to develop ever smaller structures over time.

Thus a method allowing direct computation of $\bar f$ without first computing $f$ is of practical interest.
Since $\bar f$ can be approximated by coarse graining $f_{\rm K}$ directly without taking the limit that brought us to the Vlasov equation \eqref{VlasovPoissonEq}, a straightforward method to produce an approximation to $\bar f$ is to run a N-body simulation  \cite{T02,SWF05,SWF06,SW09,AHK12,HAK13,HahnAngulo2016,SousbieColombi2016} and sample  the phase space density using the particles.\footnote{The actually achieved resolution $\sigx \sigu$ is hard to estimate since it strongly depends on the method and usually is not constant in space and time.}

\subsubsection{CDM}
The coarse-grained continuum limit of CDM \eqref{fCDM} is obtained by inserting \eqref{fCDMqspace} into \eqref{cgf}  
\begin{align} \label{fcgCDM}
\bar f_{\rm c} (t,\vx,\vu)
&= \frac{1}{ (2\pi)^3 \sigx^3 \sigu^3}\int \vol{3}{q} e^{-\frac{\left[\vx - \vX(t,\vq)\right]^2}{2 \sigx^2}}
 \, e^{-\frac{\left[\vu - \vU(t,\vq)\right]^2}{2 \sigu^2}} \,.
\end{align}
Similarly, inserting \eqref{CDMMomentsqspace} into \eqref{cgmomentsfrommoments} shows that the coarse-grained CDM moments
\begin{subequations}\label{cgCDMmoments}
\begin{align}
\bar M^{\mathrm{c}(0)}& = \frac{1}{ (2\pi)^{3/2} \sigx^3}  \int \vol{3}{q}  e^{-\frac{ \left[\vx - \vX(t,\vq)\right]^2}{2 \sigx^2}}   
\\
\bar M^{\mathrm{c}(1)}_i &=  \frac{1}{ (2\pi)^{3/2} \sigx^3}  \int \vol{3}{q}  e^{-\frac{\left[\vx - \vX(t,\vq)\right]^2}{2 \sigx^2}} U_i(t,\vq)  
\\
\bar M^{\mathrm{c}(2)}_{ij}& = \frac{1}{ (2\pi)^{3/2} \sigx^3}  \int \vol{3}{q}  e^{-\frac{\left[\vx - \vX(t,\vq)\right]^2}{2 \sigx^2}} U_i(t,\vq) U_j(t,\vq) 
\notag 
\label{cgCDM2ndmoment}
\\
& \qquad \qquad + \sigu^2 \bar M^{\mathrm{c}(0)} \delta_{ij}  \,,
\end{align}
\end{subequations}
 involve $\vX(\vq)$ and $\vU(\vq)$, just like $\bar f_{\rm c}$, 
 only in a three-dimensional integral over Lagrangian space. Alternatively one might chose not to simplify \eqref{cgf} and \eqref{cgmomentsfrommoments} and perform the smoothing only after  the sum over streams in \eqref{fCDMstreams} and \eqref{CDMMomentsstreams} has been calculated. 
It is the second method that we will use in Sec.\,\ref{sec:numerics}.

\subsection{The Schr\"odinger method} 
\label{sec:ScMtoVlasov}
The Schr\"odinger method is an alternative way to approximate $\bar f$.
In the ScM  one constructs a phase space distribution $f_\H(t,\vx,\vu)$ at a given time from a smooth complex field $\psi(t,\vx)$ 
that satisfies the Schr\"odinger-Poisson equation. The function $f_\H(t,\vx,\vu)$ is called the Husimi distribution and is the absolute square of 
the Husimi-representation of the wave function $\psi$. 
The transformation $\psi \rightarrow f_\H$ is local in time, but involves one quasi-local integral over Eulerian space.
 Thus, to evaluate $f_\H(t,\vx,\vu)$ at points $\vx$,$\vu$ one only has to consider a single time snapshot of $\psi(t,\vx')$ and only points $\vx'$ in a small neighbourhood around $\vx$.

The phase space resolution of $f_\H$ is constant in phase space and time. 
In some applications of simulations of LSS formation, for instance the investigation of voids or ray-tracing, a spatially and temporally fixed 
resolution might be desirable.  

The ScM offers a compact way to store the entire six-dimensional 
 phase space information in $f_\H(\vx,\vu)$ in  two three-dimensional degrees of freedom contained in $\psi(\vx)$. 
Although the phase space density $f_\H$ can be easily constructed from $\psi$, the most useful feature of the ScM is that 
all moments and cumulants of the phase space distribution, like the density, velocity and velocity dispersion as well as the 
stress-energy-momentum tensor, can be constructed directly from $\psi$ and its spatial derivatives without the need to first 
construct the phase space density and subsequent cumbersome projections in velocity space.
Constructing moments and cumulants involves only taking spatial derivatives of $\psi$ and using a gaussian filter,
 which is very efficient to calculate since both processes are quasi-local in space. 

This should be contrasted with the case of CDM, where the dynamics of $f(t,\vx,\vu)$ can be compressed into two three-dimensional vector field $\vX(t,\vq)$, $\vU(t,\vq)$ satisfying first order differential equations. Therefore one needs to take care of 6 degrees of freedom rather than 2.
 Similar to $\psi(t,\vx)$ the phase space density $f_{\rm c}$ and moments $M_{\rm c}$ can be directly constructed from $\vX(t,\vq)$ by a single spatial integration. The integration is, however, non-local, whereas the construction of $f_\H$ and $M_\H$ is quasi-local. 
Lastly while the reduction of degrees of freedom from $f(t,\vx,\vu)$ to $\vX(t,\vq)$ and $\vU(t,\vq)$ only works for dust initial conditions \eqref{fdust}, the ScM works for general initial $f(t_{\rm ini},\vx,\vu)$.

\label{subsection:husimivlasovmap}
\subsubsection{The Schr\"odinger-Poisson equation} \label{SPE}
The non-relativistic Schr\"odinger-Poisson equation (SPE) in a $\Lambda$CDM universe with scale factor $a$
can be derived from the following action
\begin{subequations}
\begin{align}
A[\bar \psi,\psi]  &= m \int \!\!\mathrm{d} t\,\,\vol{3}{x} \Bigg(i \thbar \bar \psi(t,\vx) \, \partial_t \psi(t,\vx) - \, \mathcal{E}[\psi] \Bigg)   \label{psiAction}\\
 \mathcal{E}[t,\psi, \bar \psi] &= \frac{\thbar^2}{2 a^2} |\vnabla_{\!\! x} \psi(t,\vx)|^2 -  \label{Hamiltondensity}\\
& \qquad - \frac{ \rho_0 G }{2 a} \int \vol{3}{x'} \frac{  |\psi(t,\vx)|^2 |\psi(t,\v{x}')|^2}{|\vx-\v{x}'|}  \notag
\end{align}
\end{subequations}
 upon variation with respect to $\bar \psi$ and is given by
\begin{subequations}
\label{schrPoissEqFRW}
\begin{empheq}[box=\mybluebox]{align}
i\thbar \del_t \psi &= - \frac{\thbar^2}{2a^2} \Delta \psi +  \varPhi_\psi \psi \label{schrEqFRW} \\
\Delta \varPhi_\psi&=\frac{4\pi G\,\rho_0}{a}\Big(|\psi|^2 -  1 \Big) \label{PoissEqFRW} \,,
\end{empheq}
\end{subequations}
see for instance \cite{WK93,AS01,UhlemannKoppHaugg2014}. 
We have defined
\begin{equation}
\thbar \equiv \frac{\hbar}{m}
\end{equation}
which simplifies the SPE and can be interpreted as the phase space resolution in coordinates $\vx$ and $\vu$. 
The SPE can also be derived by observing that from the action \eqref{psiAction} it follows that  
\begin{subequations} \label{psiHamiltonianSystem}
\begin{align}
E &= \int \vol{3}x \mathcal{E}[t, \psi, i \thbar \bar \psi] 
 \end{align}
is the Hamiltonian and  $ i \hbar \bar \psi$ is the canonically conjugate momentum of $\psi$. Following the rules of Hamiltonian mechanics the SPE is obtained as
\begin{align}
 \frac{d}{dt}  \psi &=\{E , \psi \} \label{psiHamiltonEOM}
 \end{align}
 \end{subequations}
 where $\{...,...\}$ is the Poisson bracket.\footnote{$\{A , B\} \equiv \int \vol{3}{x} \frac{\delta A}{\delta[i \thbar \bar \psi(\vx)]} \frac{\delta B}{\delta \psi(\vx) } -  \frac{\delta B}{\delta[i \thbar \bar \psi(\vx)]}  \frac{\delta A}{\delta \psi(\vx) }$.  
The time evolution of a general $A=A[t,\psi]$ is then given as $dA/dt = \partial_t A + \{E, A\}$. For $A=\psi(\vx')$ and the rules of functional derivatives one recovers \eqref{psiHamiltonEOM}.} 
This shows the close analogy to the  Hamiltonian system \eqref{HamiltonSystem} of $N$-particles\footnote{Since $\mathcal{E}(\psi, \bar \psi) = \mathcal{E}(\bar \psi, \psi)$ the second Hamiltonian equation for the canonical momentum $i \hbar \bar \psi$ is just the complex conjugate of the SPE \eqref{psiHamiltonEOM}, see \cite{AS01}.} and is of practical use since $E$ will be employed later to test the  accuracy of the numerical solution of the SPE \eqref{schrPoissEqFRW}.

The extra $-1$ in the Poisson equation \eqref{PoissEqFRW} implements the cosmological boundary conditions and may be derived 
if one starts with the action of general relativity and a real Klein-Gordon field upon taking the non-relativistic and small-scale limit
 in the conformal Newtonian gauge, see \cite{SuarezChavanis2015}.\footnote{Note that our normalization of $\psi$ is different from \cite{SuarezChavanis2015}. We chose $\langle |\psi|^2\rangle_{V}=1$ for convenience rather than the more natural $\langle |\psi|^2\rangle_{V}= \rho_0 a^{-3}$.} 
\subsubsection{Madelung representation: relation to the dust fluid}
\label{subsec:Madelung}
Using the so-called Madelung representation for the wave function 
\begin{equation}
\psi(\vx)=: \sqrt{\npsi(\vx)}\exp\left(i\phi(\vx)/\thbar\right)
\end{equation}
with $\npsi$ and $\phi$ real, and $n_\psi\neq 0$ the SPE \eqref{schrPoissEqFRW} can be transformed  into a set of fluid-like equations upon defining $\vupsi\equiv\vnabla \phi$, the Madelung-Poisson equation \cite{M27}
\begin{subequations} \label{Madelungfluideq}
\begin{align}
\del_t \npsi &= -\frac{1}{a^2}\vnabla_{\! \!  x}\cdot (\npsi\vu_
\psi) \\
 \del_t \vupsi &= -\frac{1}{a^2}(\vupsi\cdot\vnabla)\,\vupsi -\vnabla \varPhi_\psi +\frac{\thbar^2}{2a^2} \vnabla \left( \frac{\Delta\sqrt{\npsi}}{\sqrt{\npsi}} \right) \label{EulerMadelung} \\
  \hspace{-1cm} \vnabla \times \vupsi & = 0  \\
  \Delta \varPhi_\psi&=\frac{4\pi G\,\rho_0}{a}\Big(\npsi -  1 \Big) \label{Poisspsi}\,.
\end{align}
\end{subequations}
Comparing them to the dust equations \eqref{Fluideq} it becomes transparent that the only difference is the single extra term in the Euler equation proportional to $\thbar^2$, the so-called `quantum pressure' 
\begin{equation} \label{Qdef}
Q\equiv - \frac{\thbar^2}{2a^2} \frac{\Delta\sqrt \npsi}{\sqrt \npsi}\,.
\end{equation}
Thus, if  $\vnabla Q$ is small compared to the other terms in \eqref{EulerMadelung}, one can maintain  $\npsi = n_\d$ and $\phi = \phi_\d$  to arbitrary precision during time evolution choosing $\thbar$ sufficiently small.

 The dynamics of the Madelung-Poisson equation is such that sooner or later, barring fine-tuned initial conditions, $Q$ will become important, 
independent of the chosen $\thbar$.
 It turns out that for initial conditions where the quantum pressure can be neglected, the locations in time and space 
where $Q$ becomes large are close to those where shell crossing happens in the dust fluid.
Indeed, if $Q$ becomes large it signals the impending breakdown of the Madelung-Poisson equation \eqref{Madelungfluideq}:
for initial conditions that make $Q$ small, $\vu_\psi$ diverges at nearly the same locations in space and time where $n_\d$ of the dust model diverges.
 It turns out that $\vu_\psi = \infty$ events are accompanied by $\npsi=0$ and thus a vanishing $\psi$. 

\begin{figure}[t]
    \centering
    \includegraphics[width=8.8cm, angle=0]{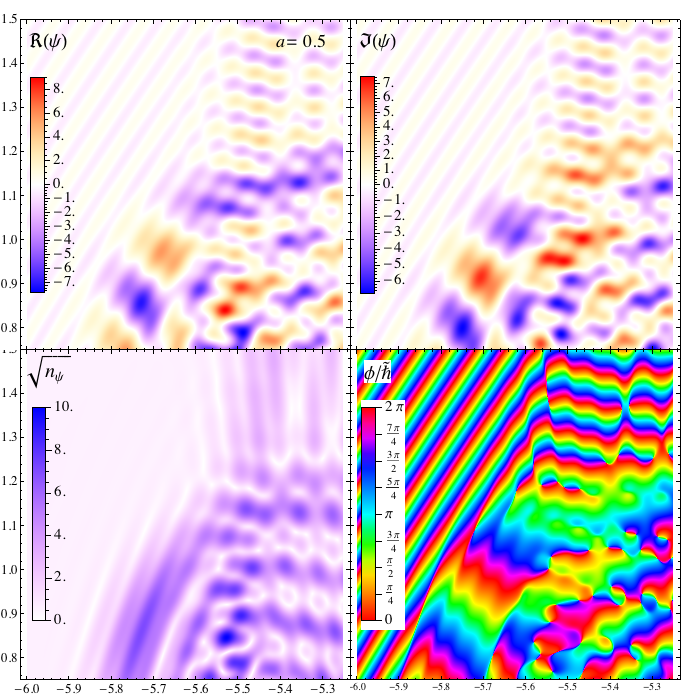}
    \caption{A small region of width $0.75\,$Mpc (309 pixels) from a $20\,$Mpc (8192 pixels) wide simulation box (the units on the axis are in Mpc). 
 The wave function $\psi$ has been evolved from linear gaussian random field initial conditions with $n_\psi = n_\d$ and $\phi=\phi_\d$ into the deeply non-linear regime, see Sec.\,\ref{sec:gaussianranfield} for details. 
    The top two panels show the dynamical variables $\Re(\psi)$ and $\Im(\psi)$, whereas the panels on the bottom show amplitude $\sqrt{n_\psi}$ and phase $\phi/\thbar$  of $\psi$.}
    \label{fig:compareRePsiImPsi}
\end{figure}

This seemingly contrived situation poses no problems in the SPE \eqref{schrPoissEqFRW} which can be numerically evolved through these events 
without any pathologies if one choses to split $\psi$ into real and imaginary parts,  $\Re(\psi)$ and $\Im(\psi)$, rather than amplitude and phase. 
The phase is ill-defined when $\psi=0$ and the Madelung equations \eqref{Madelungfluideq} no longer follow from 
\eqref{schrPoissEqFRW} if $\npsi=0$.\footnote{It is known that Madelung and Schr\"odinger equations are not necessarily 
equivalent \cite{Wallstrom1994}, see also the discussion of Sec.\,II\,A in \cite{SRvB89}. 
We therefore recommend, in contrast to a comment in \cite{HuiOstrikerTremaineEtal2016}, not to use the Madelung equations as a replacement for the Schr\"odinger equation, since the Madelung equations might not give all solutions, or even give wrong solutions  \cite{Wallstrom1994}. 
The reason is that even if the singularities \cite{SRvB89,UhlemannKoppHaugg2014} are dealt with, for instance, 
by using the momentum current $\v{j}_\psi= n_\psi \vu_\psi$ as the dynamical variable  rather than the velocity  $\vu_\psi$, 
one still has to implement the constraint \eqref{singlevaluedness} that ensures that $\vu_\psi$ is the gradient of a phase, 
or in other words that that resulting $\psi$ is a single valued complex function.  We thus expect that the applicability of a code 
based on the Madelung equations, like \cite{MoczSucci2015}, is limited to situations where nodes do not develop, and 
therefore is applicable only when the analogue of shell-crossing does not occur in the wave function. 
See Chapter 15.3 of \cite{TrahanWyatt2005} for an introduction to problems and possible solutions when one choses to stay close to the Madelung equations.}
In Fig.\,\ref{fig:compareRePsiImPsi} we show part of a simulation snapshot of the same simulation shown in Fig.\,\ref{fig:ScMteaser} but at an earlier time.
While our dynamical variables  $\Re(\psi)$ and $\Im(\psi)$, shown in the upper panel, are perfectly smooth, 
 the phase $\phi/\thbar$ of $\psi$, shown in the lower right panel, has several point-like singularities which persist in time 
and which coincide with locations where $n_\psi=0$. 
The phase has a non-zero winding number around these singularities, observed by following the color gradient around them.
 This circulation is the analogue of microscopic vorticity in the ScM and is discussed in more detail in Sec.\,\ref{Vorticity}.

We now present the definition of the ScM, that is, a procedure to extract approximate solutions to the Vlasov equation \eqref{VlasovPoissonEq} from solutions of the SPE for the most general case where $f$ is allowed to exhibit shell-crossing and mutli-streaming.

\subsubsection{The Husimi distribution: relation to the Vlasov equation}
\label{sec:Husimidist}
The so-called Husimi-representation $\psi_\H(\v{x},\v{u})$ of the wave function $\psi(\vx)$ in terms of gaussian wave packets gives rise to 
the (manifestly non-negative) Husimi phase space distribution \cite{H40}\footnote{The function $f_\H$ calculated this way is manifestly non-negative. 
It may, however, be also obtained from the Wigner quasi-probability distribution $f_\W[\psi]$
 by applying a Gaussian filter with width $\sigx$ and $\sigu=\thbar/(2 \sigx)$
such that $f_\H = e^{\frac{\sigx^2}{2}\Delta_x+\frac{\sigu^2}{2}\Delta_u} \{f_\W[\psi]\}  \label{fHfW}$.
This second route to $f_\H$ is, however, numerically inconvenient compared to \eqref{Husimi}, see \cite{UhlemannKoppHaugg2014}.
}
\begin{subequations}\label{Husimi}
\begin{empheq}[box=\mybluebox]{align}
f_\H(t, \v{x},\v{u}) &:=| \psi_\H(t, \v{x},\v{u}) |^2 \\
\psi_\H(t, \v{x},\v{u}) &:= \int \vol{d}{x'} K_\H(\v{x},\v{x}',\v{u})\, \psi(t, \v{x}')  \\
K_\H(\v{x},\v{x}',\v{u}) &:= \frac{\exp\left[-\frac{(\v{x}-\v{x}')^2}{4 \sigx^2} - \frac{i}{\thbar} \v{u}\cdot \v{x}' \right] }{\left( 2\pi \thbar \right)^{d/2}  \left(2 \pi \sigx^2 \right)^{d/4}}\,.
\end{empheq}
\end{subequations}
The spatial coarse graining scale $\sigx$
is the second parameter of the ScM. Together with $\thbar$, it  determines the resolution 
\begin{align}
\sigx\,,\qquad\sigu\equiv \frac{\thbar}{2 \sigx}
\end{align}
of the phase space density $f_\H$ in $\vx$ and $\vu$ direction. 
It can be shown that $f_\H$ approximates $\bar f$ in the sense that 
\begin{align}
\partial_t \left(  f_\H- \bar f\right) \simeq - \frac{\thbar^2}{24}(\del_{x_i}\del_{x_j}\del_{x_k}\varPhi_\H)  (\del_{u_i}\del_{u_j} \del_{u_k}   f_\H )\,, \label{fHandfcgDiff}
\end{align}
where $\varPhi_\H \equiv  e^{\frac{\sigx^2}{2} \Delta} \{\varPhi_\psi \}$, with $\varPhi_\psi$ from \eqref{PoissEqFRW}, see  \cite{T89, WK93, UhlemannKoppHaugg2014} and App.\,\ref{Diffoffs}.
Defining
\begin{subequations} \label{HusimiEqComponents}
\begin{align}
S_{\rm V} &\equiv  - \frac{\vu}{a^2}\cdot\vnabla_{\! \!  x} f_\H +  \vnabla_{\! \!  x} \varPhi_\H \cdot\vnabla_{\! \!  u} f_\H \\
S_{\rm cg V} & \equiv -  \frac{\sigu^2}{a^2 }\vnabla_{\! \!  x}\cdot\vnabla_{\! \!  u}  f_\H  +\sigx^2 (\del_{x_i}\del_{x_j}  \varPhi_\H)\, (\del_{x_i} \del_{u_j}  f_\H) \,,  \\
S_{\rm \thbar } &\equiv  -  \frac{\thbar^2}{24}(\del_{x_i}\del_{x_j}\del_{x_k}\varPhi_\H)  (\del_{u_i}\del_{u_j} \del_{u_k}   f_\H ) \, 
\end{align}
\end{subequations}
we generally have that $f_\H$ satisfies the Husimi equation  
\begin{equation}
 \partial_t  f_\H = S_{\rm V} + S_{\rm cg V} + S_{\thbar} + \mathcal{O}(\thbar^2 \sigx^2, \thbar^4)\, ,
\end{equation}
 but requiring that $f_\H$ also solves the coarse-grained Vlasov equation leads to $\partial_t  f_\H =S_{\rm V} + S_{\rm cg V} $,
 or even more optimally the Vlasov equation $\partial_t  f_\H =S_{\rm V}$. 
 Thus we know {\it i}) that we can achieve $f_\H \simeq \bar f$ only  if  quantum corrections to the Vlasov equation are smaller than coarse graining effects
 \begin{subequations}\label{fHGoodCondition}
\begin{empheq}[box=\mybluebox]{align}
| S_{\rm \thbar } | &\ll  |S_{\rm cg V}  | \,, \label{fHGoodCondition1}
\end{empheq}
 and {\it ii}) that $f_\H \simeq  f$, is possible if in addition to condition \eqref{fHGoodCondition1} the phase space smoothing scales $\sigx$ and $\sigu$ are much smaller than the smallest features of $f_\H$, or more precisely
 \begin{align}
 |S_{\rm cg V}| \ll |S_{\rm V}| \,.
 \label{fHGoodCondition2}
\end{align}
\end{subequations}
We present a  derivation in App.\,\ref{Diffoffs}.

In practice, we are more concerned with achieving $f_\H \simeq \bar f$, or \eqref{fHGoodCondition1} since it is anyway practically impossible 
to determine $f$ with sufficient spatial resolution for an arbitrary long time due to phase mixing. Regions where phase mixing 
 occurs  become  increasingly better approximated over time by a coarse-grained distribution \cite{L67}. 
Thus having $f_\H$ merely agree with $\bar f$ rather than $f$ is not only acceptable given the finiteness of computational resources, 
but also from a physical point of view.

On small scales quantum effects can become important as has been demonstrated and discussed in \cite{SchiveChiuehBroadhurst2014, SchiveLiaoWooEtal2014, SchwabeNiemeyerEngels2016, HuiOstrikerTremaineEtal2016}. 
In those studies it was assumed that dark matter is an axion-like particle in a Bose-Einstein condensate and that in the Newtonian limit this situation can be described by the SPE 
and $\psi$ more accurately  
than with the Vlasov equation.
 Independent of the physical interpretation of $\psi$, it was shown in \cite{SchiveChiuehBroadhurst2014,SchiveLiaoWooEtal2014,SchwabeNiemeyerEngels2016} that in halo centers, $|\psi|^2$ has a solitonic profile that transitions around the solitonic core radius $r_c$ into the Navarro-Frenk-White (NFW) profile. 
 Thus outside of $r_c$, the wave function behaves like collisionless matter and we expect $f_\H \simeq \bar f_c$ which is consistent with the NFW profile, 
while inside the solitonic radius we would expect $|S_{\thbar}| \simeq |S_{\rm V}|$ such that $f_\H \neq \bar f_c$. 
Since the soliton core size $r_c$ scales as $r_c \propto \thbar$, see \cite{SchiveLiaoWooEtal2014}, the breakdown of the ScM is under control.
 
Before we close the section on the Husimi distribution, we want to point out the important difference between the relation of $f_\H$ to $\bar f$ compared to the previously discussed relation between $n_d, \phi_\d$ and $\npsi, \phi$ in Sec.\,\eqref{subsec:Madelung}: the left-hand side of \eqref{fHGoodCondition} can be kept small even well after the first shell crossing, 
deep in the multi-stream regime \cite{T89,WK93, UhlemannKoppHaugg2014} by choosing a sufficiently small $\thbar$ and sufficiently large $\sigx$. 
 In contrast, the quantum pressure $Q$ in \eqref{EulerMadelung} automatically becomes large once the density contrast $\npsi -1$ becomes non-linear and hence the mapping between $n_\d, \phi_\d$ and $\npsi, \phi$ breaks down shortly before shell-crossing. A large $Q$ does not invalidate \eqref{fHGoodCondition}.

 \paragraph*{Summary for constructing the phase space density:}
\myotherbluebox{\begin{minipage}{0.48\textwidth}The ScM approximates solutions $f(t,\vx,\vu)$ to the Vlasov equation \eqref{VlasovPoissonEq} by constructing the Husimi distribution $f_\H(t,\vx,\vu)$ at a given time $t$ via \eqref{Husimi}, which involves a spatial integral of the snapshot of a wave function  $\psi(t, \vx)$ that satisfies the SPE \eqref{schrPoissEqFRW}. 
Due to the form of $K_\H$ the $\vx'$ integration is quasi-local and it is sufficient to use as integration domain a ball centered around $\vx$ with a radius of a few $\sigx$.
The accuracy of the ScM is controlled by the two parameters $\thbar$ and $\sigx$ via \eqref{fHGoodCondition}.\end{minipage}}

\subsubsection{Moments and cumulants} \label{sec:HusHierarchy}

The extraordinary feature of the ScM is that it allows to analytically evaluate the velocity space integral in \eqref{genfun}
\begin{empheq}[box=\mybluebox]{align}
 G_\H(\vx,\v{J}) &=\int \vol{d}{u}  e^{i\vu\cdot\v{J}}  f_\H \notag \\
&=e^{-\frac{1}{2}\sigu^2 \v{J}^2}e^{\frac{\sigx^2}{2} \Delta}\, \{G_\W(\vx,\v{J}) \}\label{genfuncgW}  \\ 
G_\W(\vx,\v{J})&=  \psi\left(\v{x}+\tfrac{\thbar}{2}\v{J}\right)\bar\psi\left(\v{x}-\tfrac{\thbar}{2}\v{J}\right) \notag\,,
\end{empheq}
and that this expression is quasi-local in space and, therefore, the resulting moments are efficiently evaluated numerically. This has the be contrasted to CDM, see \eqref{fCDMstreams} and \eqref{CDMMomentsstreams} where the moment generating function is non-local in $\vq$-space, or involves a sum over streams.

Using \eqref{moments}, the first three moments of the Husimi distribution are easily obtained by first evaluating the Wigner moments $M^{\W (n\leq2)}$ obtained from $G_\W$ 
\begin{subequations} \label{momentscgw}
\begin{align}
 M^{\W(0)} &=  |\psi|^2  \label{0momentw} \,, \\
 M^{\W(1)}_{ i}&= \thbar\,  \Im\left\{ \psi_{,i}\bar \psi\right\}\label{1momentw} \\
 M^{\W(2)}_{i j}&= \frac{\thbar^2}{2} \,\Re\left\{\psi_{,i}\bar \psi_{,j} -\psi_{,ij}\bar \psi\right\} \label{2momentw}
 \end{align}
and subsequent coarse graining
 \begin{align}
 M^{\H(0)} &=  e^{\frac{\sigx^2}{2} \Delta}\, \{M^{\W(0)} \}\label{0momentcgw} \,, \\
 &=:n^\H\notag\\
 M^{\H(1)}_{ i}&= e^{\frac{\sigx^2}{2} \Delta}\, \{M^{\W(1)}_{ i}\}\label{1momentcgw} \\
 &=:  n^\H  u^\H_{i} \notag\\
 M^{\H(2)}_{i j}&= e^{\frac{\sigx^2}{2} \Delta}\{M^{\W(2)}_{i j} \}+ \sigu^2 M^{\H(0)}  \delta_{ij}\label{2momentcgw} \\
 &=:  n^\H  u^\H_{i} u^\H_{j}+   n^\H\,  \Sigma^\H_{ij}  \notag \,,
\end{align}
where $\Re$ and $\Im$ denote real and imaginary part.
Comparing \eqref{2momentcgw} and \eqref{cg2ndmomentfrom2ndmoment} we recognize the same structure and  therefore it is the first term of \eqref{2momentcgw} that should be compared to $e^{\frac{\sigx^2}{2} \Delta} \{M^{(2)}_{ij}\}$.
All higher moments can be similarly calculated from \eqref{moments}. 
They are always bilinear in the wave function and its spatial derivatives with an overall spatial gaussian smoothing.
This makes it very easy to evaluate $M^{(n)}_\H(\vx)$ numerically from a time snapshot of $\psi(\vx)$. 

The $n$th cumulant $C^{(n)}_\H(\vx)$ of the Husimi distribution can be constructed from the first $n$ moments according to \eqref{cumulants}
\begin{align}
C^{\H(0)} &= \ln  n^\H \\
C^{\H(1)}_{i} &=  u^\H_i = M^{\H(1)}_{ i}/n^\H \label{C1ScM}\\
C^{\H(2)}_{ij} &=  \Sigma^\H_{ij} = M^{\H(2)}_{i j}/n^\H - u^\H_i u^\H_j \,. \label{2cumulantcgw} 
\end{align}
\end{subequations}
All higher cumulants can be constructed in a similar fashion.
If ones chooses to work directly with the moments and cumulants without constructing $f_\H$
 then one cannot test the accuracy of the ScM using \eqref{fHGoodCondition}.
We can however calculate the moments of the expressions  \eqref{HusimiEqComponents} entering the Husimi equation. The first non-vanishing moment 
of $S_{\thbar}$ is given by $\int \vol{3}{u} u_i u_j u_k\, S_{\thbar}$, such that we consider 
\begin{subequations} \label{HusimiMomentEqComponents}
\begin{align}
S^{(3)}_{\rm V}{}_{ijk} & =-\frac{1}{a^2 } \nabla_m M^{\H (4)}_{i j k m}  - \nabla_{(i} \varPhi_\H \cdot M^{\H (2)}_{j k)}\\
S^{(3)}_{\rm cg V}{}_{ijk} 
&=   \frac{\sigu^2}{a^2 }\nabla_{(i}  M^{\H (2)}_{j k)} - \sigx^2 \nabla_m \nabla_{(i} \varPhi_\H \nabla_m  M^{\H (2)}_{j k)} \\
S^{(3)}_{\rm \thbar }{}_{ijk} &   = \frac{\thbar^2}{4} n_\H \nabla_i \nabla_j \nabla_k \varPhi_\H\,,
\end{align}
\end{subequations}
where $A_{(ijk)} = A_{ijk}+A_{jki}+A_{kij}$.
The minimal requirement to satisfy the ScM condition \eqref{fHGoodCondition1} thus is
\begin{subequations}\label{M3HGoodCondition}
\begin{empheq}[box=\mybluebox]{align}
| S^{(3)}_{\rm \thbar }{}_{ijk} | &\ll  |S^{(3)}_{\rm cg V}{}_{ijk}  | \,, \label{M3HGoodCondition1}
\end{empheq}
 and similarly, a requirement for \eqref{fHGoodCondition2} is
 \begin{align}
 |S^{(3)}_{\rm cg V}{}_{ijk}| \ll |S^{(3)}_{\rm V}{}_{ijk}| \,. \label{M3HGoodCondition2}
\end{align}
\end{subequations}

 \paragraph*{Summary for constructing the moments and cumulants}
\myotherbluebox{\begin{minipage}{0.48\textwidth}The ScM allows the construction of moments $M^{\H(n)}(t, \vx)$ and cumulants  $C^{\H(n)}(t, \vx)$ using the generating function \eqref{genfuncgW} in \eqref{momentsandcumulants}, avoiding the cumbersome $2d$-dimensional phase space.
The construction of the moments from $\psi(t, \vx)$ at a given time $t$, see the results in \eqref{momentscgw}, takes place exclusively in $d$-dimensional position space and only involves calculating products of the wave function and its spatial derivates at that particular time $t$, as well as a final gaussian spatial filtering with variance $\sigx^2$. 
The spatial derivatives as well as the filtering are quasi-local processes and it is sufficient to perform the convolution in a neighborhood of $\vx$ of size of a few $\sigx$. 
The such constructed moments and cumulants approximate those of $\bar f$, see \eqref{cgmomentsfrommoments}, and therefore solve the full Boltzmann hierarchy of $\bar f$ without any truncation by virtue of \eqref{M3HGoodCondition}.
\end{minipage}}

\subsubsection{Cold initial conditions}
 The stress tensor $T_{ij}$, \eqref{stresstensor}, of a pressureless perfect fluid has the form $T^\d_{ij}= \rho_0\, n_\d\, a^{-3}\, u^\d_i u^\d_j$. 
 The resulting ScM stress tensor $T^\H_{ij}$  contains an additional contribution $\Sigma^\H_{ij}$ which can be interpreted, according to \eqref{2momentcgw} and \eqref{2cumulantcgw}, as arising solely from velocity dispersion by virtue of the correspondence $\bar f \simeq f_\H$ \eqref{fHandfcgDiff}.
One defining aspect of CDM, that it is initially a pressureless perfect fluid with $\Sigma^\d_{ij} = 0$, is not automatically implemented by the ScM. 
It can however be easily achieved, remembering that if the quantum pressure $Q$ in \ref{EulerMadelung}  is initially small 
 the SPE approximates the dust fluid \eqref{Fluideq} with arbitrarily precision.
The observation that $\nabla_i (n^\H \Sigma^\H_{ij})$ is proportional to $e^{\frac{\sigx^2}{2} \Delta}\,\{n_\psi\nabla_j Q\}$, see \cite{UhlemannKoppHaugg2014}, therefore shows us how to implement dust-like, and therefore nearly cold initial conditions.
To implement cold initial conditions, we thus choose 
 \begin{subequations}\label{inipsi}
\begin{empheq}[box=\mybluebox]{align}
\psi_{\rm{ini}}(\vx)&\equiv \sqrt{n_{\rm d}(t_{\rm ini},\vx)} \exp\left[ \tfrac{i}{\thbar} \phi_{\rm d}(t_{\rm ini},\vx) \right]  \label{inipsiFromDust}\\
\mathrm{with~\,}\thbar\mathrm{\,~such}&\mathrm{~that~}\vnabla Q(t_{\rm ini},\vx) \ll \vnabla \varPhi_\psi(t_{\rm ini},\vx) \,, \label{hbarCondition}
\end{empheq}
\end{subequations}
where  $n_{\rm d}(t_{\rm ini},\vx)$ and $\phi_{\rm d}(t_{\rm ini},\vx)$ are obtained via \eqref{Fluideq} before shell-crossing and $Q(t_{\rm ini},\vx)$ and $\varPhi_\psi(t_{\rm ini},\vx) $ are obtained by evaluating \eqref{Qdef} and \eqref{Poisspsi} with $\psi_{\rm{ini}}(\vx)$.
 This choice of $\thbar$ guarantees that $f_\H(t_{\rm ini}) \simeq f(t_{\rm ini}) = f_\d(t_{\rm ini})$.\footnote{ If $\vnabla Q \ll \vnabla \varPhi_\psi$ everywhere, then $f_\H$ underestimates the maximal achievable precision with which the SPE can model the dust fluid \eqref{Fluideq}, see \cite{UhlemannKoppHaugg2014}. In this case, $f_{\rm naive} = n_\psi( \vx)\, \delta_{\mathrm D}\big(\vu-\vnabla \phi(\vx) \big)$ is closer to $f_\d$ than $f_\H$.}
Note that ScM does not force us to use cold initial conditions and we leave it for the future to investigate $\psi_{\rm{ini}}$ that correspond to warm initial conditions.
We summarise the main features of the ScM and compare them to two other models of CDM in Table \ref{tab:tabScM}.  

\section{Numerical implementation of the 2D ScM} 
\label{sec:Implementation}
In this section we describe the algorithms used to solve the SPE \eqref{schrPoissEqFRW} and the implementation of  \eqref{momentscgw} 
for constructing the Husimi cumulants by avoiding phase space. 
The two-dimensional code can be interpreted to describe 3 dimensional dynamics in special situations where the wave function is shift symmetric 
in the $z$-direction, where $\vx=(x,y,z)$ are the cartesian coordinates. 
\subsection{SPE}
For the numerical implementation of the ScM we use $a$ as the time variable, where $da = a H dt$ and $H$ is the Hubble parameter, 
thus the Schr\"odinger equation \eqref{schrEqFRW} takes the following form
\begin{align} \label{schrEqtimea}
i\thbar \del_a \psi &= - \frac{\thbar^2}{2a^3H} \Delta \psi + \frac{\varPhi_\psi}{aH} \psi.
\end{align}
Following \cite{PeacemanRachford1955, G95, HarrisonMorozTod2003}, we use an alternating direction implicit (ADI) method to discretize and split the single two-dimensional SPE\,\eqref{schrEqtimea} into three equations: 
\begin{subequations}\label{schrbroken}
\begin{align} 
e^{-\frac{i\lambda\thbar \epsilon}{4Ha^{3}}\frac{\partial^{2}}{\partial x^{2}}}S(x,y) &=e^{\frac{i\lambda\thbar \epsilon}{4Ha^{3}}\frac{\partial^{2}}{\partial x^{2}}} \psi(a,x,y) \\
e^{-\frac{i\lambda\thbar \epsilon}{4Ha^{3}}\frac{\partial^{2}}{\partial y^{2}}}T(x,y) &=e^{\frac{i\lambda\thbar \epsilon}{4Ha^{3}}\frac{\partial^{2}}{\partial y^{2}}} S(x,y) \\
e^{\frac{i\lambda \epsilon}{2\thbar Ha}\varPhi_\psi}\psi(a+\lambda \epsilon,x,y) &= e^{-\frac{i\lambda \epsilon}{2\thbar Ha}\varPhi_\psi} T(x,y)
\end{align}
\end{subequations}
where $\epsilon$ is the spatial mesh width of the finite difference mesh, $da$ the temporal step and $\lambda=da/\epsilon$. Then, one can use the central finite difference approximation of third order accuracy 
\begin{multline} \label{centralfindiff}
\frac{\partial^{2}f(x_{0})}{\partial x^{2}} \approx  \frac{1}{12\epsilon^{2}} 
 \Big[ -f(x_{0}-2\epsilon) +16f(x_{0}-\epsilon) -30f(x_{0}) \\
+16f(x_{0}+\epsilon) -f(x_{0}+2\epsilon)\Big]
\end{multline}
to replace the second order derivatives in \eqref{schrbroken}. By defining the operator 
\begin{multline} \label{centralfindiffoperator}
\hat{\delta}_{x}f(x_{0})\equiv -f(x_{0}-2\epsilon) +16f(x_{0}-\epsilon) \\
 -30f(x_{0}) +16f(x_{0}+\epsilon) -f(x_{0}+2\epsilon),
\end{multline}

expanding the exponentials to the lowest significant order in $\epsilon$ and replacing the continuous variables with discretised  notation, we obtain
\begin{subequations}\label{schrbrokendiscret}
\begin{align} 
\left(1-\frac{i\lambda\thbar}{48\epsilon Ha^{3}}\hat{\delta}_{i}\right)S_{ij} &=\left(1+\frac{i\lambda\thbar}{48\epsilon Ha^{3}}\hat{\delta}_{i}\right)\psi_{ij}^{n} \\
\left(1-\frac{i\lambda\thbar}{48 \epsilon Ha^{3}}\hat{\delta}_{j}\right)T_{ij} &= \left(1+\frac{i\lambda\thbar}{48\epsilon Ha^{3}}\hat{\delta}_{j}\right)S_{ij} \\
\left(1+\frac{i\lambda \epsilon}{2\thbar Ha}\varPhi_{\psi\, ij}^{n}\right)\psi_{ij}^{n+1} &= \left(1-\frac{i\lambda \epsilon}{2\thbar Ha}\varPhi_{\psi\, ij}^{n}\right) T_{ij}
\end{align}
\end{subequations}
where: 
\begin{align} 
\psi_{ij}^{n}=\psi(a_{0}+n\lambda \epsilon,x_{0}+i\epsilon,y_{0}+j\epsilon)=\psi(a,x,y).
\end{align}

These difference equations are in the Crank-Nicolson form which guarantees that they are unconditionally stable and they have an error of $\mathcal{O}(\lambda^{2} \epsilon^{2})$ in time and $\mathcal{O}(\epsilon^{4})$ in space. Furthermore, the algorithm is manifestly unitary \cite{GoldbergScheySchwartz1967}, which means that like in a N-body simulation mass conservation is automatic.

We implemented the SPE on a single Nvidia K20X GPU (Kepler architecture) using CUDA C.
Due to the memory constraints of the K20X, that is $6GB$ of device memory,
we solved \eqref{schrbrokendiscret} with a maximum of $8192^2$ points on a regular grid with periodic boundary conditions. 
Finer resolution  is not possible without splitting the full simulation box into regions and 
either simply copying parts of the simulation box to CPU memory and back,
or even more efficiently, using multiple GPUs with appropriate communication between them on region boundaries.
 Substantial speed improvement over a single CPU was observed by ensuring minimal communication between the host (CPU) and the device (GPU),
 making sure that kernels had coalesced access to global device memory and
efficiently using the device local memory. More detailed description of our numerical implementation, as well as the study of 
 scaling with simulation size is left for a more dedicated publication.

To advance the wave function $\psi_{ij}^{n} \rightarrow \psi_{ij}^{n+1}$ a single time step we solved the set of cyclic penta-diagonal linear 
systems \eqref{schrbrokendiscret}  using the very efficient algorithm  described in  \cite{JiaJiang2013}. 
Note that we did not use the first order accuracy formula for the second order derivatives that was used in \cite{G95}. 
From numerical tests we found that it was not sufficient to solve \eqref{schrbrokendiscret} accurately enough for our initial conditions, 
in particular the Layzer-Irvine test, described in the following subsection, failed at significantly earlier times. 
We also did not use the midpoint value of the potential, $\varPhi_{\psi,ij}^{n+1/2}$ in \eqref{schrbrokendiscret}, as suggested in \cite{G95}, since in our case the gravitational potential is very slowly evolving. 
Indeed, although we solved the Poisson equation at every time step, 
we found very good agreement even if  $\varPhi^n_{\psi\, ij}$ was updated only every 20 time steps. 
To calculate the gravitational potential, we solved the Poisson equation \eqref{PoissEqFRW} in Fourier space, using the
CuFFT~\footnote{See \url{http://docs.nvidia.com/cuda/cufft/index.html}.} implementation of
 the  Fast Fourier Transform (FFT) method with periodic boundary conditions on the same grid.

We remind the reader that the dynamical variables are the real part $\Re(\varPsi)$ and imaginary part $\Im(\psi)$ of the wave function $\psi$. 
These functions are strongly oscillating but otherwise well behaved, see the top panels of Fig.\,\ref{fig:compareRePsiImPsi}.
While the absolute value $|\psi| = \sqrt{n_\psi}$ is also a smooth function, the phase $\phi/\thbar$ is not, see the lower panels of Fig.\ref{fig:compareRePsiImPsi}.
The phase has singular points precisely at locations at which $n_\psi=0$. 
The meaning and importance of these phase singularities is discussed in section \ref{Vorticity}.

\subsection{Layzer-Irvine test of numerical accuracy}
Since our numerical method  is manifestly unitary \cite{GoldbergScheySchwartz1967},  
testing whether the total mass $$M(t)= \rho_0 \int \vol{3}{x} | \psi(\vx)|^2 $$ is conserved at all times does not tell us how accurately we solve the SPE.
 Instead we consider the total energy per mass \eqref{psiHamiltonianSystem}
 \begin{subequations} \label{totalEnergy}
 \begin{align}
  E(t) & =:  K +  W  
  \end{align}
where we defined the kinetic $K$ and potential $W$ energy through
\begin{align}
 K(t) &= \frac{\thbar^2}{2 a^2} \int \vol{3}{x} |\vnabla_{\!\! x} \psi(\vx)|^2  \\
 W(t) &  =  - \frac{ \rho_0 G }{2 a} \int \vol{3}{x} \vol{3}{x'} \frac{  |\psi(\vx)|^2 |\psi(\v{x}')|^2}{|\vx-\v{x}'|}\\ 
& =\frac{1}{2} \int \vol{3}{x} \varPhi_\psi(t,\vx)\, |\psi(\vx)|^2 
\end{align}
\end{subequations}
where $\varPhi_\psi$ is the Newtonian potential from \eqref{PoissEqFRW}.
Since $\mathcal{E}$ depends explicitly on time through the scale factor $a$ and $\{E,E \} = 0$ we have 
\begin{equation}
\frac{d  E}{dt} = \partial_t  E \,. \label{EnergyNonCons}
\end{equation}
Inserting the expressions for $ K$ and $ W$ and taking into account that partial time derivatives act only on the scale factor $a(t)$, since $\psi(\vx)$ have to be interpreted as canonical position variables we get 
\begin{equation}
\frac{d}{dt}(  K +  W)  =-  \frac{\dot a}{a} (2  K +  W) \,.\label{EnergyNonCons2}
\end{equation}
This coincides with the Layzer-Irvine equation \cite{Layzer1963, P80, JoyceLabini2013, SousbieColombi2016}
because we have the same explicit time dependence in the Hamiltonians for $N$ particles \eqref{HamiltonSystem} and the Schr\"odinger field \eqref{totalEnergy}. 
We can rewrite  \eqref{EnergyNonCons2} as
\begin{equation}
\frac{d}{da}\left[ a E \right] =- K(a)\label{EnergyNonCons3} \,.
\end{equation}
 Since $K\geq 0$ the energy $E$ decays at least as $a^{-1}$.
We can integrate \eqref{EnergyNonCons2} to define a total conserved energy 
\begin{equation}
E_{\rm tot} := E(t) + E_{\rm exp}(t) \,, \quad E_{\rm exp} =  \int_{a_{\rm ini}}^a \frac{ 2 K(a') + W(a')}{a'} d a'  \label{EnergyNonCons4}
\end{equation}
 where we chose arbitrarliy $E_{\rm tot}(a_{\rm ini})=E(a_{\rm ini})$. 
 A useful test for the accuracy of numerical integration is thus to  check that 
 \begin{equation} \label{EnergyNonCons6}
 \delta_{E_{\rm tot}} \equiv \frac{E_{\rm tot}(a)}{E(a_{\rm ini})} -1
 \end{equation}
  remains close to zero.\footnote{If wanted the integral to contain only $W$, we could rewrite \eqref{EnergyNonCons3} as $\frac{d}{da}\left[ a^2 E \right] =aW(a)$  which leads to F.17 in \cite{SousbieColombi2016} taking into account that $E^{\texttt{ColDICE}}_{\rm k} = a^2 K$, $E^{\texttt{ColDICE}}_{\rm p} = a^2 W$. Note that there is a typo in F.20 which should read $E^{\texttt{ColDICE}}_{\rm exp} = - \int_0^a\frac{1}{a'} E_{\rm p}(a')da'$.}
Because \eqref{EnergyNonCons4} requires to calculate $K(a), W(a)$ at all times, we can alternatively test \eqref{EnergyNonCons3} 
by numerically evaluating $\frac{d}{da}\left[ a E \right]$ using a 9-point stencil, to test that
\begin{equation}
\delta_K\equiv\frac{\frac{d}{da}\left[ a E \right]}{ -K(a)} -1 \label{EnergyNonCons5}
\end{equation}
remains close to zero. 
Fig.\,\ref{fig:EnergyCons} shows $\delta_K$ for a test simulation of a 2D pancake collapse detailed in Sec.\,\ref{sec:sinewavecollapse}. 
In the range of times where our comparisons with \texttt{ColDICE} were performed ($0.02< a<0.088$), $\delta_K$ does not deviate from 0 further than $0.1\%$.
     During later times $\delta_K$ departs further from 0. This happens earlier with decreasing spatial and temporal resolution.
     By experimenting with the spatial resolution, the derivative approximation scheme, and time resolution we found 
that spatial resolution and the accuracy of the spatial derivative scheme have the largest impact on the time 
when $\delta_K$ starts to deviate from 0, see Fig.\,\ref{fig:EnergyCons}.

These findings indicate that the use of  adaptive mesh refinement (AMR) is imperative in order for a numerical implementation of the ScM 
to function as a general-purpose tool. 
We note that an efficient framework for implementing AMR in three dimentions using GPUs is publicly available.\footnote{\texttt{GAMER} is available at \url{http://iccs.lbl.gov/research/isaac/GAMER_Framework.html}.} The authors of \cite{SchiveChiuehBroadhurst2014}, have used a \texttt{GAMER}
add-on  called \texttt{ELBDM} (extremely light bosonic dark matter) which solves the SpE, however, \texttt{ELBDM} is  at present
not publicly available.

\subsection{Moments}
We use a 9-stencil finite difference method to calculate the spatial gradients entering the expressions for $M^{\W(n)}$ in \eqref{momentscgw}. 
To obtain the desired Husimi moments $M^{\H(n)}$ we apply a gaussian filter in an efficient way, described in App.\,\ref{sec:numericalimplementation}.

\subsection{The choice of $\hbar$}

In order to determine the value of $\thbar$ for cold initial conditions, we used the condition \eqref{hbarCondition} and calculated 
the $\thbar$-independent number 
\begin{equation}
\tilde{Q} =\max_{\vx}\left(\frac{|\vnabla Q(t_{\rm ini},\vx)| }{\thbar^2 | \vnabla \varPhi_\psi(t_{\rm ini},\vx) | }\right)\,,
\end{equation}
 and chosen $\thbar$ such that 
 \begin{equation}
 \thbar \ll \tilde{Q}^{-1/2} \,.
 \end{equation}
 
In practice we cannot choose arbitrary small values of $\thbar$ due numerical limitations:  spatial variations of the wave function $|\psi|/|\vnabla \psi| \simeq \thbar/|\vnabla \phi|$ need to be well resolved by the spatial mesh width $\epsilon$, such that we also require 
\begin{equation}
\thbar \gg \epsilon\, \max_{\vx} \,|\vnabla \phi (t_{\rm ini},\vx)| \,.
\end{equation}

 The closer one gets to violating these conditions, the earlier during the time evolution will the quantity $\delta_K$ deviate from 0, such that for a given problem the value of $\thbar$ has to be chosen carefully.

\section{Tests of the ScM in two dimensions}
 \label{sec:numerics}
We study two-dimensional cosmological simulations. 
In terms of three dimensional space, these simulations can be interpreted as having a phase space density that is shift symmetric in the $z$-direction, 
where $\vx=(x,y,z)$, and that vanishes for $u_z\neq0$, where $\vu=(u_x,u_y,u_z)$ are the canonical phase space coordinates.
Interpreted in three dimensions, this initial condition leads to a filament  that extends indefinitely in $z$-direction. 
This restriction, although not of much physical relevance, allows us to run two-dimensional simulations with very high resolution with the goal to disentangle any possible failure of the ScM from simple possible numerical inaccuracies.

The theoretical limitation is twofold, while $f_\c$ would indefinitely produce ever smaller structures in the winding-up phase space sheet, the ScM with a fixed value of $\thbar$ is not able to resolve structures below the phase space scale $\thbar$ even if the SPE were solved exactly.
Secondly, even on phase space scales larger than $\thbar$, the coarse-grained CDM dynamics, namely that of $\bar f_\c$, cannot exactly be described by $f^\H$ since due to \eqref{fHandfcgDiff} $$\partial_t(f_\H - \bar f_\c) \simeq  \left( \frac{\thbar}{x_{\rm typ} u_{\rm typ}} \right)^2\, \frac{\varPhi_\H f_\H}{x_{\rm typ} u_{\rm typ}}\,, $$
where typical scales, $x_{\rm typ}$ and $u_{\rm typ}$, are the scale on which $f_\H$ varies most strongly.
 Because $x_{\rm typ}$ continues to shrink over time as smaller and smaller phase space structures form, the right-hand side might become too large 
at which point the ScM breaks down.  It is this second point that requires a numerical proof and for this we perform a 
detailed comparison with the publicly available code \texttt{ColDICE} which solves the  Vlasov equation \eqref{VlasovPoissonEq}.
\begin{figure*}[t]
    \centering
    \includegraphics[width=11cm, angle=0,trim=0 0 0 0]{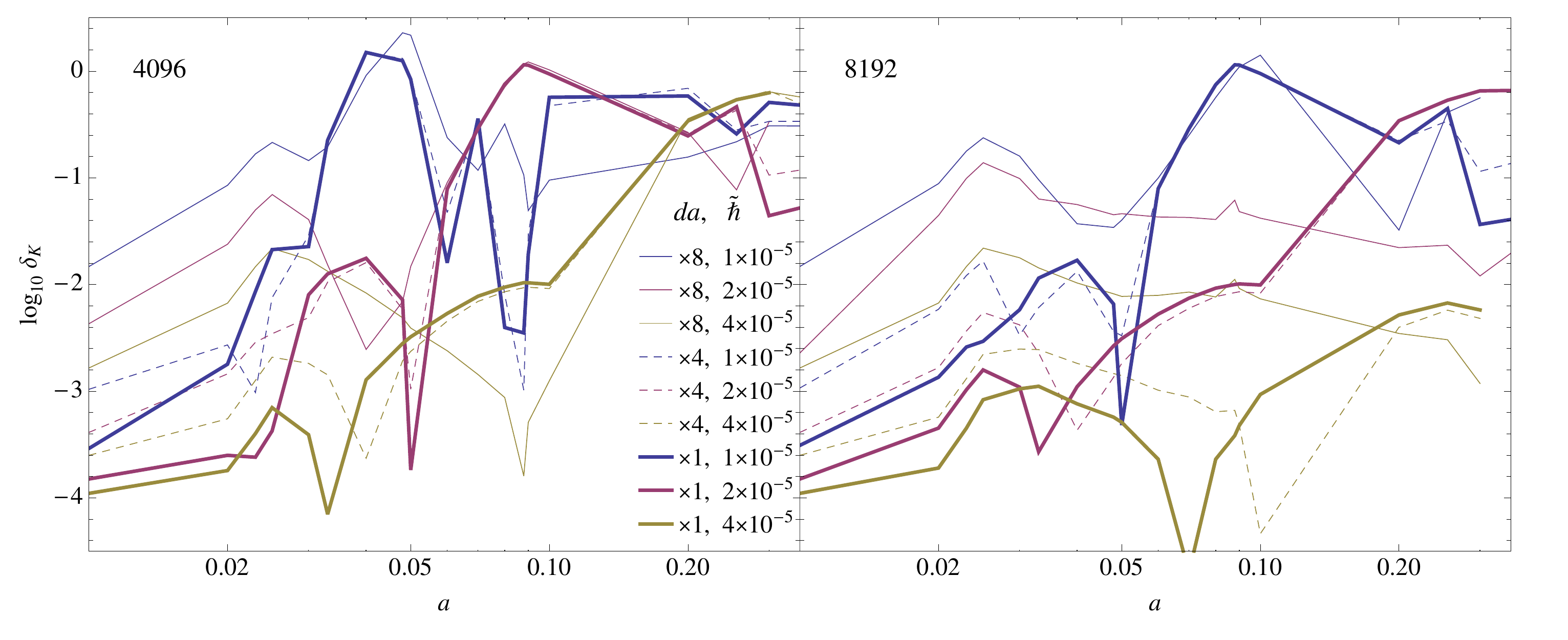}
    \includegraphics[width=6.7cm, angle=0, trim=0 -0.5cm 0 0]{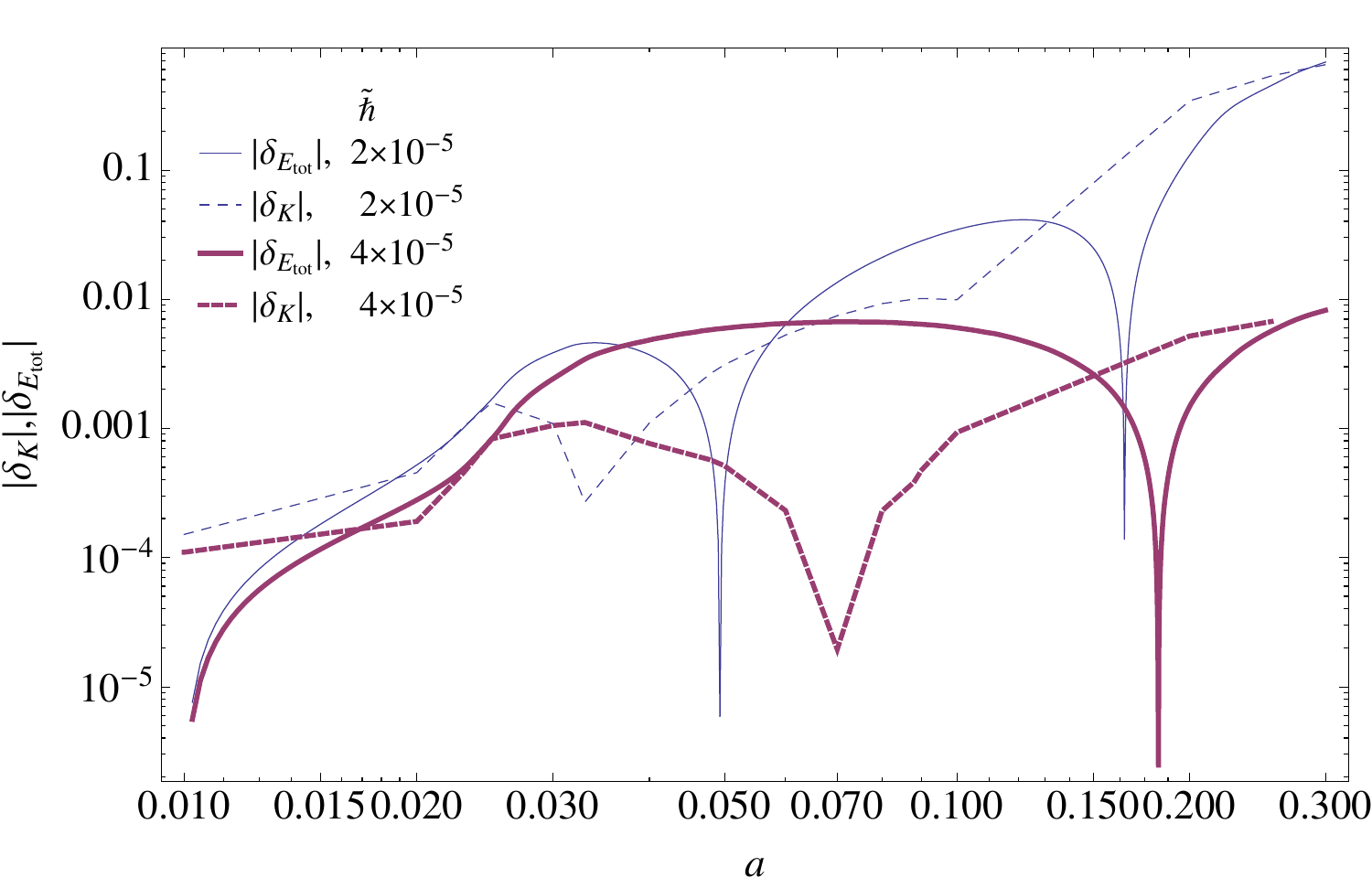}
    \caption{{\it Left}: Differential Layzer-Irvine test for the SPE solver for sine initial condition and two different spatial resolutions 4096 ({\it left}) and 8192 ({\it right}) for three values of $da$ as indicated in the legend ($\times 8$ means that d$a$ is 8 times larger) and 3 different values of $\thbar$. The value of $A$, \eqref{EnergyNonCons5}, is calculated from the snapshots of the SPE presented in Sec.\,\ref{sec:sinewavecollapse}. {\it Right}: Comparison of differential and integrated energy test for the resolution of 8192.}
    \label{fig:EnergyCons}
\end{figure*} 

We considered two kinds of cosmological simulations both in a $\Lambda$CDM universe where the Hubble parameter is 
\begin{equation}\label{HubbleLCDM}
H^2 = H_0^2 \left( \Omega_m a^{-3}+ 1-\Omega_m \right)\,.
\end{equation}
  The first test is a sine wave collapse, which is a traditional test for N-body codes \cite{KS83} for which we set  $\Omega_m=1$. 
  The second test uses a gaussian random field to generate the displacement field, and we chose the value $\Omega_m=0.312$.
  The translation of the Zel'dovich displacements used to set up the initial conditions in \texttt{ColDICE} into an initial wave function $\psi_{\rm ini}$ is detailed in App.\,\ref{sec:ZAini}.
We have chosen units where $H_0=G=c=1$.\footnote{Such that 
we measure length in units of $H_0^{-1}c$, time in units of $H_0^{-1}$ and mass in units of $H_0^{-1}c^3/G$.}

\subsection{Sine wave collapse}\label{sec:sinewavecollapse}

As we tested the ScM by comparing with \texttt{ColDICE}, 
we found it convenient to use the same parameters 
for the initial sine wave and initial integration redshift and units as in the two-dimensional sine wave collapse presented 
in the \texttt{ColDICE} article\cite{SousbieColombi2016}. 
  We checked that the initial conditions for \texttt{ColDICE} and the SPE solver agree to better 
than 0.025$\%$.\footnote{In the sense that $n(a_{\rm ini}, x,y)/ n_\psi(a_{\rm ini},x,y) -1$,   $u_{x}(a_{\rm ini}, x,y)/ \partial_x \phi(a_{\rm ini}, x,y) -1$ as well as the $y$-component of the velocity agree better than $2.5\times10^{-4}$.
  Note this test is non-trivial as it involves the extraction of $M^{\c(0)}$ and $M^{\c(1)}$ from the phase space sheet.}
Therefore, an increasing deviation signals inaccuracies in either of the two codes, or if this can be excluded, fundamental limitations of the ScM.
The initial conditions are detailed in App.\,\ref{sec:ZAplanewave}.

We ran a number of simulations by changing the number of grid points (either $4096^2$ or $8192^2$ for the same box size), 
the value of $\thbar = \{ 1  \times 10^{-5}, 2  \times 10^{-5} , 4  \times 10^{-5} \}$ 
and the temporal resolution $da/da_{{\rm base}}= \{ 1\times, 4\times, 8\times\}$.  
From this set, we performed  numerical accuracy tests in order to decide which simulation to use for our comparison, and at the same time providing
 a justification. All simulations had the same initial and final times, $a = \{0.01,0.3\}$ respectively.

\subsubsection{Accuracy considerations}

To make sure our comparison does not suffer from  numerical inaccuracies, we employed the Layzer-Irvine test \eqref{EnergyNonCons5}  
for deciding until which time the SPE solver may be trusted.  
For the particular initial conditions chosen we may observe in Fig.\,\ref{fig:EnergyCons} that the numerical SPE solution 
may be trusted until $a=0.1$ for the simulation with $4096^2$ spatial grid points and up to $a=0.3$ for the $8192^2$ grid. 
In both cases only the simulations with the largest considered value of $\thbar$  are accurate enough, that is $\thbar = 4\times 10^{-5}\,$Mpc.
Simulations with smaller values of $\thbar$ would need finer spatial resolution, that is, a larger grid of points.
Hence, we chose for our comparisons the simulation with $\thbar = 4\times 10^{-5}\,$Mpc and $8192^2$ where the differential energy 
test \eqref{EnergyNonCons5} is satisfied to better than $0.1\%$ for $a<0.1$. For this simulation one can expect that any deviations
 between the coarse-grained moments of \texttt{ColDICE}  and the corresponding ones from the ScM 
can be attributed to a failure of the ScM due to \eqref{fHandfcgDiff}, rather than numerical inaccuracy and thus  constitutes 
a test of the ScM.
In the right panel of Fig.\,\ref{fig:EnergyCons} we show that the differential energy test $\delta_K$ \eqref{EnergyNonCons5}, that can be applied to single snapshots, is a good indicator for the full energy test $\delta_{E_{\rm tot}}$ \eqref{EnergyNonCons6}, that requires evaluation of $K$ and $W$ at each time step during numerical integration of the SPE.

\begin{figure}[t]
    \centering
    \includegraphics[width=7.5cm, angle=0]{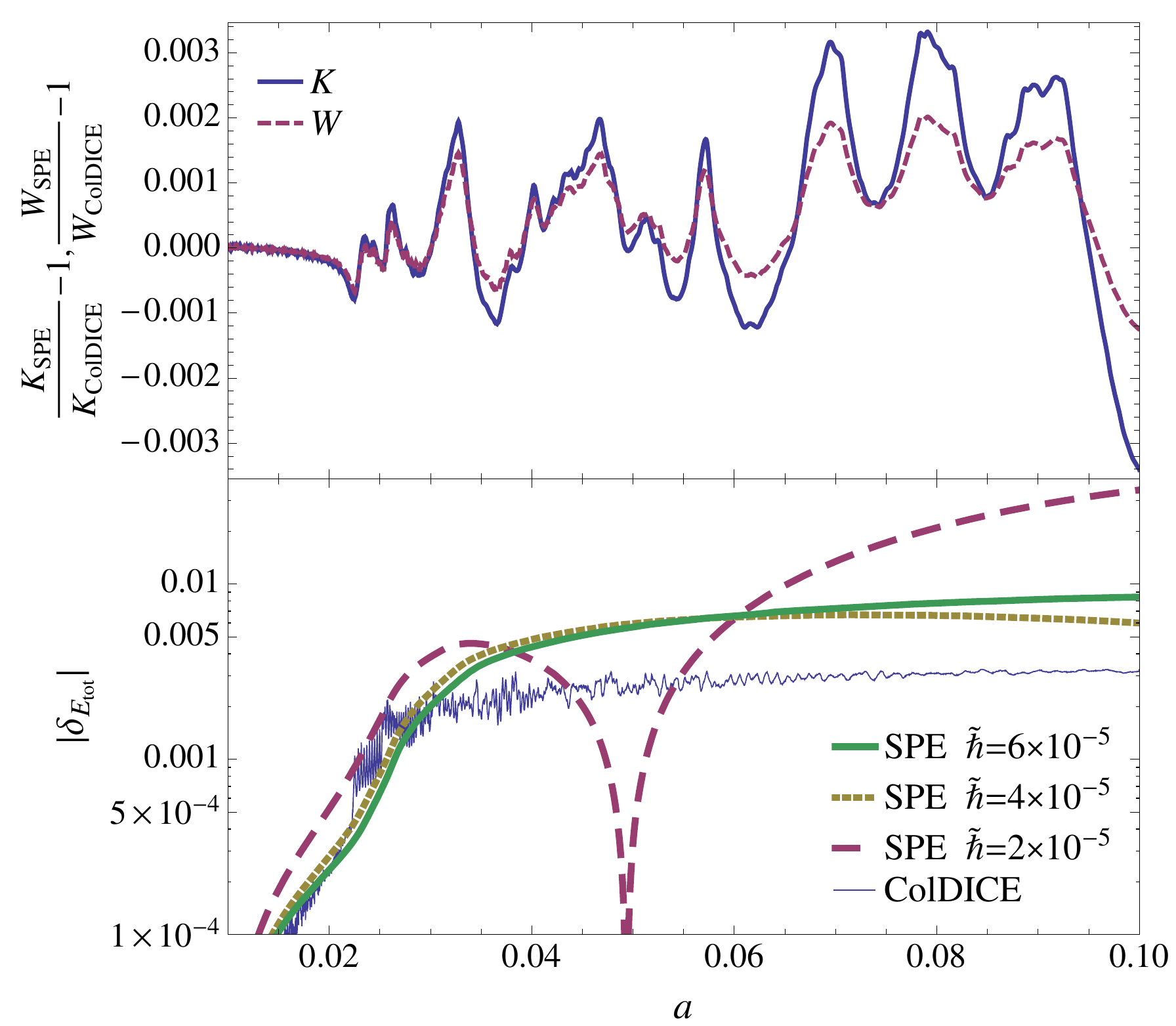}
    \caption{ {\it Upper panel}: Comparison of the kinetic ($K$) and potential ($W$) energy as obtained by \texttt{ColDICE} and the SPE solver with parameters as described in the text. 
 \\
    {\it Lower panel}: Comparison of the total energy test between \texttt{ColDICE} (thin blue) and the SPE solver with settings as described in the text (dashed), together with two other SPE solutions which differ only be the choice of $\thbar$.}
    \label{fig:CompareESPEandDice}
\end{figure}

We used the publicly available code \texttt{ColDICE} to solve the Vlasov equation \eqref{VlasovPoissonEq} for $f_\c(\vx,\vu)$ and to calculate $n(\vx)  = M^{\c(0)}(\vx) $, $M^{\c(1)}_i(\vx)$, and $M^{\c(2)}_{ij}(\vx)$ from it.\footnote{To calculate $M^{\c(1)}_i(\vx)$ and $M^{\c(2)}_{ij}(\vx)$ from $f_\c(\vx,\vu)$ we use an add-on for \texttt{ColDICE} that has been kindly provided to us by T.\,Sousbie.}
These quantities were then coarse-grained according to \eqref{cgmomentsfrommoments} with the same values of $\sigx$ and $\sigu=\thbar/(2\sigx)$ used in the ScM and then compared to $n^\H(\vx)$, $M^{\H (1)}_{i}(\vx)$ and $M^{\H (2)}_{ij}(\vx)$, \eqref{Husimi} and \eqref{momentscgw}, with  $\psi$ satisfying the SPE  \eqref{schrPoissEqFRW}.
We ran \texttt{ColDICE} with a precision setting for the invariant threshold\footnote{This number determines the refinement with which the phase space sheet is sampled.} $\epsilon_I = 10^{-8}$, and a force resolution grid with $1024^2$ pixels, for which equation \eqref{EnergyNonCons4} is satisfied to better than $0.2\%$, see the thin jagged curve Fig.\,\ref{fig:CompareESPEandDice}.
It it thus justified to take the output of \texttt{ColDICE} to represent $f_\c$, \eqref{fCDM} and its moments \eqref{CDMMoments}.

A comparison of $K$ and $W$ between \texttt{ColDICE} and our SPE solver is shown in the upper panel of Fig.\,\ref{fig:CompareESPEandDice}.
The agreement between the codes  is better than $0.2\%$ in the range $0.01<a<0.09$ for both variables.
In the lower panel we compare the total energy conservation of \texttt{ColDICE} and the SPE solver for three different values of $\thbar$. 
We observe that $E_{\rm tot}$ is conserved to a satisfactory level for $\thbar = 4\times10^{-5}$ and $\thbar = 6\times10^{-5}$, but not so for $\thbar = 2\times10^{-5}$, where numerical 
errors start to accumulate too early in the evolution, causing $\delta_{ E_{\rm tot}}$ to change sign around $a=0.05$.
We conclude from Fig.\,\ref{fig:CompareESPEandDice} that the SPE is solved with sufficient accuracy such that our subsequent comparison 
of cumulants $\bar C^{\c (n)}$ and $ C^{\H (n)}$ is a test of the ScM itself, in the sense that observed 
deviations between $\bar C^{\c (n)}$ and $ C^{\H (n)}$ are not due to numerical inaccuracies in either \texttt{ColDICE} or our 
SPE solver, but due quantum artifacts discussed in Sec.\,\ref{sec:HusHierarchy}.

\FloatBarrier
\begin{figure*}[!]
\includegraphics[width=0.90\textwidth]{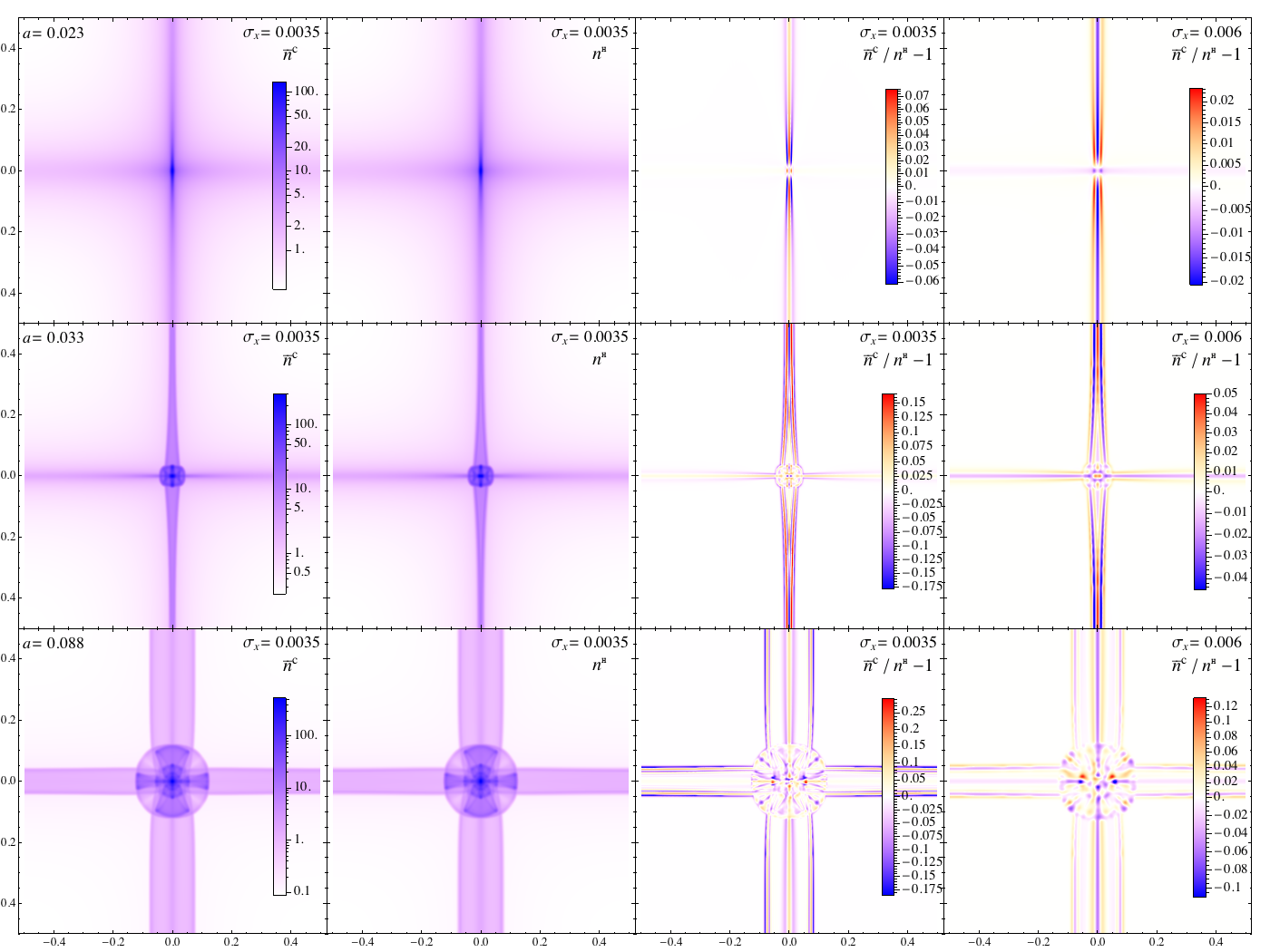}

\vspace{-0.27cm}

\includegraphics[width=0.90\textwidth]{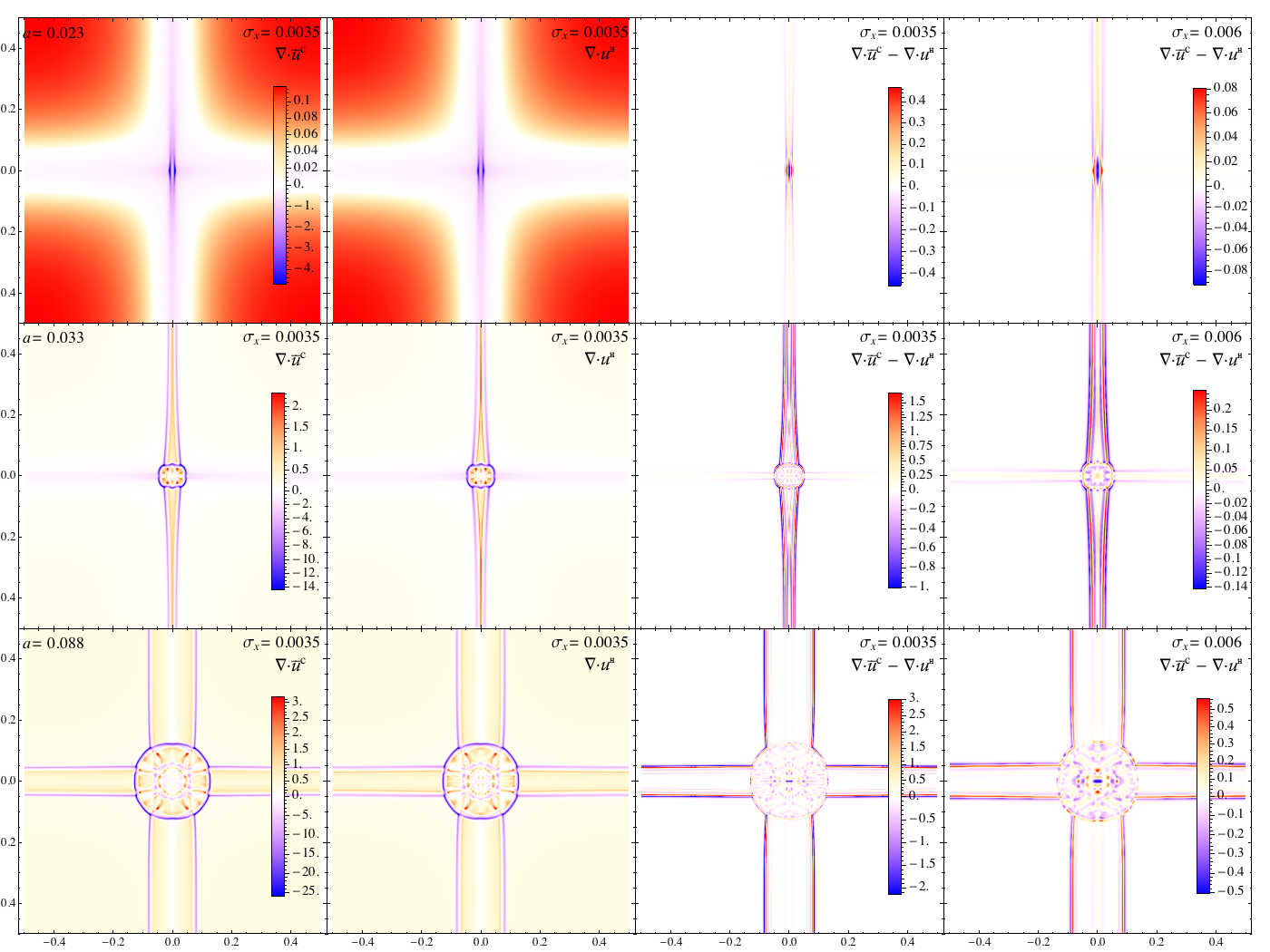}
\caption{Sine wave collapse: density (top panel) and velocity divergence (bottom panel) at three different times.}
\label{fig:alldensdiv}
\end{figure*}

\begin{center}
\begin{figure*}[!]
\includegraphics[width=0.90\textwidth]{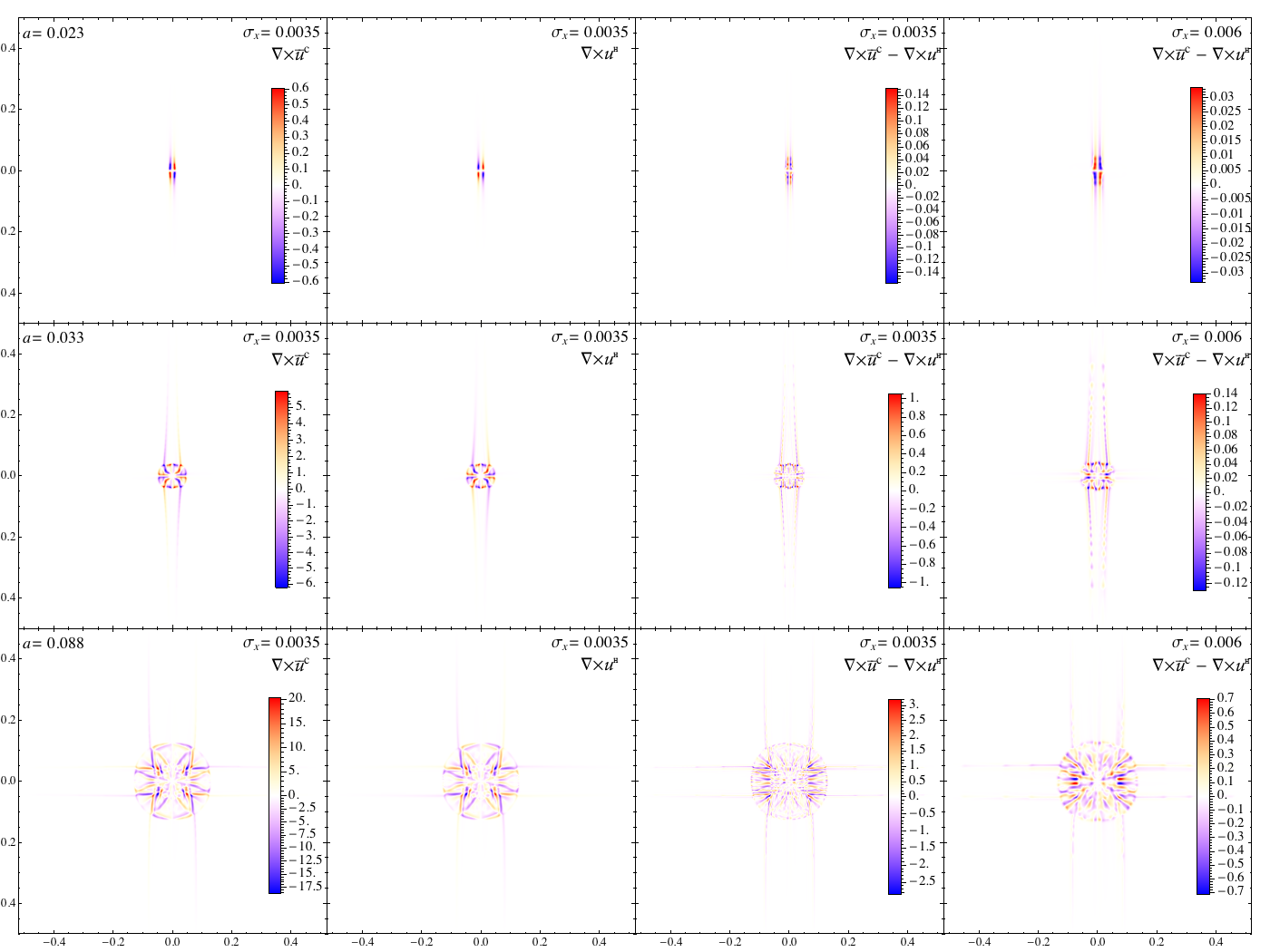}

\vspace{-0.27cm}

\includegraphics[width=0.90\textwidth]{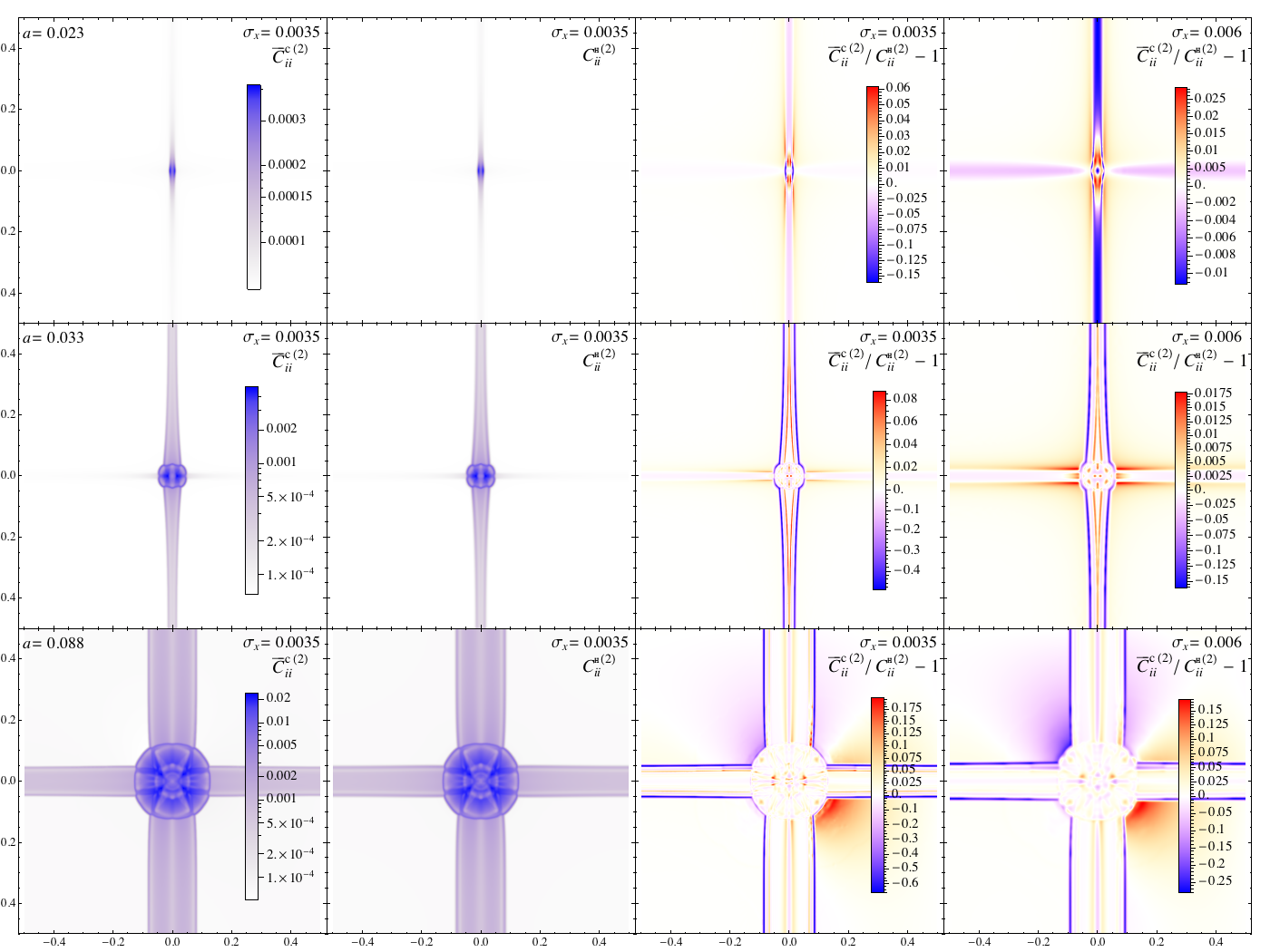}
\caption{Sine wave collapse: velocity curl (top panel)  and trace of the velocity dispersion (bottom panel) at three different times.}
\label{fig:allrotdisp}
\end{figure*}
\end{center}

\vspace{-1.2cm}
\subsubsection{Density}
\label{sec:numericsSineDensity}
We coarse-grained the output snapshots of the density field $n^\c(\vx)$ from \texttt{ColDICE} to obtain $\bar n^\c(\vx) = e^{\frac{\sigx^2}{2} \Delta} \{ n^\c \}$.
We then compared this to $n^\H(\vx)$ at various times during the simulation.
Based on our discussion in Sec.\,\eqref{sec:ScMtoVlasov}, we expect these quantities to coincide provided 
we have chosen the same $\sigx$ in the calculation of $n^\H$ and $\bar n^\c$, chosen a sufficiently small $\thbar$ \eqref{hbarCondition} and 
constructed the initial wave function according to \eqref{inipsiFromDust}.

The top panel (top $3\times 4$ matrix of figures) of Fig.\,\ref{fig:alldensdiv} exhibits 
the results for 3 snapshots at times $a=0.023$, $a=0.033$ and $a=0.088$ respectively.
In the left column we show $\bar n^\c(\vx)$ of \texttt{ColDICE}, and in second column $n^\H(\vx)$, both with $\sigx=0.0035$. 
In the third and fourth column we show the fractional difference for $\sigx=0.0035$ and $\sigx=0.006$ respectively.
The first snapshot at $a=0.023$ is taken shortly after the first shell-crossing in the $y$-direction.
By $a=0.033$ the phase space sheet has wounded round ten times.
This process is fully resolved by \texttt{ColDICE} with its adaptive mesh refinement. 
 Although the density is only the projection of the phase space density one can get a feeling for its structure at $a=0.088$ 
where several caustics are clearly visible as spherical shells and lobes.
The differences between  the first and second columns of  Fig.\,\ref{fig:alldensdiv}, 
i.e. the coarse-grained result of the Vlasov and the ScM solvers respectively,  are impossible to discern by eye.

 The accuracy by which the ScM code agrees with the coarse-grained Vlasov solver depends on $\thbar$ and $\sigx$ and the integration time $a-a_{\rm ini}$. 
 For fixed $\thbar$ the agreement becomes better if $\sigx$ is increased above a minimal value that avoids interference fringes to become visible. 
 The value of $\sigx=0.0035$ is already slightly too small in order to hide all quantum artifacts.
 Although they are not visible by eye in the second column, they can be seen when looking at the ratios $\bar n^\c/n^\H\,-1$ in the third column.
 Increasing the coarse graining scale to $\sigx=0.006$, gives a far better quantitative agreement, at
 the price of sacrificing some of the small scale structures in the density fields.
 The fractional difference in the smoothed density field increases from the initial 0.035\% to about 10\% at the final time for $\sigx=0.006$, or about twice that much for the smaller smoothing scale $\sigx=0.0035$. This seems to be mostly due to small offsets in the positions of high density regions and might have partially a numerical origin. 
We also tested the larger value $\thbar=6\times10^{-5}$ and found a maximum deviation of 20\% at the final time for $\sigx=0.006$.

\begin{figure}[t]
    \centering
    \includegraphics[width=7cm, angle=0]{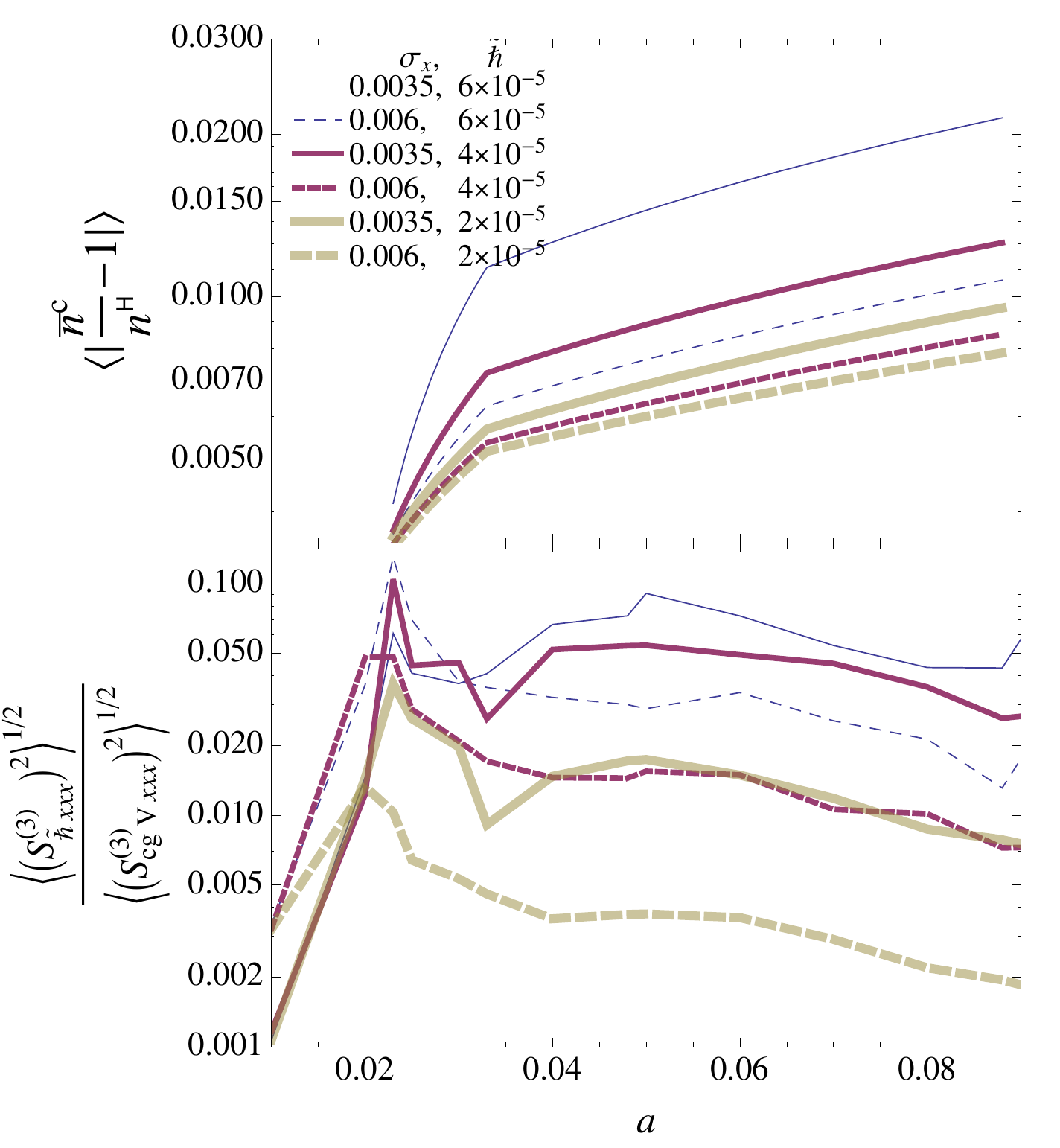}
    \caption{Upper panel:  Mean of the fractional difference of the smoothed density fields $\bar n^c$ and $n^\H$ as obtained from \texttt{ColDICE} and the ScM, respectively.
\\
  Lower panel: Testing for quantum artifacts using \eqref{M3HGoodCondition1} for the component $i=j=k=x$
 for the same parameters as for the top panel.}
    \label{fig:ShbarcgV}
\end{figure}

 Since for larger $\thbar$ the numerical accuracy is better, the loss of agreement is likely due to the quantum artifacts of the ScM 
which can be quantified through the magnitude 
of $\langle (S^{(3)}_{\thbar\,ijk})^2 \rangle^{1/2}/\langle (S^{(3)}_{{\rm cg V}\,ijk})^2\rangle^{1/2}$ to test the condition \eqref{M3HGoodCondition1}.
 We plot this quantity for $i=j=k=x$ in the bottom panel of Fig.\,\ref{fig:ShbarcgV}.
Comparing this to the upper panel, showing the mean of $\bar n^\c/n^\H -1$, we see that there is good correspondence between the scaling of $\langle \bar n^\c/n^\H -1 \rangle$ with time and $\thbar$ and the corresponding scaling of our measure of quantum artifacts.
In order for the ScM to be in good correspondence with the coarse-grained Vlasov equation we require $ | S^{(3)}_{\thbar}| \ll |S^{(3)}_{\rm cg V}| $. 
We achieve for $\langle (S^{(3)}_{\thbar\,xxx})^2 \rangle^{1/2}/\langle (S^{(3)}_{{\rm cg V}\,xxx})^2\rangle^{1/2}$ only about $10^{-2}$ for $\thbar=4\times 10^{-5}$ and a factor 10 worse for $\thbar=6\times10^{-5}$. 
Yet, as can be seen in the upper panel of Fig.\,\ref{fig:ShbarcgV}, for $\thbar=4\times 10^{-5}$ and $\sigx\geq 0.0035$, we find $\langle \bar n^\c/n^\H -1 \rangle \lesssim 1\%$ at all times, such that $\langle (S^{(3)}_{\thbar\,xxx})^2 \rangle^{1/2}/\langle (S^{(3)}_{{\rm cg V}\,xxx})^2\rangle^{1/2} \lesssim 10^{-2}$ seems to indicate acceptable results and allows us to judge whether we approximate the coarse-grained Vlasov from within the ScM. 

For even small values of $\thbar < 4\times10^{-5}$ the energy test begins to fail, see the dashed curve in the lower panel of Fig.\,\ref{fig:CompareESPEandDice}.
Thus smaller values cause our SPE solver to have too large numerical inaccuracies, 
while larger values lead to too large quantum artifacts and degrade the accuracy with which the coarse-grained Vlasov equation is solved, 
see the top panel of Fig\,\ref{fig:ShbarcgV}.
Hence, for this particular initial condition, SPE integrator and integration time interval, the value $\thbar \simeq 4\times10^{-5}$ is optimal.

Note that $|S_{\thbar}| \ll |S_{\rm cg V}|$ only guarantees that the source term for $\partial_t f^\H(t) - \partial_t \bar{f}(t)$ is small. This 
implies that, in principle, a nonzero yet small value for that source would always accumulate errors through  time evolution so that to make the ScM fail at some point in time.

\subsubsection{Velocity}
We used a code provided to us by T.\,Sousbie that calculates  the first moment $M^{\c(1)}_i(\vx)$ from the output phase space density of \texttt{ColDICE}.
We coarse-grained it, $\bar M^{\c(1)}_i(\vx) = e^{\frac{\sigx^2}{2} \Delta} \{ M^{\c(1)}_i \}$, and constructed
 the coarse-grained velocity field  $\bar u^\c_i(\vx) = \bar M^{\c(1)}_i/ \bar n$ from it.
We split the latter into a divergence $\vnabla \cdot \bar {\vu}^\c$ and a rotation $\vnabla \times \bar {\vu}^\c$ component.

We compare the divergence $\vnabla \cdot \bar {\vu}^\c$  obtained from the Vlasov solver \texttt{ColDICE}
 to that obtained from the ScM in the bottom panel of Fig.\,\ref{fig:alldensdiv} for the same three snapshots as with the density.
In the left column we show results of the coarse-grained Vlasov, in second that of ScM and in the third and fourth column the difference for 
$\sigx=0.0035$ and $\sigx=0.006$.
The rotation $\vnabla \times \bar {\vu}^\c$ has only a non-vanishing $z$-component  which we plot in the top panel of Fig.\,\ref{fig:allrotdisp}.
As was the case for the density, we again observe an astonishing visual agreement between the Vlasov code and the ScM for both 
of these velocity components.
Once more, the quantitative agreement for fixed $\thbar$ may be improved by increasing the amount of spatial smoothing as
 can be seen in the 4th column that compares ScM to Vlasov for a coarse graining scale of $\sigx=0.006$.

It is worth noticing that the wave function with its mere two degrees of freedom is able produce ``vorticity without vorticity.''
What we mean by this is that $\vnabla \times \vu^\H$ is non-zero, while the naive estimate of the velocity $\vu_\psi = \vnabla \phi$ has vanishing vorticity. 
This by itself might not seem surprising as $\vu^\H = e^{\frac{\sigx^2}{2} \Delta} \{n_\psi \vu_\psi \}/ n^\H$ 
is the mass-weighted $\vu_\psi$ and thus trivially has some vorticity.
The remarkable aspect of this vorticity is that no extra degree of freedom is necessary to correctly describe the vortical degree of freedom of the velocity field present in the CDM in two dimensions.
Although this has been anticipated from our theoretical discussion, this is the first time that it has been actually demonstrated.
We discuss this and the microscopic origin of vorticity in the ScM further in Sec.\,\ref{Vorticity}.  
In a similar fashion despite having only two degrees of freedom in the ScM, all 
higher Husimi cumulants $C^{\H (n)}$ will automatically be present and should agree with their coarse-grained Vlasov counterparts $\bar C^{\c(n)}$.

\subsubsection{Velocity dispersion}
In the bottom panel Fig.\,\ref{fig:allrotdisp}  we compare the trace of the velocity dispersion $\bar C^{\c(2)}_{ii}$ to $C^{\H (2)}_{ii}$. 
The asymmetry seen in the ratios $\bar C^{\c(2)}_{ii}/C^{\H (2)}_{ii}$ is due to \texttt{ColDICE}. 
We tested that our results are perfectly mirror-symmetric with respect to the axes $x=0$ and $y=0$.

\FloatBarrier
\begin{center}
\begin{figure*}[!]
\includegraphics[width=0.895\textwidth]{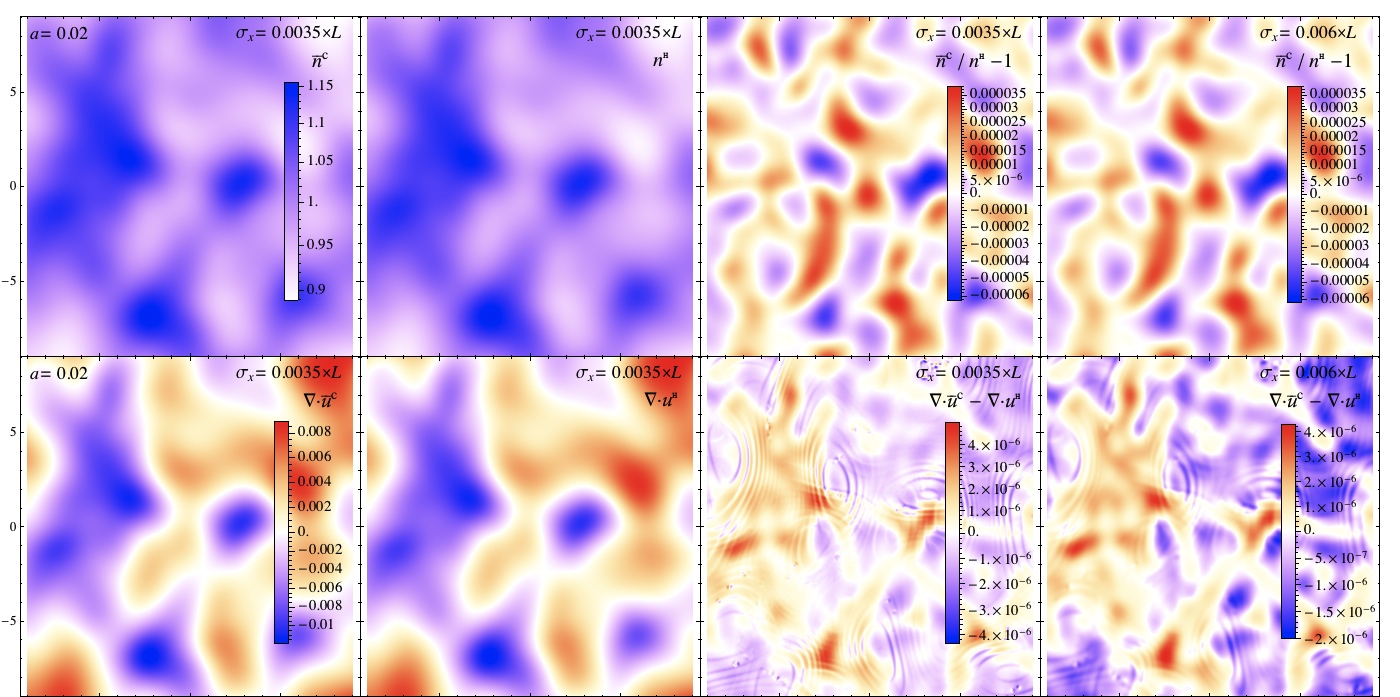}


\includegraphics[width=0.90\textwidth]{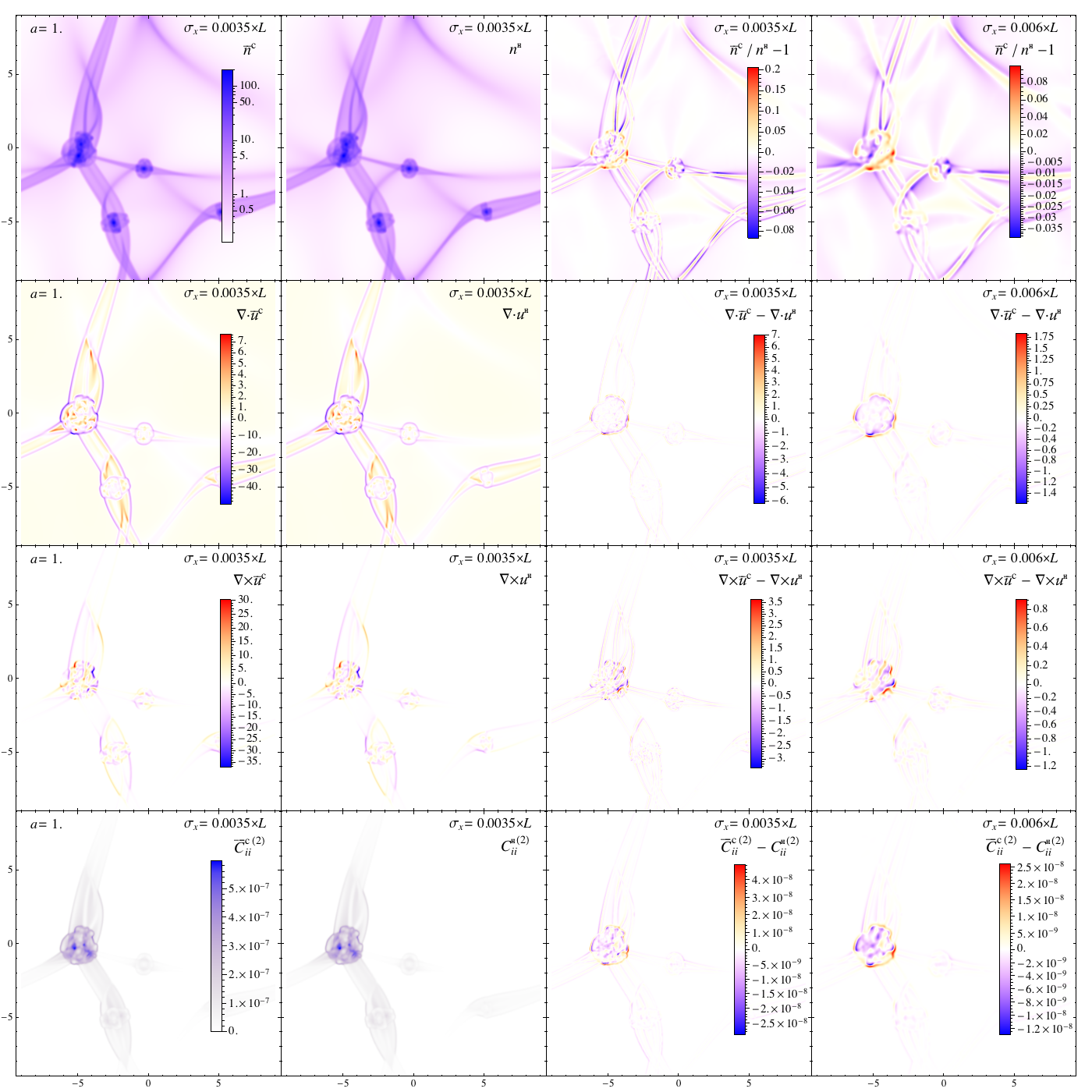}
\caption{Gaussian random field collapse: Top two panels at $a=0.02$ shortly after initial time $a_{\rm ini}=1/51$, lower panels at $a=1$}
\label{fig:gaussian}
\end{figure*}
\end{center}

\subsection{Gaussian random field}
\label{sec:gaussianranfield}
For this test we assumed that the primordial curvature perturbation is a gaussian random field with a
 power spectrum generated with the publicly available Boltzmann code \texttt{CLASS}, \cite{Lesgourgues2011}. 
We smoothed the matter power spectrum with a gaussian filter of width $1$\,Mpc in order to make the numerical evolution better behaved. 
The erasure of initial power on the smallest scales postpones the occurrence of shell crossings, reducing in turn the amount of necessary 
refinements in \texttt{ColDICE} and allowing the numerical solution of the SPE to be accurate over longer time periods.

The relation between conventional initial conditions for N-body simulations and the initial wave function $\psi_{\rm ini}(\vx)$ is 
described in App.\,\ref{ZeldoCosmo}.  The initial density $M^{\c(0)}(t_{\rm ini})$ and velocity fields $M^{\c(1)}_i/M^{\c(0)}(t_{\rm ini})$ 
constructed from the \texttt{ColDICE} phase space sheet agree with $n_\psi(t_{\rm ini})$ and $\nabla_i \phi(t_{\rm ini})$ better than $0.01\%$.

We followed the evolution from $a_{\rm ini}=1/51\simeq 0.0196$ to $a=1$. We present here the first snapshot at $a=0.02$ shortly after the start
 of the simulation and at $a=1$, the final time.  The results for $\thbar=5\times10^{-10}$ are shown in Fig.\,\ref{fig:gaussian}.
The maximal value of the fractional difference in the smoothed density field increases from 0.01\% to about 10\% at the final time for $\sigx=0.006 L$, or about twice that 
for the smaller smoothing scale $\sigx=0.0035 L$. This seems to be mostly due to small offsets in the positions of high density regions and might partially 
have numerical origin. 

We also tested the larger value $\thbar=10^{-9}$ and found a maximum deviation of 20\% at the final time for $\sigx=0.006 L$. 
Since for larger $\thbar$ the numerical accuracy is better, the loss of agreement is due the degrading of the ScM with increasing $\thbar$,
 reflected in an increased value of quantum artifacts $S_{\thbar}/S_{\rm cg V}$. 
 For the gaussian random field we found values of  $\langle (S^{(3)}_{\thbar\,xxx})^2 \rangle^{1/2}/\langle (S^{(3)}_{{\rm cg V}\,xxx})^2\rangle^{1/2} \simeq 10^{-2}$ for $\thbar=5\times10^{-10}$ and about 0.05 for $\thbar=1\times10^{-9}$, which are  very similar to the sine collapse.
For smaller $\thbar < 5\times10^{-10}$ the energy test failed before the end of the simulation
so that the optimal value of $\thbar$ is $\sim 5\times10^{-10}$ for this particular set of simulation parameters and initial conditions. For this value and $\sigx\geq 0.0035 L$, we find $\langle \bar n^\c/n^\H -1 \rangle < 1\%$ at all times.

We conclude that the ScM can be successfully used to solve the coarse-grained Vlasov equation. The accuracy of that solution is driven by $\thbar$. Joint optimization of $  |S^{(3)}_{\thbar}| \ll |S^{(3)}_{\rm cg V}| $ and $|\delta_{E_{\rm tot}} | \ll 1 $ lead to an optimal value of $\thbar$.

\section{Discussion} \label{sec:Discussion}

  \subsection{Vorticity without vorticity}
 \label{Vorticity}
   \begin{figure}[t]
\includegraphics[width=0.48\textwidth]{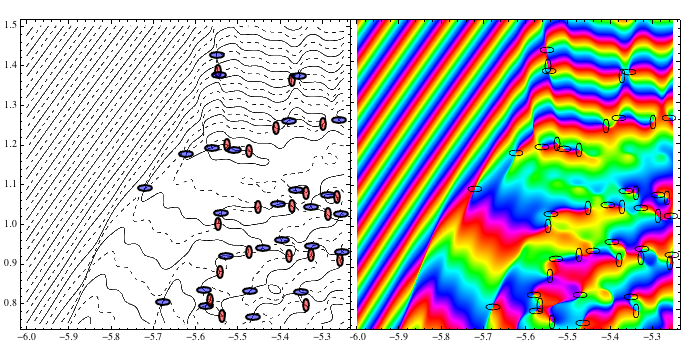}\\
 \caption{The \textit{solid} curves show the zeros of $\Re(\psi)$  and  \textit{dashed} 
 curves the zeros of $\Im(\psi)$. $\Re(\psi)$ and $\Im(\psi)$ are depicted in Fig.\,\ref{fig:compareRePsiImPsi} upper panels. The zeros of $\psi$ are the crossing of both types and encircled by ellipses. The same ellipses are shown on to right panel, which is a reproduction of Fig.\,\ref{fig:compareRePsiImPsi} showing the phase $\phi/\thbar$ the wave function. The color and orientation of the ellipses correspond
 to the orientation of the circulation: blue/horizontal (red/vertical) corresponds to negative (positive) winding number.}
\label{fig:compareVorticity}
\end{figure}
 It might appear surprising how the two degrees of freedom contained in the wave function include vorticity. 
 First of all, vorticity is not associated with a new degree of freedom, in contrast to CDM where vorticity is a degree of freedom that cannot be 
constructed from density and velocity divergence,  \NEW{since it is a multi-stream phenomenon \cite{PichonBernardeau1999,PueblasScoccimarro2009, HahnAnguloAbel2015}}, and  moreover satisfies its own equation of motion.
 Second, the coarse-grained velocity $\vu_\H =  e^{\frac{1}{2}\sigx^2 \Delta} \{n_\psi \vnabla \phi \}/n_\H$, has some trivial vorticity due to the involved coarse graining which can be seen by rewriting the smoothing as 
 \begin{equation}
 n^\H \vu_\H = n^\H \exp\left(\sigx^2\overleftarrow{\vnabla}_{\! \!  x}\overrightarrow{\vnabla}_{\! \!  x}\right) \vnabla \bar \phi\,, \label{Husimiconstr}
  \end{equation}
  where $\bar \phi = e^{\frac{1}{2}\sigx^2 \Delta} \phi $ is the coarse-grained phase.
 Taylor expanding  $\vnabla \times \vu_\H$  to leading order in $\sigx^2$ and using that 
 \begin{equation}
 \vnabla \times (\vnabla \phi) =0  \label{novorticityofgrad}
 \end{equation}
  we find
 \begin{align} \label{curlofv}
 (\vnabla \times \vu_\H)_i = \sigma_x^2 \left(\vnabla\frac{  n_{,i}^\H}{n^\H}\times  \vnabla \bar \phi_{,i}  \right)\,,
\end{align} 
 suggesting that the vorticity is a purely smoothing related byproduct in contrast to CDM where $\vnabla \times \vu_\c$ has a component that survives the limit $\sigx \rightarrow 0$ given by the above mentioned vortical degree of freedom. 
However it turns out that the innocent looking \eqref{novorticityofgrad} is in fact not the whole story 
and there is indeed a microscopic seed for vorticity in the ScM that survives the limit  $\sigx \rightarrow 0$.

  \begin{figure}[t]
\includegraphics[width=0.48\textwidth]{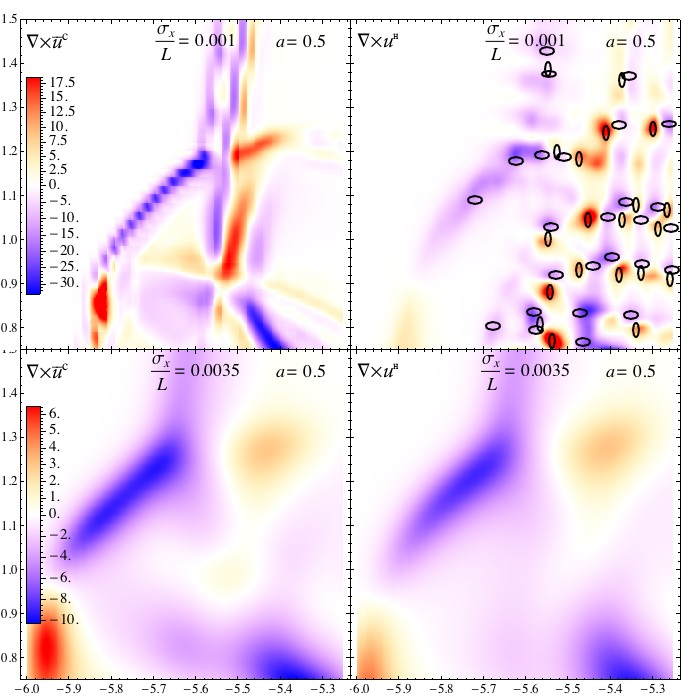}\\
 \caption{Coarse-grained vorticity. Upper panels are for $\sigx=0.001\times 20$\,Mpc= 0.02\,Mpc, lower panels for $\sigx=0.0035\times 20$\,Mpc = 0.07\,Mpc. Left panels show the results of \texttt{ColdDICE}, right panels those of the ScM. Overplotted on the right are the locations of the vortices identified in Fig.\,\ref{fig:compareVorticity}.}
\label{fig:compareVorticity2}
\end{figure}

The right-hand side of \eqref{novorticityofgrad} does not vanish in general if the phase has singular gradients $\vnabla \phi$ which is known to generally appear during time evolution \cite{UhlemannKoppHaugg2014}. 
 If a closed curve $C$ with unit tangent $\v{l}$ encloses a singularity,\footnote{In two dimensions they are point-like whereas in three dimension
 they are line-like and thus a network of vortex lines will form for $d=3$. 
 In one dimension, although pointlike phase gradient singularities arise, as shown in  \cite{UhlemannKoppHaugg2014}, they are not persistent 
in the sense that they appear only at isolated points in time at which the phase jumps by $2\pi$.} the circulation 
 \begin{equation} \label{singlevaluedness}
 \frac{1}{ 2 \pi \thbar }\oint_{C} \vnabla \phi \cdot d\v{l} =m
 \end{equation}
  is a non-zero integrer $m$ \cite{PitaevskiiStringari2003,Wallstrom1994}. In that case the right-hand side of \eqref{novorticityofgrad} consists of a sum of Dirac delta functions positioned at the singularities of the phase, which in two dimensions is given by
  \begin{equation}
 \vnabla \times (\vnabla \phi) = \hat{\v{z}} \,2\pi\, \thbar \,\sum^{N_{\rm vort}}_{i} m_i \delta_{\rm D} (\v{x}_i) \,,  \label{withvorticityofgrad}
 \end{equation}
 with $\hat{\v{z}}$ the unit vector perpendicular to the two dimensional surface \cite{PitaevskiiStringari2003}.
Thus, this contributes to $\vnabla \times \vu_\H$ even at zeroth order in $\sigx$
  \begin{align} \label{curlofv2}
 \vnabla \times \vu_\H = \hat{\v{z}}\, 2\frac{\sigu}{\sigx} \,\sum^{N_{\rm vort}}_{i} m_i e^{\frac{(\v{x}-\v{x}_i)^2}{2 \sigx^2}} + \mathcal{O}(\sigx^2)\,,
\end{align} 
with the $\mathcal{O}(\sigx^2)$ part given by \eqref{curlofv}.
 In practice it is not necessary to keep track of the creation, motion and merging of quantum vortices located at the $\vx_i(t)$ 
as well as the varying of their total number $N_{\rm vort}(t)$ 
and their individual winding numbers $m_i(t)$: the real and imaginary part of $\psi$ 
encode these objects but the vortices are not independent vortical degrees of freedom.

  In Fig.\,\ref{fig:compareVorticity} we show that vortices are identified with phase singularities, which in turn are identified with zeros of the amplitude.
  The first panel shows in dashed and dotted the zeros of $\Re(\psi)$ and $\Im(\psi)$. 
Points where both types of zeros (zeros of $\Re(\psi)$ and $\Im(\psi)$) cross are places where $\psi$ is also zero. We mark the zeros of $\psi$ with ellipses. 
Drawing the same ellipses on the right panel of  Fig.\,\ref{fig:compareVorticity}, that is the plot of the phase $\phi/\thbar$,
   we observe that all these points carry non-zero circulation with precisely $|m|=1$.
 Color and orientation denote the sign of $m$, with blue/horizontal $m=-1$ and red/vertical $m=1$.

  In Fig.\,\ref{fig:compareVorticity2} we overplot the vortices on the coarse-grained vorticity obtained by the ScM. In the top panel we use a very small smoothing scale, which is clearly too small to be in good correspondence with the CDM one shown on the left. 
 However, it clarifies that reducing $\sigx$ to ever smaller values makes  $\vnabla \times \vu^\H $ more and more dominated by the vortices \eqref{curlofv}, whereas letting $\sigx$ to flow to large enough values 
that give good correspondence to the coarse-grained CDM, the vortices are not visible and loose their apparent correlation with $\vnabla \times \vu^\H$.
  This is yet another variation of our general theme that CDM and ScM are very different  UV-completions of the dust fluid, but that once 
UV physics is smeared out sufficiently, the two theories are in direct correspondence with each other.

\subsection{Pressure of CDM, EFTofLSS and backreaction}
\label{sec:Pressure}
\begin{figure}[t]
    \centering
    \includegraphics[width=8.5cm, angle=0]{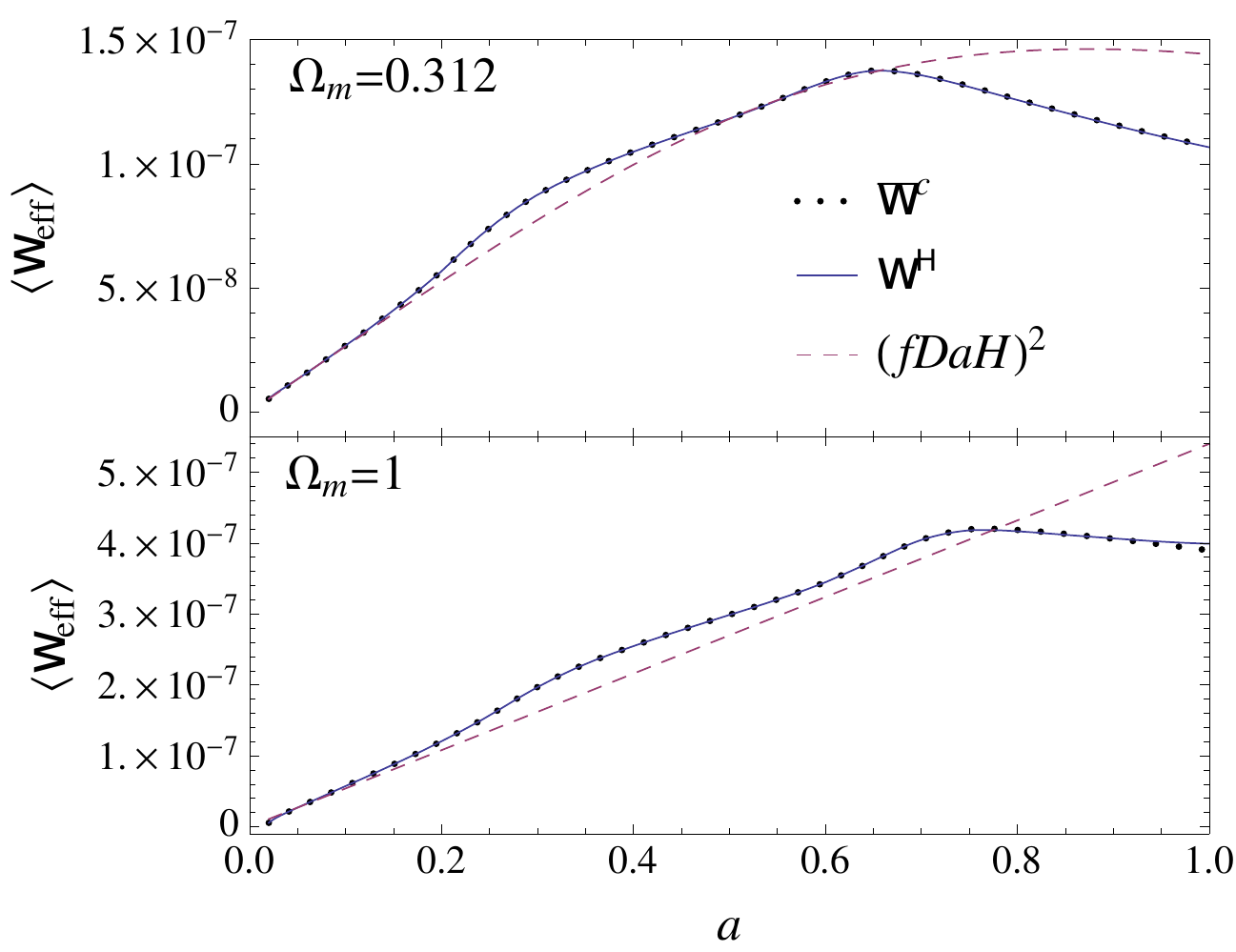}
    \caption{Effective CDM background equation of state \eqref{weff}. Dotted curves show the result from the Vlasov solver, the solid curve
 that of the ScM. The dashed curve shows the time dependence suggested by simple perturbative estimates within the EFTofLSS.
 The upper and lower panel show the result for different cosmologies.}
    \label{fig:weff}
\end{figure} 
The stress tensor is given by $T_{ij}=\rho_0 \bar M^{(2)}_{ij}/a^3$ which vanishes in linear perturbation theory $T_{ij}=0$, such that we can interpret $T^i{}_{i} = \rho_0 M^{(2)}_{ii}/a^5 = 2 P_{\rm eff} $ as the effective pressure $P_{\rm eff}$ \cite{B10}.\footnote{In the effective stress tensor $\tau^i{}_j$, (3.26)  of \cite{B10}, one should replace $\rho v_i v_j$ by $T^{i}{}_j$ at the right-hand side, see footnote 15 of \cite{B10}. 
The effective stress tensor $\tau^i{}_j$ contains also gradients of the Newtonian potential. 
These terms cancel however in the two-dimensional trace that we consider here, such that $\tau^i{}_i = T^i{}_i=2 P_{\rm eff}$.}

 Furthermore, let us define the effective equation of state as $w_{\rm eff} \equiv P_{\rm eff} a^3/\rho_0$ such that, when applied to quantities smoothed in phase space we get
\begin{equation} \label{weff}
\bar w_{\rm eff} = a^{-2} (\bar M^{(2)}_{ii} - 2 \sigu^2 \bar n) /2\,,
\end{equation}
and analogously for $w^\H_{\rm eff}$. We subtract the `trivial' part given by $\sigu^2 \bar n$, since the effective field theory of large scale structure (EFTofLSS), usually involves only spatial averages \cite{BaumannNicolisSenatoreEtal2012}.
The cosmological average value $\langle \bar w_{\rm eff} \rangle$ that we measure from the Vlasov and Schr\"odinger simulations as a function of time gives us an estimate of  backreaction on the cosmological background using the second Friedmann equation
\begin{equation} \label{effectiveFriedmann}
 3H_{\rm eff}^2 +2\dot H_{\rm eff} = - 8 \pi G \langle P_{\rm eff} \rangle+  \Lambda \,.
\end{equation}
In Fig.\,\ref{fig:weff} we show the time evolution of $\langle w_{\rm eff} \rangle$ for a $\Lambda$CDM and Einstein-de Sitter universe. We also compare 
in both cases ScM (solid) with the direct Vlasov solution (dotted), which agree extremely well. 
Furthermore we plot in dashed const$\times(fDaH)^2$, where $f=d\ln D/d \ln a $ and $D$ is the linear growth, which is the estimate for the time dependence of this quantity from perturbation theory. 
The constant has been fit to match $\langle w_{\rm eff} \rangle$ for $a<0.1$, where the perturbation theory estimate should be valid.

\begin{figure}[t]
    \centering
    \includegraphics[width=8.5cm, angle=0]{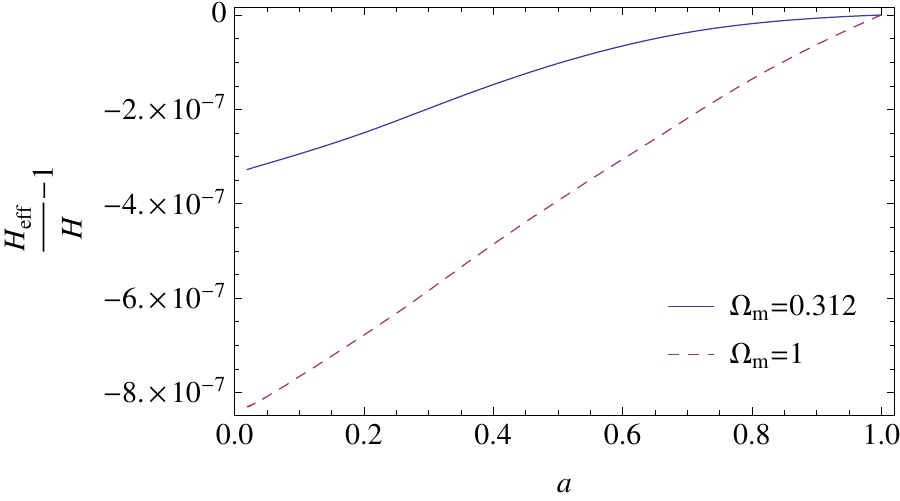}
    \caption{Comparison of the time evolution of $H(a)$ and $H_{\rm eff}(a)$ for the different cosmologies shown in Fig.\,\ref{fig:weff}.}
    \label{fig:Heff}
\end{figure}

In Fig.\,\ref{fig:Heff} we compare the result of integrating \eqref{effectiveFriedmann} with and without $\langle P_{\rm eff} \rangle$. We denote the result $H_{\rm eff}$  when including $\langle P_{\rm eff} \rangle$ and the standard $\Lambda$CDM Hubble parameter \eqref{HubbleLCDM} by $H$ which is a solution to \eqref{effectiveFriedmann} without $\langle P_{\rm eff} \rangle$.
We find that the backreaction on the expansion is a small and a likely a negligible effect.

We observe in Fig.\,\ref{fig:weff} that $w_{\rm eff}$ as calculated from the UV complete Vlasov equation is the same as the one from the UV complete ScM. This exemplifies once more that the two degrees of freedom contained in the complex wave function which in the IR are simply related to those of the dust fluid can be used to derive parameters entering an effective field theory of large scale structure (EFTofLSS) based on the ScM. Hence such an EFTofLSS based on the ScM will not require operators associated with missing small scale physics and the field theory itself can be used to measure or derive its EFT parameters.

\subsection{Entropy production in LSS formation}

Lynden-Bell suggested in his seminal work \cite{L67} that the entropy functional of the coarse-grained phase space density $\bar f$ can be used to understand the virialization process as ``violent relaxation'' and the resulting states as those maximizing the entropy functional.
Since the ScM offers an approximation to $\bar f$ it is interesting to study the entropy density $s(\vx)$ and entropy $S$ of $f^\H$, 
which can be defined respectively as
\begin{align}
s(\vx) = &- \frac{1}{\sigx^d} \int \vol{d}{u} f_\H \ln \left[ f_\H\, \sigu^d \right]\notag \\
S =& \int \vol{d}{x} s(\vx) \,.
\end{align}
Another interesting quantity is the Massieu function \cite{Katz1978}
\begin{equation}
J = S - \frac{\delta S}{ \delta E} E \,,
\end{equation}
closely related to the free energy  $F= - \left(\frac{\delta S}{ \delta E} \right)^{-1} J$ and temperature $T\equiv \left(\frac{\delta S}{ \delta E} \right)^{-1}$ in thermal systems.
We expect $S$ and $J$ to increase until virialization ends when a quasi equilibrium state is reached for which $S$ and $J$ are extremised. 
There are no truly stationary states for self-gravitating systems\footnote{There are truly stable static configurations for the SPE \cite{L77,MPT98,AS01}, 
but these are purely quantum pressure supported and deep in the regime where the right-hand side of \eqref{fHandfcgDiff} 
can no longer be expanded in powers of $\thbar$, such that the correspondence between $\bar f$ and $f_\H$ is lost.} 
but only long-lived quasistationary states \cite{StanisciaChavanisdeNinno2011}.  The free energy functional was also used 
in \cite{Chavanis2006, StanisciaChavanisdeNinno2011,Wei2016} to study non-equilibrium phase transitions in systems with long range interactions like gravity.
We leave investigating the entropy functional for a future work.

\section{Conclusion and Prospects}\label{sec:conclusion}

We showed that the  Schr\"odinger method (ScM) can be used to solve the Vlasov equation in $d=2$ dimensions by comparing our ScM code 
 to the state of the art Vlasov code \texttt{ColDICE}, see Figs.\,\ref{fig:alldensdiv} and \ref{fig:allrotdisp}.
We found excellent agreement between the two methods in the cases where an agreement was expected for the ScM to work.

The advantages of the ScM over standard methods of solving the Vlasov equation are:
\begin{itemize}
 \item  the information about the phase space distribution in $2d$ dimensions is compressed in a complex wave function in the $d$ dimensions of Eulerian space. 
\item The moments of the phase space density, which are of direct relevance for observations, can be constructed from the wave function without dealing with the complications of the $2d$-dimensional phase space.
\end{itemize}
This procedure is summarised in Sec\,\eqref{sec:HusHierarchy} and comprises the definition of the ScM as used in this article.

Another difference compared to conventional methods that solve the Vlasov equation is that the ScM approximates the coarse 
grained Vlasov equation with fixed phase space resolution.
N-body simulations usually have phase space resolutions that depend on time and position in the simulation box as the phase space density is sampled by the clustering particles. 
In particular, for applications ranging from ray tracing through the simulations, measurement of topological properties of the cosmic web, evaluation of entropy density, or the extraction of metric components, a constant spatial resolution could be a desirable feature.
Furthermore, for applications regarding voids or warm initial conditions, which are naturally poorly sampled by N-body simulations, 
 the ScM with its constant resolution might be advantageous. 

Another attractive feature of the ScM is that the wave function is a UV completion of the dust fluid \cite{UhlemannKoppHaugg2014}, in the sense that that two degrees of freedom $n_\d(t,\vx),\, \phi_\d(t,\vx)$ that run into infinities and thus cannot describe CDM in the UV, are replaced by $\psi(t,\vx)$, again 2 degrees of freedom that are finite in the UV and contain all the physics. More importantly, once the finite UV behaviour is coarse-grained over, the ScM produces an approximate solution to the Vlasov equation.
This means that an effective field theory of large scale structure  (EFTofLSS) based 
on $n_\d(t,\vx),\, \phi_\d(t,\vx)$ has to introduce operators that take into account the 
fact that  $n_\d(t,\vx),\, \phi_\d(t,\vx)$ do not actually describe  physics in the UV  \cite{FloerchingerGarnyTetradisEtal2017}. 
On the other hand, since $\psi(t,\vx)$ is UV complete, an EFTofLSS based on the ScM, will only 
require operators that arise through integrating out small scales; the type and number of degrees of freedom does not change. 
We corroborated that claim in Sec.\,\ref{sec:Pressure} where we measure one parameter of the EFTofLSS $w_{\rm eff}$ 
within ScM which agrees with the same quantity measured using the Vlasov solver.

We note that replacing the dust energy momentum tensor $T^\d_{\mu \nu}$ in the formalism explored in \cite{KUH14} by the ScM counterpart $T^\H_{\mu \nu}$, 
using \eqref{EMtensor} and \eqref{momentscgw} allow a well defined extraction of the non-Newtonian metric quantities 
like the gravitational slip, vector and tensor perturbations.  We leave this for a future study.
Further future tests of the ScM may involve warm initial conditions, further investigating
 vorticity, entropy built-up, extending the method to three-dimensions and in addition implementing adaptive mesh refinement.

\section*{Acknowledgements}
We thank Thierry Sousbie for making \texttt{ColDICE} publicly available, helping us with the use of \texttt{ColDICE} and also for giving us the projection code for obtaining 
the higher moments from \texttt{ColDICE} snapshots.  
The research leading to these results 
has received funding from the European Research Council under the European Union's Seventh Framework 
Programme (FP7/2007-2013) / ERC Grant Agreement n.\,617656 ``Theories and Models of the Dark Sector: Dark Matter, Dark Energy and Gravity''.

\appendix

\section{Zel'dovich initial conditions}
\NEW{In this Appendix we describe how we set up the initial conditions for our comparison of ScM and \texttt{ColDICE} described in Sec.\,\ref{sec:numerics}. 
We use the so-called Zel'dovich approximation (ZA), which is the first order within Lagrangian Perturbation Theory (LPT).
For the purpose of comparing two Vlasov solvers, it does not matter that ZA is not as accurate as 2LPT, the second order LPT solution \cite{CroccePueblasScoccimarro2006,TatekawaMizuno2007}.
For future applications of the ScM in $d=3$ dimensions we plan to use \texttt{MUSIC} \cite{HahnAbel2011}, to set up initial conditions with 2LPT.}
 \label{sec:ZAini}
\subsection{Lagrangian formulation basics}
For CDM in the single stream regime, there exists a diffeomorphism between Lagrangian coordinate $\vq$ of a CDM particle (or dust fluid element) and its Eulerian coordinates 
\begin{equation} \label{EulerLagrangeDef}
\v{x}= \vX(t,\vq)= \vq + \v{\varPsi}(t,\vq)\,,
\end{equation}
 where the trajectories $\vX(t,\vq)$, or alternatively the displacement field $\v{\varPsi}(\vq,\tau)$, are the integral curves of the Eulerian velocity field
\begin{align}
\vu_\d(\vX) = \partial_{\eta}\big|_{q}\v{\varPsi}(\vq) \,, \label{defPsi}
\end{align}
where $\eta$ is superconformal time $a^2 d \eta = a d\tau = d t$.
The Jacobian of the coordinate transformation is
\begin{align} \label{Fij}
 F_{ij}=\frac{\del  X^i}{\del q^j} = \delta_{ij}+ \partial_{q^j}\varPsi_{i}\,, \quad  J_{ F}=\det\left[\delta_{ij}+ \partial_{q^j}\varPsi_{i}\right] \,,
\end{align}
and conservation equation \eqref{Contidust} can be integrated exactly to give
\begin{align}
n_\d(\vX)=J^{-1}_{ F}(\vq) \,.\label{JFtoDeltaCG}
\end{align}
In order to evaluate $n_\d$ and $\vu_\d$ at the Eulerian position $\v{x}$ one has to invert \eqref{EulerLagrangeDef} and express $\vq=\vQ(t,\vx)$ and insert this into  $J^{-1}_{ F}$ and $ \partial_{\eta}\big|_{q}\v{\varPsi}(\vq)$, respectively.

\subsection{ZA: general initial conditions}
In the Zel'dovich approximation $\v{\varPsi}(t,\vq)$ separates into 

\begin{subequations}
\label{ZeldoDispl}
\begin{equation} \label{ZADispla}
\v{\varPsi}=D(a)\, \v{P}(\vq)  
\end{equation}
  with a purely time-dependent linear growth function, with $D(a\!\!=\!\!0)=0$ and $D(a\!\!=\!\!1)=1$, as well as a time-independent gradient 
  \begin{equation}
\v{P}(\vq) = \vnabla_{\!\! q}\, \phi_P(\vq)
\end{equation} 
\end{subequations}
of some displacement potential $\phi_P(\vq)$. 
First we explore the consequences of \eqref{ZeldoDispl}, without taking into account further properties of $\phi_P(\vq)$ and $D(a)$, which will be the subject of the following paragraphs.
Relevant for us are expressions for the Eulerian velocity potential $\phi_\d(\vx)$ and density $n_\d(\vx)$ as they enter the initial wave function $\psi_{\rm ini}(\vx) = \sqrt{n_\d^{\rm ini}} \exp(i \phi^{\rm ini}_\d / \thbar)$.
In the remaining section we approximate $n_\d = n_{\rm ZA}$ and $\phi_\d = \phi_{\rm ZA}$, such that the dust solution is initially given by
\begin{subequations}\label{Zeldonandupot}
\begin{align}
n_\d(t,\vx)  &= \left\{ \det\left[\delta_{ij}+ D(a) \partial_{q^i} \partial_{q^j}\phi_P(\vq) \right]  \right\}^{-1} \Big|_{\vq = \vQ(t,\vx)} \label{ndZA}\\
 \phi_\d(t,\vx)&  = a^2 H f D\, \left( \phi_P(\vq) + \frac{1}{2} D(a)\, |\v{P}(\vq)|^2 \right) \bigg|_{\vq = \vQ(t,\vx)} \,,
 \end{align}
\end{subequations}
where $f = d \ln D/d \ln a$ is the linear growth rate. 
The derivation of $n_d$ simply follows by substitution of \eqref{ZeldoDispl} into \eqref{Fij}.
The Eulerian velocity potential $\phi_\d$,  defined by $\vu_\d = \vnabla_{\!\! x} \phi_\d$, is obtained by first writing  the left-hand side of \eqref{defPsi} with the help of \eqref{Fij}  as $ \partial_{\! x^i} \phi_\d =  F^{-1}_{i j} \partial_{q^j} \phi_\d$ and then multiplying both sides by $F_{k i}$. Using again \eqref{Fij} and substituting for $\v{\varPsi}$ on the right-hand side \eqref{ZeldoDispl}, the resulting equation for $\partial_{q^k} \phi_\d$ is
\begin{equation}
\partial_{q^k} \phi_\d =   a^2 H f D\,(\delta_k^i + D\,\partial_{q^k} P^i(\vq) )\, \partial_{q^i}\, \phi_P(\vq) 
\end{equation}
which is  a total derivative and can be integrated to give $\phi_\d(\vq)$ in \eqref{Zeldonandupot}.
The necessary inversion of $\v{x} = \vq +  D(a)\,\v{P}(\vq)$ for $ \v{Q}(t,\vx) $ appearing in \eqref{Zeldonandupot} has to be done numerically given the specific $\v{P}(\vq)$.
We thus use on the one hand the field $\phi_{P}(\v{q})$ to create initial amplitude $n^{\rm ini}_\d(\v{x})$ and phase $\phi_\d^{\rm ini}(\v{x})$ of the wave function in Eulerian space, and on the other hand we calculate its gradient \eqref{ZeldoDispl} to obtain $\v{\varPsi}_{\rm ini}(\v{q})$ to get the initial Lagrangian displacements \eqref{EulerLagrangeDef} and velocities \eqref{defPsi}.
The latter are used as the initial conditions for \texttt{ColDICE} 
and requires evaluation of $\v{\varPsi}_{\rm ini}(\v{q})$ on a regular $\v{q}$ grid, whereas $n^{\rm ini}_\d(\v{x})$ and $\phi_\d^{\rm ini}(\v{x})$ are evaluated on  regular $\v{x}$ grid. The such created functions can be turned into an initial wave function $\psi_{\rm ini}(\v{x})= \sqrt{n^{\rm ini}_\d(\v{x})} e^{i \,\phi_\d^{\rm ini}(\v{x})/ \thbar}$ for any value of $\thbar$.

\subsection{ZA: cosmological simulation initial conditions}\label{ZeldoCosmo}
In the Zel'dovich approximation $\vnabla_{\!\! q}\cdot\v{P}(\vq) = - \delta_{\rm lin} (\v{x}\!\!=\!\!\vq) $  where $\delta_{\rm lin} (\vq)$ is the initial linear density field, linearly extrapolated to $a=1$.
From \eqref{Poissdust} it then follows that the velocity potential is a gaussian random field with power spectrum
\begin{equation}
P_{\phi_P}(k)  = \frac{P_{\rm lin}(k,z\!=\!0)}{k^4}
  \end{equation}
where $P_{\rm lin}(k,z\!=\!0)$ is the linear matter power spectrum at $a=1$. For cosmological simulations $ \phi_P(\vq)$ is therefore a gaussian random field and can be easily generated from $P_{\rm lin}$ which in turn can obtained from Einstein-Boltzmann codes like \texttt{class}. We choose the following parameters and units
\begin{align}
L &= 20\, \rm{Mpc}\\
\Omega_m &= 0.312046 \\
h&=0.67556\\
a_{\rm ini} &=\tfrac{1}{51}
\end{align}
Where $L$ is the size of the periodic box. We also apply a gaussian filter with width $R=1\,$Mpc to the linear power spectrum $\bar P_{\rm lin}(k) =  P_{\rm lin}(k) e^{- R^2 k^2}$ to prevent the formation of structure on very small scales.
In order to avoid sampling a three-dimensional $\phi_{P}(\v{q})$ and considering then only some two-dimensional restriction $\phi_{P}(q_x,q_y,q_z\!=\!0)$ relevant for our two-dimensional simulations, we can directly sample such a two-dimensional realisation with the same statistics if we use instead a two-dimensional version of that power spectrum
\begin{equation}
P^{2\rm D}_{\phi_P}(k)  = 2 \int_0^\infty \mathrm{d} p\, \frac{\bar P_{\rm lin}(\sqrt{k^2+p^2})}{(k^2+p^2)^2} \,.
  \end{equation}

\subsection{ZA: plane wave initial conditions} \label{sec:ZAplanewave}
For numerical  tests we have assumed instead that $\v{P}(\vq)$ consists of 3 perpendicular sine waves with a periodicity of $2L$ and constant amplitude $\v{A}$ such that
\begin{equation}
 \phi_P(\vq)=  \sum_{i=1}^3 A_i  \frac{L^2}  {\pi^2}\cos\left( \frac{q_i \pi}{L} \right) \,.
\end{equation}
The 2-dimensional case studied in this article then corresponds to $A_3=0$.
 With this definition \eqref{Zeldonandupot} simplify to
\begin{align}
n_\d(t,\vx)  &= \left\{ \prod_{i=1}^3 \left[ 1- D(a)\, A_i \cos\left( \frac{q_i \pi}{L} \right)   \right]  \right\}^{-1} \Bigg|_{\vq = \vQ(t,\vx)} \\
 \phi_\d(t,\vx)&  = a^2 H f D\, \sum_{i=1}^3 \Bigg( A_i  \frac{L^2}  {\pi^2}\cos\left( \frac{q_i \pi}{L} \right) +  \\
 & \qquad \qquad \quad + \frac{1}{2} D(a)\, \left( A_i  \frac{L}  {\pi}\sin\left( \frac{q_i \pi}{L} \right)  \right)^2 \Bigg) \bigg|_{\vq = \vQ(t,\vx)} \notag\,.
 \end{align}
 For our two-dimensional setup we have chosen
\begin{align}
A^i&=(30,40,0) \\
L &= H^{-1}_0 c= 1\\
\Omega_m &= 1 \\
D(a) &= a \\
a_{\rm ini} &=0.01
\end{align}
to match initial conditions presented in \cite{SousbieColombi2016}.

\section{Gaussian filtering}
\label{sec:numericalimplementation}

A convolution of a function $f$ with a filter $W^{(d)}$ 
 $$(W^{(d)}*f)(\v{x}) = \int_{L^d} \vol{d}{x} x\, W^{(d)}(\v{x}-\v{x}', \sigx)\, f(\v{x}') $$ in $n$-dimensional space can be approximated by  $$(W^{(d)}*f)(\v{x}) \simeq  \int_{\v{x}\pm 5\sigx} \vol{d}{x} W^{(d)}(\v{x}-\v{x}', \sigx)\, f(\v{x}') $$ involving only a neighbourhood of a few $\sigx$ around the point of interest $\v{x}$, if the filter has an effective support only in a small  region $5\sigx$ compared to the size $L$ of the periodic space. This makes the filtering process quasi-local and allows a significant speed up by a factor $\frac{10 \sigx}{L}$ in one dimension. 
Furthermore, since we use an isotropic gaussian $$W^{(d)}(\v{x})=\frac{1}{(2\pi \sigx^2)^{d/2}}\exp\left( - \frac{|\v{x}|^2}{2 \sigx^2} \right)$$ that decomposes like $$W^{(d)}(\v{x}) = W^{(1)}(x^1)* ... * W^{(1)}(x^d)\,,$$
we can apply the one-dimensional filter sequentially which again allows a significant speed up such that we reduce the total operations per pixel from the a priori $N$ to $ \frac{10 n \sigx}{L} N^{1/d}$, where $N$ is the total number of grid points. For our two-dimensional case with $\sigx/L = 0.0035$ and $ N=8192^2$ we reduce the number of computations by a factor $0.4\times10^{-5}$.

\section{1D pancake collapse}
\label{sec:CDM1D}

In this section we extend the analysis of App.B of \cite{UhlemannKoppHaugg2014} for the 1-dimensional pancake collapse \cite{Zeldovich70,KS83,GoudaNakamura1989,YanoGouda1998,TatekawaMaeda2001}. 
In \cite{UhlemannKoppHaugg2014} the ``Bohmian'' trajectories, the closest analogy to the concept of trajectories that exists in the ScM, was compared only to ZA trajectories. Here we also compare it to CDM.
We also focus on the difference between ZA and CDM, suggesting a straightforward algorithm to improve the ZA beyond the occurrence of shell crossing.

For plane-parallel initial conditions $$\vX(t,\vq) =: (X(t,q), q_y, q_z)$$ and $\vq=:(q,q_y,q_z)$. 
 Without loss of generality we consider the trajectories $(X(t,q),0,0)$ such that \eqref{CDMtrajectoriesEOM} becomes
\begin{align} \label{1DCDMtrajectoriesEOM}
\ddot X(q)  &=  G\,\rho_0 a \int \d{q'} \d q'_y \d q'_z \left\{ \frac{X(q) - q'}{\left[(X(q)- q')^2 + q_y'^2 + q_z'^2\right]^{3/2}}\right. \notag\\  
& \qquad \qquad   \left.
- \frac{X(q) -X(q')}{\left[ \left(X(q)- X(q')\right)^2 + q_y'^2 + q_z'^2\right]^{3/2}}  \right\}\,, 
 \end{align}
 where $d \eta = a^2 dt $ is superconformal time and the dot denotes throughout this section a derivative w.r.t. $\eta$ rather than $t$. Performing the $q_y'$ and $q_z'$ integrals gives
 \begin{align}
 \ddot X(q)  &=  2 \pi G\,\rho_0 a \int \d q'\,\,  \Big[ \mathrm{sgn}\big(X(q) - q'\big) \notag \\
 & \qquad \qquad \qquad \qquad - \mathrm{sgn}\big(X(q) - X(q')\big)   \Big]\,,
 \end{align}
 where $\mathrm{sgn}$ is $1$ if the argument is positive, $-1$ if it is negative and $0$ if the argument vanishes. 
 The first integral can be performed and we arrive at 
  \begin{align}
 \ddot X(q)  &=  4 \pi G\,\rho_0 a\, \Bigg\{ 
  X(q) -  \frac{1}{2}\int \d q'\,\,  \mathrm{sgn}\big[X(q) - X(q') \big] \Bigg\} 
\label{CDM1DEOM}
 \end{align}
 Before shell-crossing, that is, before any $q'$ other than $q$ exists such that $X(q)=X(q')$, 
the quantity $\mathrm{sgn}(X(q) - X(q'))$ is constant in time and therefore equals $\mathrm{sgn}(q - q')$. Then the second term can be integrated and we obtain 
   \begin{align} \label{ZeldoPancake}
 \ddot \varPsi^{\rm ZA}(q)  &=  4 \pi G\,\rho_0\, a\, \varPsi^{\rm ZA}(q) \,,
 \end{align}
where we used the displacement field $\varPsi(q) \equiv X(q)-q$ and attached the label ZA to make clear that the solution to \eqref{ZeldoPancake} is the Zel'dovich approximation.\footnote{It is an approximation to \eqref{CDMtrajectoriesEOM} in three dimensions and arises in Lagrangian perturbation theory. 
Only before shell-crossing $\varPsi^{\rm ZA}(q)$ is the exact solution in the one-dimensional case.} 
\begin{figure}[t]
\includegraphics[width=0.48\textwidth]{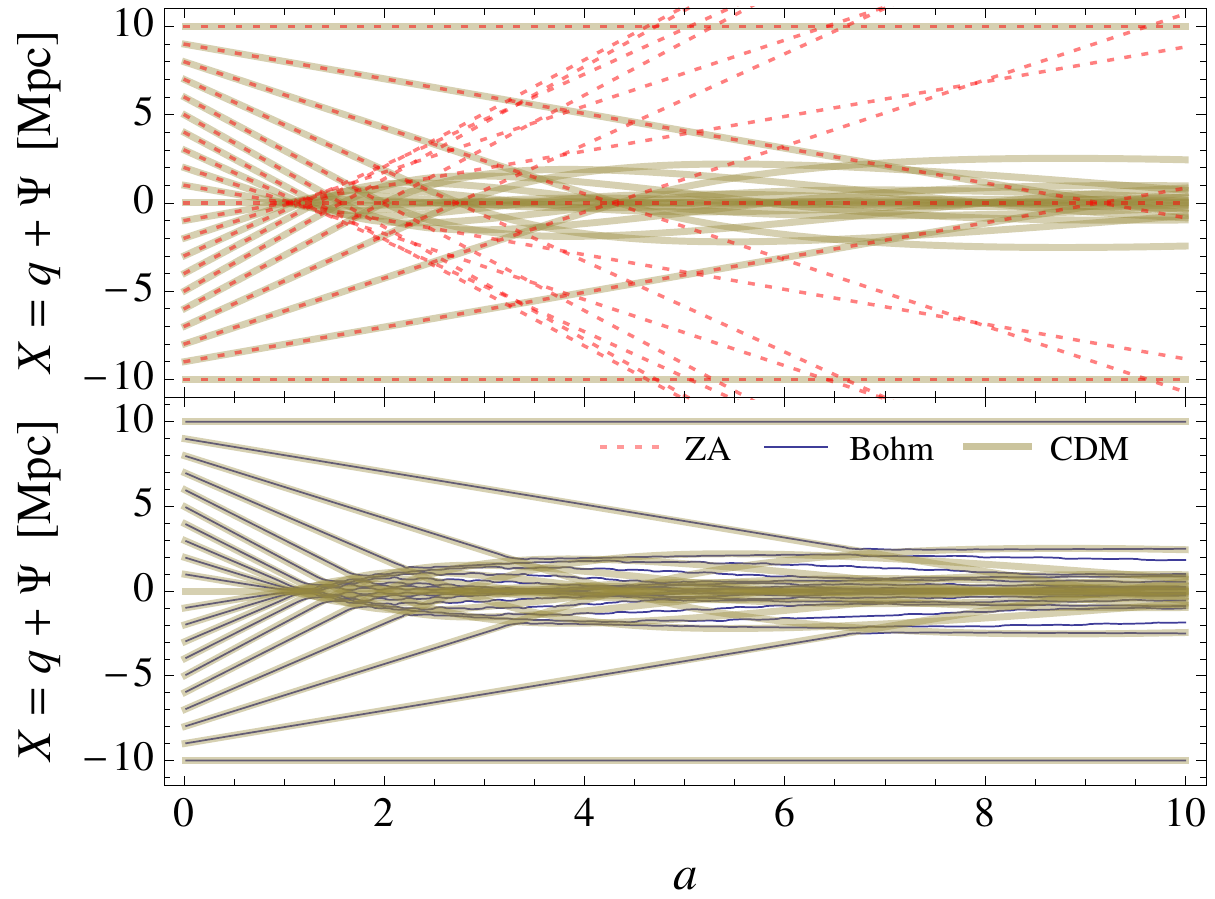}\\
\caption{Upper panel: \textit{red dotted} Zel'dovich trajectories \eqref{ZeldoPancake} and \textit{thick yellow} CDM trajectories  \eqref{CDMPancake}. 
 \\ 
 Lower panel: \textit{blue} Bohmian trajectories (B2) of \cite{UhlemannKoppHaugg2014} \textit{thick yellow} CDM trajectories \eqref{CDMPancake}. }
\label{fig:trajectories}
\end{figure}
 We can rewrite \eqref{CDM1DEOM} in terms of $\varPsi$ 
   \begin{align} \label{CDMPancake}
 \ddot \varPsi(q)  &=  4 \pi G\,\rho_0 a\ \Big(\varPsi(q)  + g[\varPsi](q) \big) \Big) \\
 g[\varPsi](q) & = q- \frac{1}{2}\int \d q'\,\,  \mathrm{sgn}\big(X(q) - X(q') \big) \notag
 \end{align}
 which makes it manifest that before shell-crossing, when  $g$ identically vanishes,  we recover the ZA,
 but once CDM undergoes shell-crossing and mixing, $g$ becomes non-zero and the ZA is no longer a solution  to \eqref{CDMPancake}. 
 
 We could have also arrived to \eqref{ZeldoPancake}  following the standard treatment of the Lagrangian formulation of CDM that assumes the existence of the inverse of $\partial_{q^i} X^j$ and makes use of it, see i.e.\,App.\,B of \cite{UhlemannKoppHaugg2014} or \cite{Buchert1996}. 
Such an assumption, however, does exclude the multi-stream regime from the outset 
and is therefore useless for our purposes. We thus see that the $g$-term missing  in the Zel'dovich approximation is the ``glue'' 
that distinguishes the dispersing and physically wrong behavior of the ZA trajectories (red dotted) from those of 
CDM (yellow thick) in the upper panel of Fig.\,\ref{fig:trajectories}.
\NEW{To solve \eqref{CDMPancake} we discretised the equation using 512 particles. An alternative method that relies on piecewise analytic solutions has been proposed and employed in one-dimensional simulations in \cite{GoudaNakamura1989,YanoGouda1998,TatekawaMaeda2001}.}

 In the lower panel we reproduced Fig.\,8 from \cite{UhlemannKoppHaugg2014} showing the Bohmian trajectories, the integral curves
 of the $\vnabla \phi(t,\vx)$, overplotted with CDM trajectories.
Note that although the Bohmian trajectories approximate those of the ZA 
before shell crossing and stick together in a similar fashion as CDM after shell crossing, 
these non-intersecting Bohmian trajectories are not of any direct physical relevance and in 
particular do not correspond to CDM trajectories after shell-crossing.
What the Bohmian trajectories reveal is that the ScM produces ``shell-crossing without shell-crossing'' in a similar fashion as Newton's cradle allows a ball to apparently cross the center of the cradle.
\begin{figure}[t]
\includegraphics[width=0.48\textwidth]{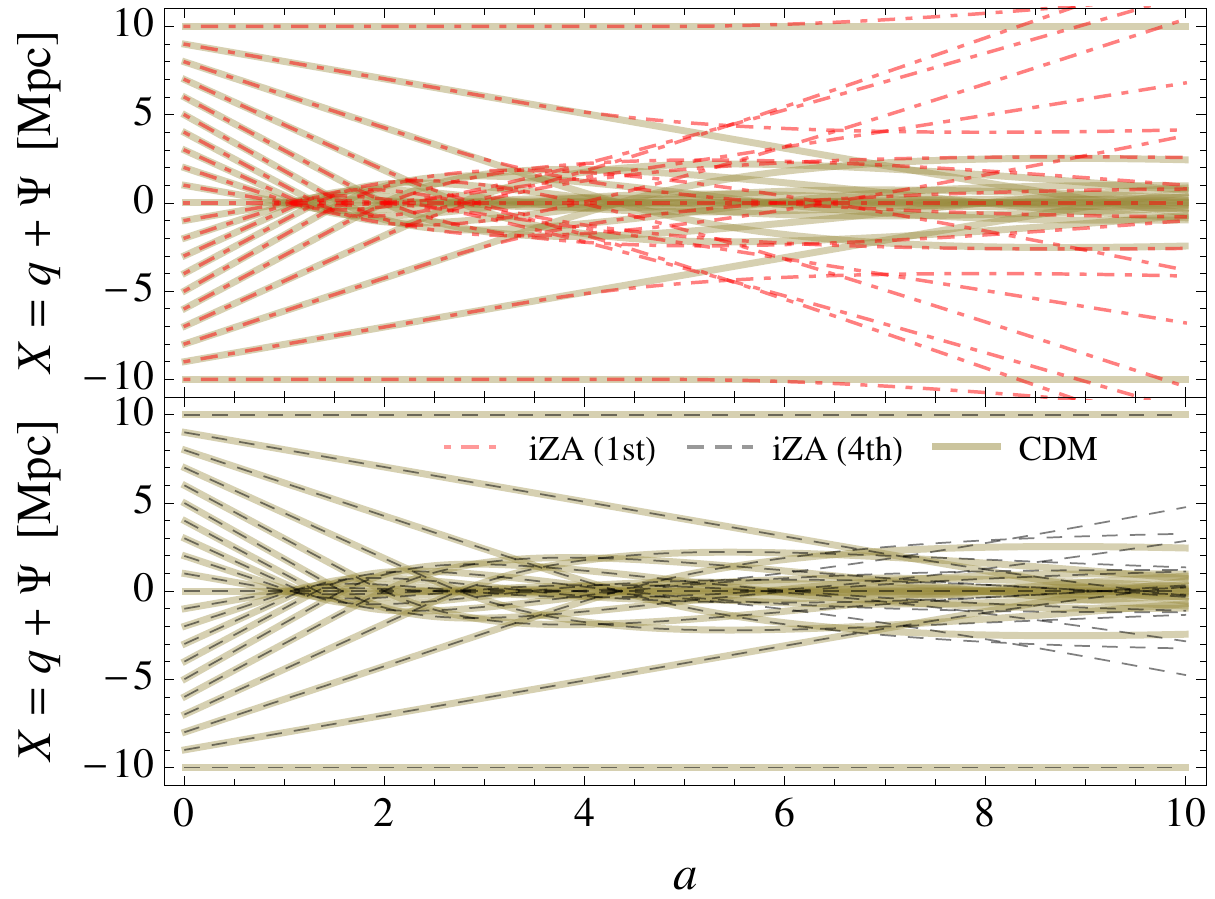}\\
Upper panel: \caption{\textit{red dot-dashed} is the 1st iteration of the  iZA, \eqref{CDMPancake} solved with fixed $g[\varPsi^{\rm ZA}]$. 
 \\ Lower panel: \textit{black dashed} is fourth iteration of the iZA, \eqref{CDMPancake} solved with fixed $g[\varPsi^{\rm iZA (3rd)}]$. In both panels are the \textit{thick yellow} 
curves CDM trajectories of  \eqref{CDMPancake}. }
\label{fig:iZA}
\end{figure}

Since very shortly after shell-crossing $\varPsi^{\rm ZA}$ and $\varPsi$ do not deviate much from each other it is interesting to investigate an iterated ZA (iZA) in which $g[\varPsi]$ in \eqref{CDMPancake} is replaced by $g[\varPsi^{\rm ZA}]$. 
This function can be pre-calculated and stored in an interpolation table $g[\varPsi^{\rm ZA}](t,q)$ such that the now again local equation 
   \begin{align} \label{iZAPancake}
 \ddot \varPsi^{\mathrm{iZA}(n)}(q)  &=  4 \pi G\,\rho_0 a\ \Big(\varPsi^{\mathrm{iZA}(n)}(q)  + g[\varPsi^{\mathrm{iZA}(n-1)}](q)  \Big) \,,
 \end{align}
 where $\varPsi^{\mathrm{iZA}(0)} = \varPsi^{\mathrm{ZA}}$,
 can be solved for $\varPsi^{\rm iZA (1st)}$. 
The such obtained $\varPsi^{\rm{iZA(1st)}}$ could then be used to calculate $\varPsi^{\rm{iZA(2nd)}}$, leading to the hierarchy \eqref{iZAPancake}. This iterative scheme to improve the ZA has been suggested before \cite{Tassev2011}, but to our knowledge, has never been explicitly tested.
A similar scheme in one dimension has been proposed in \cite{Colombi2015} and   successfully tested \cite{TaruyaColombi2017}.

We plot the first and fourth iteration of the iZA in Fig.\,\ref{fig:iZA}.
 Each iteration significantly delays the time at which iZA trajectories deviate from CDM in the multi-stream regime, such that this method indeed seems to converge to CDM.
It should be noted that this improvement of the ZA in one dimension is completely independent from any perturbative improvement of the ZA in three dimensions.
Indeed, if we expanded  \eqref{CDMPancake} perturbatively, the $g$-term vanishes at each order in perturbation theory.
Non-perturbative corrections mimicking the glue term $g$ have been suggested to improve mock simulations \cite{GurbatovSaichevShandarin1989,WeinbergGunn1990,GurbatovSaichevShandarin2012} and also to improve perturbation theory \cite{BuchertDominguezPerez-Mercader1999,Valageas2011,ValageasNishimichiTaruya2013}. 
 It would be interesting to investigate whether the iZA is a feasible approach to improve three-dimensional mock simulations or Lagrangian perturbation theory, and how this relates to other Lagrangian schemes that improve the ZA beyond shell-crossing like \texttt{COLA} \cite{TassevZaldarriagaEisenstein2013,TassevEisensteinWandeltEtal2015} and \texttt{PINOCCHIO} \cite{MonacoTheunsTaffoniEtal2002,MonacoTheunsTaffoni2002}.
 Recently an approach to calculate statistical quantities of the displacement field  based on the CDM \eqref{CDMtrajectoriesEOM} has been developed in \cite{McDonaldVlah2017} which does neither assume invertibility of $\partial_{q^i} X^j$ nor a Taylor expansion in $\v{\varPsi}$, and thus allows for a proper inclusion of multi-streaming effects.


\section{Derivation of \eqref{CDMtrajectoriesEOM}}
\label{sec:DerivationofCDMtrajectories}
The CDM potential is given by
\begin{equation}
\varPhi_\c(\vx) = - \frac{ G \rho_0}{a} \int \vol{3}{x'} \frac{n_\c(\vx') -1}{|\vx -\vx'|}
\end{equation}
and its gradient is
\begin{equation}
\vnabla_{\!\!x} \varPhi_\c(\vx) = \frac{ G \rho_0}{a} \int \vol{3}{x'} \left[n_\c(\vx') -1\right] \frac{\vx -\vx'}{|\vx -\vx'|^3}\,.
\end{equation}
We then split the integral into two parts as
\begin{equation} \label{acceleration1}
\vnabla_{\!\!x} \varPhi_\c(\vx) = \frac{ G \rho_0}{a} \left( \int \vol{3}{x'} \!\!n_\c(\vx')  \frac{\vx -\vx'}{|\vx -\vx'|^3} -\int \vol{3}{q'} \frac{\vx -\vq'}{|\vx -\vq'|^3} \right)\,,
\end{equation}
where we simply renamed the dummy variable $\vx'$ in the second integrand to $\vq'$.
Next, we switch to Lagrangian coordinates in the first integral. One has to be very careful with this 
variable transformation since it is not a diffeomorphism, because we allow for multi-streaming. 
We assume that Eulerian space $\mathbb{X}= \bigcup_{\mu=1}^{M} \mathbb{X}_\mu$ can be decomposed into $M$ subsets $\mathbb{X}_\mu$ in each of which $\vX_\mu$, the restriction of $\vX$ that takes values in $\mathbb{X}_\mu$, has a fixed number of streams $N_\mu$, neglecting the zero-measure subsets that are the caustics between those regions. Furthermore, the inverse image $\mathbb{Q}_\mu$ in Lagrangian space, defined by $\vX_\mu(\mathbb{Q}_\mu) = \mathbb{X}_\mu$  decomposes into $\mathbb{Q}_\mu =\bigcup_{\alpha_\mu=1}^{N_\mu}   \mathbb{Q}_{\alpha_\mu}$, such that  $\vX_{\alpha_\mu}$, the restriction of $\vX_\mu$ that has inverse image $\mathbb{Q}_{\alpha_\mu}$, is a diffeomorphism for all $\alpha_\mu$ and $\mu$. Therefore, the inverse of $\vX_{\alpha_\mu}(\vq)$ exists and is denoted by $\vQ_{\alpha_\mu}(\vx)$.

We now consider the integral
\begin{equation}
 \int_{\mathbb{X}} \vol{3}{x} \!\!n_\c(\vx)\,  h(\vx)
\end{equation}
with some function $h(\vx)$ and split it into the sum of $M$ Eulerian regions with fixed number of streams
\begin{equation}
 \sum_{\mu =1}^M \int_{\mathbb{X}_\mu} \vol{3}{x} \!\!n_\c(\vx)  h(\vx)\,.
\end{equation}
Next we insert the expression of $n_\c(\vx)$ in terms of sum over streams, \eqref{CDMMomentsstreams}, such that
\begin{equation}
 \sum_{\mu =1}^M  \sum_{\alpha_\mu=1}^{N_\mu}  \int_{\mathbb{X}_\mu} \vol{3}{x} h(\vx) \frac{1}{|\det \partial_{q^i} X^j_{\alpha_\mu}(\vq)|}\Bigg|_{\vq=\vQ_{\alpha_\mu}(\vx)}  \,,
\end{equation}
where we pulled out of the integral the sum over streams. Then we change coordinates from $\vx$ to $\vq$ in the integral. In the Lagrangian region $\mathbb{Q}_{\alpha_\mu}$, the volume element transforms as $\mathrm{d}^3\! x =\mathrm{d}^3 \!q \det\left(\partial_{q^i} X^j_{\alpha_\mu}(\vq) \right)$, such that we are left with  
\begin{equation}
 \sum_{\mu =1}^M  \sum_{\alpha_\mu=1}^{N_\mu}  \int_{\mathbb{Q}_{\alpha_\mu}} \vol{3}{q} h(\vX_{\alpha_\mu}(\vq))\ \mathrm{sgn}\left(\det \partial_{q^i} X^j_{\alpha_\mu}(\vq) \right)\,.
\end{equation}
Next, we push the $\alpha_\mu$-sum back into the integral 
\begin{equation}
 \sum_{\mu =1}^M  \int_{\mathbb{Q}_{\mu}} \vol{3}{q} h\left[\vX_\mu(\vq)\right] 
  \sum_{\alpha_\mu=1}^{N_\mu} \mathrm{sgn}\left(\det \partial_{q^i} X^j_{\alpha_\mu}(\vq) \right)\,,
\end{equation}
after which we pulled out of the sum $h[\vX_\mu(\vq)]$, since $h$ is a function of $\vx$ and therefore the same for all $\alpha_\mu$. 
The function $\mathrm{sgn}$ equals $1$ if the argument is positive, $-1$ if it is negative and $0$ if the argument vanishes. 
Since $\vX(t,\vq)$ belongs to the homotopy class of the identity as $\vX(t\!=\!0,\vq)=\vq$, 
the $\alpha_\mu$-sum in the integral is 1, see  \S4 of \cite{Milnor1988}.
We can finally absorb the sum over $\mu$ to get the desired result
\begin{equation} \label{EulerToLagrange}
 \int_{\mathbb{X}}  \vol{3}{x} n_\c(\vx) \, h(\vx)=  \int_{\mathbb{Q}} \vol{3}{q} h(\vX(\vq)) \,.
 \end{equation}
Applying this result to the first term in  \eqref{acceleration1} we obtain
\begin{equation}
\vnabla_{\!\!x} \varPhi_\c(\vx) = \frac{ G \rho_0}{a}  \int \vol{3}{q'} \left( \frac{\vx -\vX(t,\vq')}{|\vx -\vX(t,\vq')|^3} - \frac{\vx -\vq'}{|\vx -\vq'|^3} \right)\,,
\end{equation}
which leads to \eqref{CDMtrajectoriesEOM} upon evaluating the gravitational acceleration $-\vnabla_{\!\!x} \varPhi_\c$ at $\vx=\vX(t,\vq)$.

\section{Proof that $f_\c$ satisfies the Vlasov equation}
\label{sec:CDMsolvesVlasovProof}
Multiply $f_\c(t,\vx,\vu)$,  \eqref{fCDM}, by a smooth test function $h(\vu)$ and consider the time derivative of $\int \vol{3}{u} f_\c(t,\vx,\vu) h(\vu)$. 
With the help of \eqref{EulerToLagrange} and \eqref{CDMtrajectoriesEOM} it is easy to see  that $f_\c$ satisfies the Vlasov equation \eqref{VlasovPoissonEq}.

 \section{Differences of phase space distributions}
 \label{Diffoffs}
 \begin{widetext}
 
 Here we collect some results from \cite{UhlemannKoppHaugg2014} relevant for testing within the ScM,
 how well the ScM is expected to approximate the coarse-grained Vlasov or the Vlasov equation.  The coarse-grained Vlasov equation is
 \begin{subequations}
\label{cgVlasovEq}
\begin{eqnarray} 
\partial_t  \bar f&&= -\frac{\vu}{a^2 }\vnabla_{\! \!  x}  \bar f -\frac{\sigu^2}{a^2 }\vnabla_{\! \!  x}\vnabla_{\! \!  u}  \bar f  +\vnabla_{\! \!  x} \bar \varPhi \exp(\sigx^2\overleftarrow{\vnabla}_{\! \!  x}\overrightarrow{\vnabla}_{\! \!  x})\vnabla_{\! \!  u}  \bar f \label{cgVlasovEqa} \,
\end{eqnarray}
\end{subequations}
whereas evolution of the Husimi distribution is given by
\begin{subequations}
\label{cgWignerVlasov} 
\begin{align}
\partial_t   f_\H &= -\frac{\vu}{a^2}\vnabla_{\! \!  x}   f_\H -\frac{\sigu^2}{a^2 }\vnabla_{\! \!  x}\vnabla_{\! \!  u}   f_\H  +   \varPhi_\H \exp(\sigx^2\overleftarrow{\vnabla}_{\! \!  x}  \overrightarrow{\vnabla}_{\! \!  x})\frac{2}{\thbar}\sin\left(\frac{\thbar}{2}\overleftarrow{\vnabla}_{\! \!  x}  \overrightarrow{\vnabla}_{\! \!  u}\right)   f_\H \label{cgWignerVlasov1} \,.
\end{align}
\end{subequations}
If initially $f_\H(t_{\rm ini}) = \bar f(t_{\rm ini})$, then 
\begin{align}
\partial_t \left(  f_\H- \bar f\right)_{\rm ini} &=  \varPhi_\H(t_{\rm ini}) \exp(\sigx^2\overleftarrow{\vnabla}_{\! \!  x}  \overrightarrow{\vnabla}_{\! \!  x})\left(\frac{2}{\thbar}\sin\left(\frac{\thbar}{2}\overleftarrow{\vnabla}_{\! \!  x}  \overrightarrow{\vnabla}_{\! \!  u}\right)  -\overleftarrow{\vnabla}_{\! \!  x}  \overrightarrow{\vnabla}_{\! \!  u}\right) f_\H(t_{\rm ini}) \\
& = - \frac{\thbar^2}{24} \varPhi_\H(t_{\rm ini}) \left(\overleftarrow{\vnabla}_{\! \!  x}  \overrightarrow{\vnabla}_{\! \!  u}\right)^3 f_\H(t_{\rm ini}) + \mathcal{O}(\thbar^2 \sigx^2, \thbar^4)  \label{fHquantumCorr}
\end{align}
which we assumed to also hold at later times in \eqref{fHandfcgDiff} leading to the condition \eqref{fHGoodCondition1}. The smallest non-vanishing moment of the right-hand side of \eqref{fHquantumCorr} is for $n=3$. 
Therefore, once the simulation is in the multi-stream regime and condition \eqref{hbarCondition} no longer applies,
we can use the evolution equation of the third moment as an indicator of the goodness of the approximation 
of the coarse-grained Vlasov equation as in Sec.\,\ref{sec:HusHierarchy}.  If this condition does not hold then $\partial_t f_\H$ 
is sourced by ``quantum'' corrections \eqref{fHquantumCorr} and the correspondence with the coarse-grained Vlasov equation is lost.

Similarly from \eqref{VlasovEq} we see that 
\begin{align} \label{fcgCorr}
\partial_t \left( \bar f-  f\right)_{\rm ini} &=   -\frac{\sigu^2}{a^2 }\vnabla_{\! \!  x}\vnabla_{\! \!  u}  \bar f(t_{\rm ini})  +\vnabla_{\! \!  x} \bar \varPhi(t_{\rm ini}) \left( \exp(\sigx^2\overleftarrow{\vnabla}_{\! \!  x}\overrightarrow{\vnabla}_{\! \!  x}) -1\right)\vnabla_{\! \!  u}  \bar f(t_{\rm ini})  \\
&=  -\frac{\sigu^2}{a^2 }\vnabla_{\! \!  x}\vnabla_{\! \!  u}  \bar f(t_{\rm ini})  +\sigx^2 \nabla_{x_i}\nabla_{x_j}  \varPhi_\H \,\nabla_{x_i} \nabla_{u_j}  \bar f(t_{\rm ini})   + \mathcal{O}(\sigx^4)
\end{align}
leading to the condition \eqref{fHGoodCondition2}.
\end{widetext}

\newcommand{\apjl}{Astrophys. J. Letters}
\newcommand{\apjs}{Astrophys. J. Suppl. Ser.}
\newcommand{\mnras}{Mon. Not. R. Astron. Soc.}
\newcommand{\pasj}{Publ. Astron. Soc. Japan}
\newcommand{\apss}{Astrophys. Space Sci.}
\newcommand{\aap}{Astron. Astrophys.}
\newcommand{\physrep}{Phys. Rep.}
\newcommand{\mpla}{Mod. Phys. Lett. A}
\newcommand{\jcap}{J. Cosmol. Astropart. Phys.}

\normalem
\bibliography{2DVlasovScMbib}

\end{document}